\documentclass[12pt,epsf]{article}
\usepackage{amssymb,amsmath,amsbsy}
\usepackage{graphicx}

\newcommand{\be}{\begin{equation}}
\newcommand{\ee}{\end{equation}}
\newcommand{\bea}{\begin{eqnarray}}
\newcommand{\eea}{\end{eqnarray}}
\newcommand{\beas}{\begin{eqnarray*}}
\newcommand{\eeas}{\end{eqnarray*}}
\newcommand{\ba}{\begin{array}}
\newcommand{\ea}{\end{array}}
\newcommand{\tr}{\operatorname{tr}}

\def\identity{{\rlap{1} \hskip 1.6pt \hbox{1}}}

\def\href#1#2{#2}

\begin{document}
\begin{titlepage}
\hfill
\vbox{
    \halign{#\hfil         \cr
           } 
      }  
\vspace*{20mm}
\begin{center}
{\Large \bf Lectures on Gravity and Entanglement}

\vspace*{15mm}
\vspace*{1mm}
Mark Van Raamsdonk
\vspace*{1cm}
\let\thefootnote\relax\footnote{mav@phas.ubc.ca}

{Department of Physics and Astronomy,
University of British Columbia\\
6224 Agricultural Road,
Vancouver, B.C., V6T 1W9, Canada\\
\vspace*{0.2cm}}

\vspace*{1cm}
\end{center}
\begin{abstract}
The AdS/CFT correspondence provides quantum theories of gravity in which spacetime and gravitational physics emerge from ordinary non-gravitational quantum systems with many degrees of freedom. Recent work in this context has uncovered fascinating connections between quantum information theory and quantum gravity, suggesting that spacetime geometry is directly related to the entanglement structure of the underlying quantum mechanical degrees of freedom and that aspects of spacetime dynamics (gravitation) can be understood from basic quantum information theoretic constraints. In these notes, we provide an elementary introduction to these developments, suitable for readers with some background in general relativity and quantum field theory. The notes are based on lectures given at the CERN Spring School 2014, the Jerusalem Winter School 2014, the TASI Summer School 2015, and the Trieste Spring School 2015.
\end{abstract}

\end{titlepage}

\vskip 1cm

\section{Introduction}

Quantum mechanics is believed to be the basic underlying framework for the physics of our universe. It is the foundation for quantum field theory, which successfully describes the physics of electromagnetic, strong and weak interactions through the Standard Model of Particle Physics.
One of the great challenges in theoretical physics over the past decades has been trying to fit gravity, the most universal of all interactions, into this quantum mechanical framework. While apparently of little relevance for understanding everyday gravitational physics, coming up with a quantum theory of gravity is essential in order to understand some of the most fundamental questions in physics, including the physics of the big bang and the nature of black holes.

In our modern understanding provided by Einstein's general relativity, gravity refers to the dynamics of spacetime and its interaction with matter as governed by the Einstein Equation. To obtain a quantum version of the theory, the most direct route would seem to be applying standard quantization rules directly to the variables describing spacetime geometry. However, this approach typically runs into various troubles, and there are now deep reasons to believe that the correct route to a quantum theory of gravity must be fundamentally different. Foremost among these is the notion that gravity is ``holographic'' \cite{'tHooft:1993gx,Susskind:1994vu}. That is, given a region of spacetime, the number of degrees of freedom is not proportional to the volume of the region (as is the case in conventional quantum field theories) but rather to the area of the region's boundary. In this case, quantizing gravity as a local field theory would seem bound to fail, since these local fluctuations of the spacetime geometry cannot be the fundamental degrees of freedom in the same way that local fluctuations of the electromagnetic field represent the fundamental degrees of freedom in electromagnetism.

Great progress in our understanding of quantum gravity has come over the past few decades from string theory, culminating in the first complete non-perturbative models of quantum gravity provided by {\it the AdS/CFT correspondence} \cite{maldacena1997large}. These realize the holographic principle directly: the quantum gravitational theories are defined as ordinary non-gravitational quantum theories (typically quantum field theories) on a fixed lower-dimensional spacetime. How does this work? The basic idea is that each state of the ordinary quantum system encodes all the information about the state of the higher-dimensional gravitational system. As examples, the vacuum state typically corresponds to an empty spacetime, states with some low-energy excitations might correspond to the spacetime with a few gravitational waves, while very highly excited states of the quantum system might correspond to a spacetime with a massive black hole.

Since the correspondence was proposed by Maldacena in 1997, there has accumulated a great deal of evidence that the conjecture is correct. However, no proof for the correspondence exists, and there remain a number of very basic questions, such as
\begin{itemize}
\item
How and why do spacetime and gravity emerge from CFT physics?
\item
Precisely how is the spacetime geometry and other local gravitational physics encoded in the CFT state?
\item
Which CFT states correspond to spacetimes with a good classical description (as opposed to e.g. quantum superpositions of different spacetimes).
\item
What are necessary and sufficient conditions for a theory to have a gravity dual?
\end{itemize}
In recent years, it has become clear that to better understand these questions, it is very useful to think about the CFT from the perspective of quantum information theory. There is now a significant amount of evidence that the structure of quantum entanglement in the CFT state is related directly to the geometrical structure of the dual spacetime. Natural quantum-information theoretic quantities such as entanglement entropy and relative entropy map directly over to natural physical quantities in the gravitational theory. Even aspects of gravitational dynamics can be seen to emerge directly from the physics of entanglement. In these lectures, I will attempt to provide an elementary introduction to some of these exciting recent developments.

\section{The very basics of AdS/CFT}

For completeness, we will start with a very rudimentary introduction to AdS/CFT. Some useful reviews include \cite{aharony2000large, McGreevy:2009xe} but there are many others. The basic idea here is that certain non-gravitational quantum systems, defined on fixed spacetimes, are equivalent to quantum gravitational theories whose states correspond to different spacetimes with specific asymptotic behavior.\footnote{Alternatively, we could say that they nonperturbatively {\it define} these dual gravitational theories, since there is generally not a complete alternative definition available.} Each state in the non-gravitational system corresponds to a state in the dual gravitational theory, and each observable in the non-gravitational system corresponds to some observable in the gravitational theory. For both states and observables, the interpretation on the two sides of the correspondence can be completely different. One important exception is total energy: the energy of a CFT state corresponds to the total energy of the spacetime (measured at the classical level by the ADM mass).

\begin{figure}
\centering
\includegraphics[width=0.5 \textwidth]{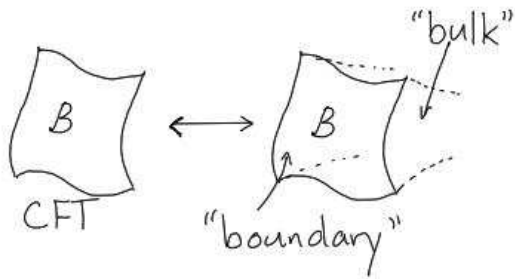}
\caption{Basic AdS/CFT. States of a CFT on some fixed spacetime ${\cal B}$ correspond to states of a gravitational theory whose spacetimes are asymptotically locally AdS with boundary geometry ${\cal B}$.}
\label{adscft}
\end{figure}

In the examples that we will consider throughout these lectures, the non-gravitational system will be a conformal field theory (CFT), i.e. a quantum field theory with conformal invariance, defined on some fixed spacetime ${\cal B}$.\footnote{There are other examples in which the non-gravitational theory is a non-conformal field theory, or even an ordinary quantum mechanical theory with a large number of quantum variables.} Often ${\cal B}$ will be Minkowski space $R^{d-1,1}$ or a sphere plus a time direction $S^{d-1} \times R$, but we are free to choose any geometry.

\begin{figure}
\centering
\includegraphics[width=\textwidth]{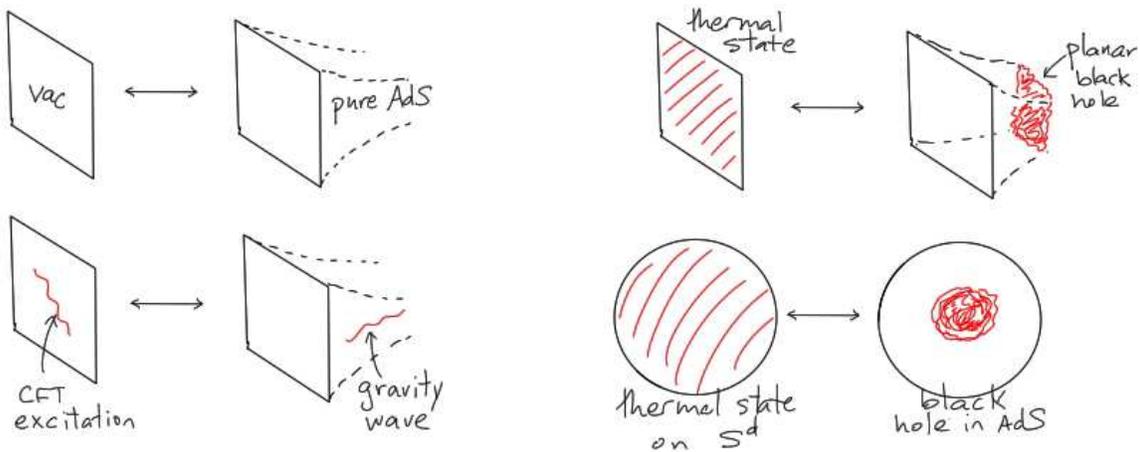}
\caption{Different CFT states correspond to different asymptotically AdS geometries.}
\label{adscft2}
\end{figure}

If the CFT is ``holographic,'' i.e. if there is a dual gravitational theory, the various quantum states of the CFT are each associated with some state of this dual theory, as illustrated in figure \ref{adscft2}. The various states may describe different spacetime geometries, but for a specific CFT, the asymptotic behavior of each of these spacetimes is the same. For a CFT on Minkowski space $R^{d-1,1}$, the vacuum state of the theory corresponds to $(d+1)$-dimensional Anti-de-Sitter spacetime (AdS) with a Minkowski space boundary, which may be described by the metric
\be
ds^2 = {\ell^2 \over z^2} ( dz^2 + dx_\mu dx^\mu) \; .
\ee
This is a maximally symmetric negatively-curved spacetime with curvature set by the length scale $\ell$. The spatial geometry is hyperbolic space. The spacetime has a boundary at $z=0$ which lies at an infinite proper distance from any point in space, but which light rays can reach and return from in a finite proper time. More general excited states of the CFT are dual to different geometries which approach this geometry as $z \to 0$.\footnote{More precisely, we expect that only a subset of CFT states correspond to states of the gravitational theory with a simple classical geometrical description. For example, a quantum superposition of two states corresponding to two different geometries would describe a quantum superposition of geometries.} Explicitly, we can describe these more general geometries as
\be
\label{FG}
ds^2 = {\ell^2 \over z^2} ( dz^2 + \Gamma_{\mu \nu}(x,z) dx^\mu dx^\nu) \; ,
\ee
where for small $z$,
\be
\label{asympt}
\Gamma_{\mu \nu}(x,z) = \eta_{\mu \nu} + {\cal O}(z^d) \; .
\ee
This description represents a choice of coordinates known as Fefferman-Graham coordinates. For states of a CFT defined on a more general spacetime ${\cal B}$, the story is similar, except that $\eta_{\mu \nu}$ in (\ref{asympt}) is replaced by the metric describing ${\cal B}$.

For small perturbations to the vacuum state of the CFT, the corresponding geometries should be represented by small perturbations to AdS, while for high-energy excited states, the corresponding spacetimes can have significantly different geometry and even topology. One important example is the case of a thermal state of the CFT. For the Minkowski-space CFT, the geometry corresponding to a thermal state is the planar $AdS$ black hole. The story is a little more interesting for CFTs defined on a sphere. In this case, there is a deconfinement phase transition in the CFT as the temperature is increased, with the low-temperature phase dual to a gas of particles in AdS and the high-temperature phase dual to a spherically-symmetric AdS-Schwarzschild black hole \cite{Witten:1998zw}.

It is interesting in general to understand which CFTs are holographic. There are specific examples (e.g. maximally symmetric large $N$ Yang-Mills theories) for which lots of evidence for a dual gravitational description exists. However, we don't have a set of necessary and sufficient conditions to tell us whether any particular CFT is holographic. It is believed that having a gravity dual that looks like Einstein gravity coupled to matter requires a CFT with a large number of degrees of freedom (``large N'') and strong coupling. There are also more detailed conditions on the spectrum of states/operators of the theory; roughly, these conditions say that the CFT should have only as many low-energy states as we would expect for a theory of gravity on asymptotically AdS spacetime.  On the other hand, it is plausible that any UV-complete theory of quantum gravity on AdS can be associated with a CFT, because the gravitational observables could be used to define a conformal field theory.

\section{Entropy and geometry}

The recent connections between quantum information theory and gravitational physics actually have their roots in the early 1970s with the work of Jacob Bekenstein. In thinking about black holes in classical general relativity, Bekenstein and others \cite{bekenstein1973black,bardeen1973four} realized that the physics of black hole horizons shares qualitative features with the physics of entropy in thermodynamics. Specifically, horizon area is non-decreasing with time (in the classical limit), and it obeys a relation akin to the first law of thermodynamics, $dE = T dS$ with black hole mass playing the role of energy, $2 \pi T$ identified with the surface gravity $\kappa$ (a measure of curvature at the horizon) and entropy identified precisely with area as
\be
\label{entropy_area}
S \leftrightarrow {\rm Area \over 4 G_N} \; .
\ee

Based on these observations, Bekenstein made the bold suggestion that the qualitative similarities were not a coincidence; he postulated that black holes are thermodynamic systems and that the area of the horizon {\it is} the entropy of the black hole. This was quickly confirmed by Hawking's demonstration \cite{hawking1975particle} that black holes radiate a thermal spectrum of particles exactly consistent with the predicted temperature $T = \kappa / 2 \pi$. However, the underlying statistical interpretation of the entropy as a count of states remained mysterious for decades, since there did not exist a framework to understand the microstates of black holes.

Jumping ahead to the present time, we can now understand the microscopic interpretation of black hole entropy for black holes in Anti-de-Sitter spacetime, since the AdS/CFT correspondence gives us a complete definition of the underlying quantum theory. From the previous section, a Schwarzschild black hole in Anti-de-Sitter space is identified with a high-energy thermal state of the corresponding CFT on a sphere. Such a theory has a discrete spectrum of energy eigenstates $|E_i \rangle$, and the thermal state corresponds to the usual canonical ensemble. This has a well-defined entropy counting these microstates and (up to a numerical factor which is difficult to compute due to the strong coupling in the CFT) the relation between entropy and temperature for the CFT system is exactly what was predicted for the black hole.

To summarize, thermal states of holographic CFTs are dual to Schwarzschild black holes in AdS, and the CFT entropy $S$ corresponds to the horizon area of the black hole (divided by $4 G_N$). As it turns out, this particular connection between entropy and geometry is just the tip of the iceberg. It admits a massive and beautiful generalization proposed in 2006 by Ryu and Takayanagi. According to our present understanding, for {\it any} CFT state corresponding to some asmptotically AdS spacetime with or without a black hole, and for {\it any subsystem} of the CFT, the entropy of the subsystem corresponds to a the area of a particular surface in the corresponding spacetime. Before describing this relation in detail and discussing its implications, it will be useful to review in more detail the description of quantum subsystems, and the associated measures of entropy.

\subsection{Quantum subsystems and entanglement}

Consider a quantum system with a subsystem $A$.\footnote{For a more complete review of the basics of quantum information theory, see \cite{nielsen2010quantum}.} We will denote the complement of this subsystem by $\bar{A}$. Then the Hilbert space may be decomposed as
\be
{\cal H} = {\cal H}_A \otimes {\cal H}_{\bar{A}}
\ee
Given a state $|\Psi \rangle \in {\cal H}$ of the full system, we can ask: ``What is the state of the subsystem $A$?'' Naively, we might think that it is possible to find some state $|\psi^A \rangle \in {\cal H}_A$ that captures all information about the subsystem. We might demand that for every operator ${\cal O}_A$ acting on ${\cal H}_A$ alone,
\be
 \langle \psi^A |{\cal O}_A |\psi^A \rangle = \langle \Psi |{\cal O}_A \otimes \identity |\Psi \rangle  \; .
\ee
However, for general $|\Psi \rangle$, there does not exist such a state $|\psi^A \rangle$. In the context of the larger system, the state of the subsystem is not described by any single state in the Hilbert space ${\cal H}_A$.

\subsubsection*{Ensembles of quantum states}

To properly describe the subsystem, we need to use the idea of an ENSEMBLE of states, alternatively known as a MIXED STATE (as opposed to a PURE STATE i.e. a single Hilbert space vector). That is, we consider a collection $\{ (|\psi_i \rangle, p_i)\}$ of orthogonal states and associated probabilities. We define the expectation value of an operator ${\cal O}$ in the ensemble to be the average of the expectation values for the individual states, weighted by the probabilities,
\be
\langle {\cal O} \rangle_{ensemble} \equiv \sum_i p_i \langle \psi_i |{\cal O} |\psi_i \rangle \; .
\ee
We can think of the probabilities $p_i$ as representing some classical uncertainty about the state of the system.

It turns out that for a multipart system, given any state $|\Psi \rangle$ of the entire system (or, more generally an ensemble of states for the full system), we can always find an ensemble of states for subsystem $A$ such that all expectation values of operators ${\cal O}_A$ are reproduced, i.e. such that
\be
\label{dmcondition}
\langle \Psi |{\cal O}_A \otimes \identity |\Psi \rangle = \sum_i p_i  \langle \psi_i^A |{\cal O}_A |\psi_i^A \rangle \; .
\ee
We will prove this below by explicitly constructing the ensemble from $|\Psi \rangle$ .

\subsubsection*{The density matrix}

Given such an ensemble, we can define an associated operator
\be
\label{density_matrix}
\rho_A \equiv \sum_i p_i |\psi_i^A \rangle \langle \psi_i^A | \;
\ee
known as the DENSITY OPERATOR or DENSITY MATRIX for the subsystem. The density matrix is a Hermitian operator with unit trace and non-negative eigenvalues $p_i$. In fact, any operator with these properties can be used to define an ensemble by taking $|\psi_i \rangle$ and $p_i$ to be the orthogonal eigenvectors and eigenvalues of the matrix.\footnote{When some of the $p_i$s coincide, we need to make a choice for the orthogonal eigenvectors in the subspace with eigenvalue $p_i$. Expectation values of operators in the ensemble do not depend on this choice, so all such ensembles are equivalent.} Thus, the density matrix provides an equivalent mathematical representation of an ensemble. To compute the expectation value of an operator using the density matrix, we simply take a trace:
\be
\langle {\cal O}_A \rangle  = \tr({\cal O}_A \rho_A ) = \sum_i p_i  \langle \psi_i^A |{\cal O}_A |\psi_i^A \rangle \; .
\ee

\subsubsection*{Calculating the density matrix for a subsystem}

 Starting from a state $\Psi$ for the full system, it is straightforward to determine the density matrix corresponding to a subsystem and thus the associated ensemble. Given some basis $\{|\psi_n \rangle \}$ for $A$ and $\{|\psi_N \rangle \}$ for $\bar{A}$, we can represent the state of the full system as
\be
|\Psi \rangle = \sum_{n,N} c_{n,N} |\psi_n \rangle \otimes |\psi_N \rangle \; ,
\ee
where $c_{c,N}$ are complex coefficients satisfying the normalization condition
\be
\sum_{n,N} |c_{n,N}|^2 = 1 \; .
\ee
Then the operator $\rho = |\Psi \rangle \langle \Psi |$ represents the density matrix for the whole system. Taking the expectation value of an operator ${\cal O}_A \otimes \identity$ acting on our subsystem, we have
\bea
\tr(({\cal O}_A \otimes \identity) \rho ) &=&  \tr(\sum_{n,N} \sum_{m,M} c^*_{m,M} c_{n,N} {\cal O}_A |\psi_n \rangle \langle \psi_m | \otimes |\psi_N \rangle  \langle \psi_M | )  \cr
&=& \tr( \sum_{n,m} \sum_N c^*_{m,N} c_{n,N} {\cal O}_A |\psi_n \rangle \langle \psi_m |)  \cr
&=& \tr({\cal O}_A \rho_A)
\label{dmcalc}
\eea
where
\be
\label{partial_trace}
\rho_A  \equiv \sum_{n,m} \sum_N c^*_{m,N} c_{n,N} |\psi_n \rangle \langle \psi_m | \equiv \tr_{\bar{A}} \rho \; .
\ee
The calculation (\ref{dmcalc}) shows that for the density matrix $\rho_A$ defined by the operation (\ref{partial_trace}), known taking the PARTIAL TRACE over the subsystem $\bar{A}$, the property (\ref{dmcondition}) holds. This proves the claim that any quantum subsystem can always be represented by an ensemble.

It is easy to check that equation (\ref{partial_trace}) also defines a subsystem density matrix (or REDUCED DENSITY MATRIX) with the desired properties in the case when the full system is in an ensemble.

\subsubsection*{Thermodynamic ensembles and entropy}

The notion of an ensemble is familiar from quantum statistical mechanics.

The MICROCANONICAL ENSEMBLE is defined as $\{(|E_i \rangle, p_i = 1/n)\}$, consisting of all energy eigenstates of a system within some small range of energies $[E, E + dE]$, each weighted with equal probability. We consider this ensemble when it is desired to understand the expected values of various physical quantities when only the overall energy for some closed system is known, but not the precise state. Our ignorance of this ``microstate'' can be quantified by ENTROPY as\footnote{We use units for which the Boltzmann constant $k_B$ is equal to one.}
\be
S = \log n \; .
\ee
This serves as a measure of the number of microstates in the given energy range; the logarithm is chosen so that the entropy of a non-interacting multipart system is the sum of the entropies for the parts. We can think of each state as giving an individual contribution
\be
S_{state} = {1 \over n} \log n = - p \log p
\ee
to the entropy. Assuming that this contribution to the entropy from such a state is the same for more general ensembles where the probabilities are not all equal (i.e. that the entropy is extensive in the space of states), we obtain the more general definition\footnote{The latter formula for the entropy in terms of the density matrix is usually referred to as the VON NEUMANN ENTROPY of a density matrix.}
\be
\label{entropy}
S = - \sum_i p_i \log p_i = -\tr(\rho_A \log \rho_A)\;
\ee
valid for any ensemble. This general formula allows us to associate an entropy to the ensemble describing any quantum subsystem. This serves as a measure of the classical uncertainty arising from the mixed nature of the subsystem state.

Another ensemble familiar from quantum statistical mechanics is the CANONICAL ENSEMBLE or THERMAL STATE of a system. This represents a system $A$ weakly coupled to a heat bath which makes up the remainder $\bar{A}$ of the full system. The ensemble can be defined by maximizing the entropy (\ref{entropy}) subject to some fixed expectation value for the energy of the subsystem,
\be
\label{enexp}
\tr(\rho_A H_A) = E \; .
\ee
In terms of the energy eigenstates $|E_i \rangle$ of the Hamiltonian $H_A$, the resulting ensemble is $\{(|E_i \rangle, p_i = e^{-\beta E_i}/Z)\}$ where $Z = \sum_i e^{-\beta E_i}$ for $\beta$ chosen to ensure (\ref{enexp}). This parameter defines the temperature of the system via $\beta = 1/T$.

\subsubsection*{Entanglement}

The need to invoke ensembles as a description of quantum subsystems is directly linked to the notion of QUANTUM ENTANGLEMENT. Indeed, we can define entanglement by saying that a subsystem $A$ is entangled with the rest of the system if the ensemble describing it has probabilities $\{ p_i \} \ne {1}$. When the subsystem is not entangled, there is a single pure state $|\psi^A \rangle \in {\cal H}_A$ that describes it. In this case, we can write the state of the full system (assuming it is pure) as
\be
|\Psi \rangle = |\psi^A \rangle \otimes |\psi^{\bar{A}} \rangle\; .
\ee
Thus, entanglement can alternatively be defined as the failure of the full system to be representable as a product state.

\subsubsection*{Measures of entanglement}

Some entangled states are more entangled than others. For example, in a system of two spins, the state
\be
A |\uparrow \rangle \otimes |\uparrow \rangle + B | \downarrow \rangle \otimes | \downarrow \rangle
\ee
is not entangled if $A$ or $B$ vanish, and entangled otherwise, but it is sensible to say that the state is more entangled when the magnitudes of $A$ and $B$ are similar than when the state is very close to one of the unentangled states. It is useful to come up with measures to quantify the degree of entanglement for a subsystem.

Since having entanglement is the same as having classical uncertainty about the state of the subsystem, one natural measure of entanglement is simply the subsystem entropy (\ref{entropy}) that quantifies this classical uncertainty. Since it also serves as a measure of entanglement, this subsystem entropy is alternatively known as ENTANGLEMENT ENTROPY.

A more general set of quantities useful in characterizing entanglement are the RENYI ENTROPIES, defined as
\be
\label{Renyi}
S_\alpha = {1 \over 1 - \alpha} \log \sum_i p_i^\alpha = {1 \over 1 - \alpha} \log \tr( \rho^\alpha) \; .
\ee
These are typically defined for integer $\alpha$, but by considering general real values of $\alpha$, we can recover the entanglement entropy in the limit $\alpha \to 1$. The Renyi entropies for the integer values of $\alpha$ tend to be easier to compute than entanglement entropy. Knowing the Renyi entropies for all integer values of $\alpha$ up to the dimension of the Hilbert space is equivalent to knowing the full set of probabilities $\{p_i\}$, also known as the ENTANGLEMENT SPECTRUM.\footnote{The explicit relation is via the characteristic polynomial $\prod_i (\lambda - p_i) = \det( \lambda \identity - \rho) = \lambda^D - \lambda^{D-1} \tr (\rho) + \dots$.}

\subsubsection*{Purifications}

We have seen that given any pure state $|\Psi \rangle$ of a quantum system, any subsystem $A$ can be described by a density matrix $\rho_A$ or equivalently an ensemble $\{(|\psi^A_i \rangle, p_i)\}$. It is sometimes useful to consider the reverse question: given an ensemble $\rho_A$ for a quantum system, can we find a pure state of some larger system such that $\rho_A$ is the reduced density matrix for the subsystem $A$?

In general there are an infinite number of such PURIFICATIONS. For an ensemble $\rho = \{(|\psi^A_i \rangle, p_i)\}$ in a Hilbert space ${\cal H}_A$, we can describe a general purification by
\be
\label{Schmidt}
|\Psi \rangle = \sum_i \sqrt{p_i} |\psi^A_i \rangle \otimes |\psi^B_i \rangle
\ee
where $\{|\psi^B_i \rangle\}$ are an orthogonal set of states in some Hilbert space ${\cal H}_B$ whose dimension is at least as large as the number of non-zero eigenvalues of $\rho$. The state represented in this form is known as a SCHMIDT DECOMPOSITION; it is possible to represent any state of a combined system in this way.

Starting from the expression (\ref{Schmidt}) it is useful to note that the ensemble describing the subsystem $B$ has precisely the same probabilities (i.e. entanglement spectrum) as the ensemble describing $A$. It follows immediately that the entanglement entropy is the same for a subsystem and its complement when the full system is in a pure state.

One example of the idea of a purification is the idea that a thermal state (i.e. a system in the canonical ensemble) arises by considering the system weakly coupled to a much larger system known as a heat bath. The full system including the bath is taken to be a pure state and the entropy of the thermal ensemble can be understood as measuring entanglement with the bath.

It is sometimes useful to consider a simpler purification of the canonical ensemble, obtained by choosing the purifying system to be a copy of the original system and considering the state
\be
\label{TFD}
{1 \over Z} \sum_i e^{- \beta E_i /2} |E_i \rangle \otimes |E_i \rangle \; .
\ee
This state, known as the THERMOFIELD DOUBLE state, is precisely symmetrical between the two subsystems, giving a thermal state with the same temperature upon reduction to either subsystem.

\subsection{Two-sided black holes in AdS/CFT}

Let's now return to thinking about gravity and the AdS/CFT correspondence. We mentioned earlier that a Schwarzschild black hole in AdS is described via AdS/CFT by a high-energy thermal state of the CFT on a sphere, and that the area of the black hole horizon can be identified with the entropy of the CFT.

Various observables computed in the CFT thermal state tell us about the black hole spacetime. It is interesting to be more precise and understand which part of the black hole we can learn about. Is it just the region outside the horizon, some of the physics behind the horizon, or the entire  maximally extended Schwarzschild geometry (which includes two asymptotic regions connected by a wormhole)?

\begin{figure}
\centering
\includegraphics[width=\textwidth]{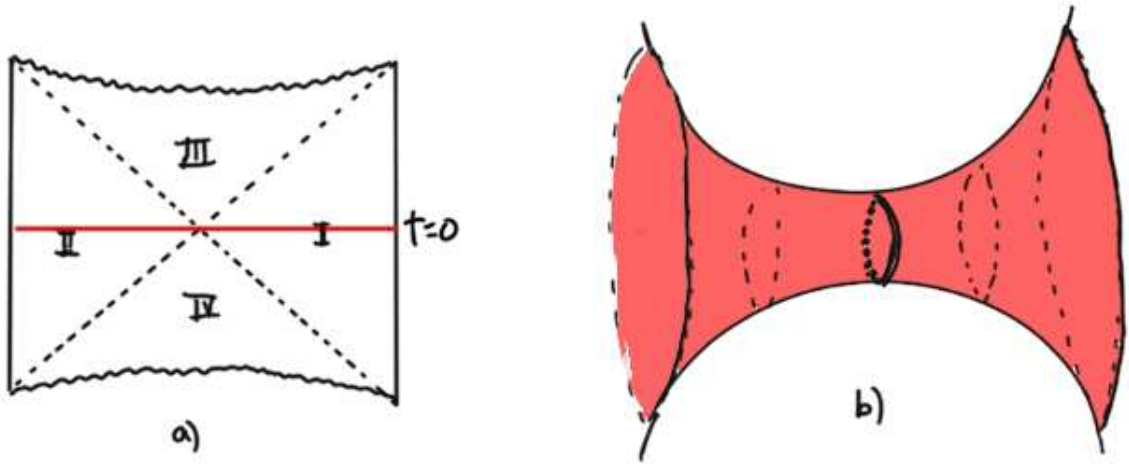}
\caption{Depictions of the maximally extended AdS-Schwarzschild black hole: a) Penrose (conformal) diagram for the spacetime, with exterior regions I and II and interior regions III and IV behind the horizon (dashed); b) spatial geometry of the $t=0$ slice (shown in red in a)), showing the horizon as the minimal area surface dividing the space into two parts each with one asymptotically AdS region. }
\label{MaxBH}
\end{figure}

In a 2001 paper \cite{Maldacena:2001kr}, Maldacena argued that the maximally extended spacetime, depicted in figure \ref{MaxBH}, is most naturally associated not with the thermal state of a single CFT, but rather with the thermofield double state (\ref{TFD}) of a two-CFT system.\footnote{It is possible to give a direct path-integral argument for this. See \cite{Balasubramanian:2014hda} and references therein.} The geometry has two asymptotic regions, each with its own boundary sphere and its own black hole exterior. Correspondingly, the thermofield double state invoves two separate CFTs on a sphere, each defined on a sphere and in a thermal state. Further, this special purification of the thermal state is symmetrical between the two systems, like the extended black hole geometry.

\begin{figure}
\centering
\includegraphics[width = 0.7 \textwidth]{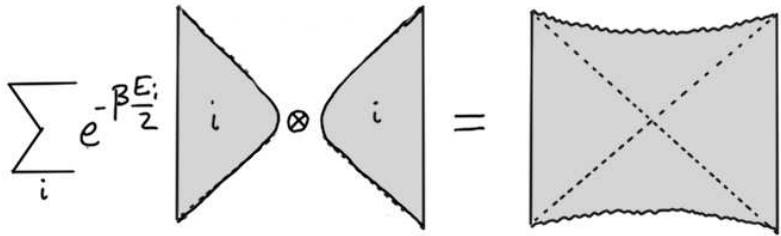}
\caption{Gravity interpretations for the thermofield double state in a quantum system defined by a pair of noninteracting CFTs on $S^d$ times time. A particular quantum superposition of disconnected spacetimes gives a connected spacetime.}
\label{qsup}
\end{figure}

The proposal of Maldacena is very natural but has dramatic implications. The individual terms in the superposition (\ref{TFD}) are product states in a system of two non-interacting CFTs. In these states, the two theories have absolutely nothing to do with one another, so in the gravity picture, these states must correspond to two completely separate asymptotically AdS spacetimes. On the other hand, the quantum superposition of these states in (\ref{TFD}) apparently corresponds to the extended black hole, where the two sides of the geometry are connected by smooth classical spacetime in the form of a wormhole. The remarkable conclusion (emphasized in \cite{VanRaamsdonk:2009ar,VanRaamsdonk:2010pw}) is that by taking a specific quantum superposition of disconnected spacetimes, we obtain a connected spacetime, as depicted in figure \ref{qsup}. Alternatively, we can say that by entangling the degrees of freedom underlying the two separate gravitational theories in a particular way, we have glued together the corresponding geometries!

The extended black hole picture gives us a new way to think about the entropy-area connection for black holes. In the thermofield double state, the black hole entropy is the entropy of a single CFT subsystem or the entanglement entropy measuring the entanglement of the two subsystems with each other. On the gravity side, the horizon is a surface that divides the spatial geometry into two parts, with each part containing one boundary sphere. It is the unique such surface in the spacetime that extremizes the area functional. Thus, denoting the two CFT subsystems as $A$ and $\bar{A}$, we can restate the entropy-area connection for the black hole as follows: {\it the entropy of the subsystem $A$ (or the entanglement entropy of $A$) corresponds to the area of the extremal-area surface dividing the bulk geometry into two parts with boundaries $A$ and $\bar{A}$}. This version leads us immediately to the remarkable generalization by Ryu and Takayanagi, which postulates this to hold for any region $A$ in any state with a dual geometry.

\subsection{The Ryu-Takayanagi formula}

In the context of AdS/CFT, the Bekenstein formula provides a geometrical interpretation for the total entropy of a CFT in a high-energy thermal state, identifying it with the area of the horizon of the black hole in the dual spacetime.  The formula proposed by Ryu and Takayanagi \cite{ryu2006holographic} (and generalized to a covariant version by Hubeny, Rangamani, and Takayanagi \cite{hubeny2007covariant}) suggests an interpretation for the entropy of any spatial subsystem of the CFT, for any CFT state associated with some classical spacetime.

To state the proposal, consider a holographic CFT defined on a spacetime geometry ${\cal B}$. We suppose that the CFT is in a state $|\Psi \rangle$ associated with a classical dual geometry $M_\Psi$. We now consider an arbitrary spatial subsystem $A$ of the CFT, defined by first choosing a spatial slice $\Sigma_{\cal B}$ of ${\cal B}$ and then choosing a subset $A \subset \Sigma_{\cal B}$ of this slice. The spatial region $A$ can be connected or a union of disconnected regions. Since the boundary geometry $\partial M_\Psi$ of $M_\Psi$ is the same as ${\cal B}$, we can define regions on $\partial M_\Psi$ corresponding to $\Sigma_{\cal B}$, $A$, and $\bar{A} \subset \Sigma_{\cal B}$ (and we will use the same letters to refer to these).

Now, let $S_A$ be the entropy of the subsystem $A$ i.e. the entanglement entropy measuring the entanglement of fields in $A$ with the the rest of the system. The covariant version of the Ryu-Takayanagi proposal states that this entropy equals the area of a certain codimension-2 surface $\tilde{A}$ in $M_\Psi$ (i.e. $d-1$-dimensional for a geometry that is asymptotically $AdS_{d+1}$)
\be
\label{HRT}
S(A) = {1 \over 4 G_N} {\rm Area}(\tilde{A}) \; .
\ee
The surface $\tilde{A}$ is defined by the following conditions:
\begin{itemize}
\item
The surface $\tilde{A}$ has the same boundary as $A$.
\item
The surface $\tilde{A}$ is homologous to $A$. This means that $A$ and $\tilde{A}$ together form the boundary of some $d$-dimensional spacelike surface in $M_\Psi$. This condition together with the previous condition are equivalent to saying that the surface $\tilde{A}$ divides some bulk spatial slice ending on $\Sigma_{\cal B}$ into two parts, such that $\Sigma_{\cal B}$ splits into $A$ and $\tilde{A}$.
\item
The surface $\tilde{A}$ extremizes the area functional. If there are multiple such surfaces, $\tilde{A}$ is the one with least area.
\end{itemize}

\begin{figure}
\centering
\includegraphics[width = 0.5 \textwidth]{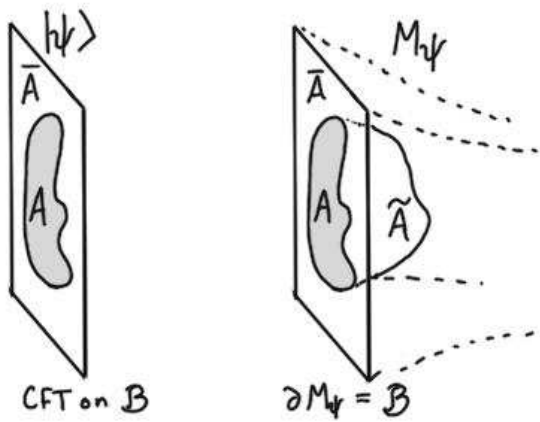}
\caption{Geometrical features relevant to the Ryu-Takayanagi proposal. In the diagram, the time direction has been suppressed. The left side shows a spatial slice $\Sigma_{\cal B}$ of the spacetime ${\cal B}$ on which the CFT lives. The right side shows a spatial slice of the spacetime $M_\Psi$ dual to the state $|\Psi \rangle$, containing $\Sigma_{\cal B}$ and the extremal surface $\tilde{A}$.}
\label{rt}
\end{figure}

The proposal is depicted in figure \ref{rt}. A nice aspect of the proposal is that it relates completely universal quantities on the two sides of the correspondence. Entanglement entropy can be defined for any CFT, while on the gravity side, the area of extremal surfaces is purely geometrical and thus relevant in any gravitational theory.

\subsubsection*{Minimal area surfaces}

In the special case of a static geometry\footnote{More generally, these comments apply to a geometry with time reflection symmetry about a spatial slice ending on $\Sigma_{\cal B}$ when we restrict to this slice.}, we can ignore the time direction and say that the entanglement entropy for a boundary region $A$ is computed as the area of the minimal-area surface in the bulk space with the same boundary as $A$. This was the original Ryu-Takayanagi proposal before the covariant generalization above.

More generally, assuming the geometry $M_\Psi$ satisfies the null energy condition, the covariant prescription above is equivalent finding the minimal-area surface on a spatial slice $\Sigma$ bounded by $\Sigma_{\cal B}$, but then maximizing this area over all possible slices $\Sigma$ \cite{Wall:2012uf}. This ``maximin'' construction turns out to be very useful in proving certain results about holographic entanglement entropy.

\subsubsection*{QFT entanglement and divergences}

The Ryu-Takayanagi formula as it is usually stated is somewhat ill-defined, since both sides of the equation actually represent divergent quantities. On the gravity side, the area $\tilde{A}$ is divergent because there is an infinite proper distance to the boundary of AdS.

To understand the divergence on the field theory side, note that the subsystem we are talking about (i.e. the subset of field degrees of freedom living in some spatial region $A$) actually contains an infinite number of degrees of freedom, since we have field modes with arbitrarily short wavelengths. The field modes on either side of the region boundary $\partial A$ are coupled to each other by the field theory Hamiltonian, and are entangled with each other in the field theory ground state. Summing the contribution to the entanglement entropy from these infinite number of modes, we obtain a divergence typically proportional to the area of the boundary of $A$ for two or more spatial dimensions.

To make a sense out of the RT formula, we have several options. First, we can work with a UV cutoff in the field theory at some high scale $1/\epsilon$. In the AdS/CFT correspondence, this corresponds to an IR cutoff, where we keep only the $z > \epsilon$ part of the geometry. With the explicit cutoff, both the CFT entanglement and the area on the gravity side become finite, and when explicit calculations are possible, the two results can be shown to match, up to terms which vanish as the cutoff is removed. We will give one explicit example of this matching below.

Alternatively, we can take the usual approach in quantum field theory and work with quantities that remain finite as the cutoff is removed. There are several options
\begin{enumerate}
\item
We can consider certain combinations of entanglement entropies for which the divergences cancel. For example $S(A) + S(B) - S(A \cup B)$, which defines the MUTUAL INFORMATION between $A$ and $B$. We can use this to obtain a regulated version of $2 S(A)$ by choosing $B$ to be all the points with distance $\ge \epsilon$ from $A$.
\item
For excited states, we can consider the entanglement entropy relative to the vacuum entanglement entropy, $S_A(|\Psi \rangle) - S_A(|vac \rangle)$.
\item
We can look at certain derivatives of the entanglement entropy with respect to some parameters describing the region. For example, for an interval of length $L$ in 1+1 dimensions, we can look at $dS/dL$.
\end{enumerate}
In each of these cases, we can work with a cutoff, perform the calculations, and then remove the cutoff in the end to obtain a finite result.

\subsubsection*{Example}

As a simple example, we can consider the calculation of entanglement entropy for a ball-shaped region $B$ for a CFT in the vacuum state on $R^{d-1,1}$. In this case, the dual geometry is Poincar\'e AdS
\be
{\ell^2 \over z^2} (dz^2 -dt^2 + dx^i dx^i) \equiv G_{\mu \nu} dx^\mu dx^\nu \; .
\ee
We need to find the extremal-area $(d-1)$-dimensional surface in the geometry whose boundary is the same as the boundary of the ball $B$, which we choose to be at $(x^i)^2 = R^2$ and $t=0$. Since the geometry is static, we expect that the bulk extremal surface should lie in the $t=0$ slice.

If we parameterize a $(d-1)$ dimensional surface as $X^\mu(\sigma)$, the $(d-1)$-dimensional area functional is
\be
\label{Area_explicit}
{\rm Area} = \int d^{d-1} \sigma \sqrt{\det g_{ab}}
\ee
where
\be
\label{Induced}
g_{ab} = G_{\mu \nu}(X(\sigma)) {\partial X^\mu \over \partial \sigma^a} {\partial X^\nu \over \partial \sigma^b}
\ee
is the induced metric on the surface. We can take the coordinates $\sigma$ on the surface to be the spatial coordinates $x^i$, so that the surface is parameterized as $Z(x^i)$ with  $T(x^i)=0$. In this case, the induced metric reduces to
\be
g_{ij} = {\ell^2 \over Z^2} \left(\delta_{ij} + {\partial Z \over \partial x^i} {\partial Z \over \partial x^j}\right)
\ee
and the area functional becomes
\be
{\rm Area} = \int d^{d-1} x \left( {\ell \over z} \right)^{d-1} \sqrt{1 + {\partial Z \over \partial x^i} {\partial Z \over \partial x^i}}
\ee
It is straightforward to write down the corresponding Euler-Lagrange equations and check that the minimal area solutions ending on the spheres $(x^i)^2 = R^2$ are the bulk hemispheres\footnote{For the special case we are considering, there are more elegant ways of obtaining this result. For example, by a conformal transformation, the boundary of the ball can be mapped to the line $x^1=0$. The extremal surface ending on this line is simply the bulk surface $x^1 = 0$. Applying the bulk coordinate transformation corresponding to the reverse conformal transformation, this surface maps to a hemisphere.}
\be
(x^i)^2 + z^2 = R^2 \; .
\ee
Now, we need to calculate the area of the extremal surface. We will work in $d=2$ as an explicit example. To regulate the divergence associated with the infinite distance to the AdS boundary, we will calculate the area of the surface in the region $z > \epsilon$ for some small $\epsilon$. We have
\be
\label{gravS}
S_B = {1 \over 4 G_N} {\rm Area} = \int_{z>\epsilon}  {\ell \over z}  (dx^2 + dz^2) = {\ell \over 2 G_N} \ln \left({L \over \epsilon}\right)
\ee
where we have defined $L = 2R$ since the ``ball'' of radius $R$ is simply an interval of length $2L$.

We can now compare this to a direct calculation of the entanglement entropy in the vacuum state of a two-dimensional CFT. We recall that CFTs are characterized by a central charge $c$ which serves as a measure of the number of degrees of freedom. It turns out that there is a universal formula for the entanglement entropy of an interval in the CFT vacuum state, depending on the central charge and no other property of the CFT (see, for example, \cite{Calabrese:2004eu}). In terms of the UV cutoff $1/\epsilon$ the result is
\be
\label{S2D}
S = {c \over 3} \ln \left({L \over \epsilon}\right) \; ;
\ee
the calculation that gives rise to this result is reviewed in appendix C.
Comparing with (\ref{gravS}), we see that the two results agree precisely, including the cutoff dependence, so long as we identify
\be
c = {3 \over 2} {\ell \over G_N} \; .
\ee
This is indeed the known relation between the central charge and the gravitational parameters for examples of the $AdS_3/CFT_2$ correspondence. It can be established independently by comparing other physical observables such as the relation between entropy and temperature in thermal states.

The agreement here is actually more than we would have expected in general. In general, we would expect gravity calculations to be reproduced only for CFTs with a gravity dual description. It just happens that the result for the entanglement entropy of a single interval takes the same form for all CFTs. For more general quantities, such as the entanglement entropy of a union of disjoint intervals, the CFT result depends on the details of the CFT, and only for special holographic CFTs does the result match with a gravity calculation \cite{Hartman:2013mia,Faulkner:2013yia}.

\begin{figure}
\centering
\includegraphics[width=0.4\textwidth]{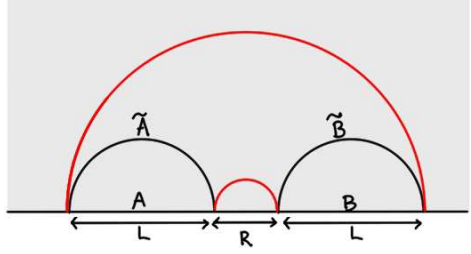}\\
\caption{Extremal surfaces for computing the mutual information between disjoint intervals in a holographic two-dimensional CFT. For sufficiently large $R$, the minimal-area extremal surface with boundary $\partial (A \cup B)$ is simply $\tilde{A} \cup \tilde{B}$, so the mutual information vanishes at leading order in $1/N$. For smaller $R$, the mutual information is the difference in area between the red surfaces and the black surfaces $\tilde{A}$ and $\tilde{B}$.}
\label{mutual}
\end{figure}

As we discussed earlier, while the result (\ref{S2D}) is cutoff-dependent, there are various UV finite quantities that we can derive from this:

\begin{enumerate}
\item
First, we can take combinations of entanglement entropies for which the divergences cancel. An example is provided by the mutual information for a holographic CFT in the vacuum state between two intervals $A$ and $B$, which we will take to be of length $L$ separated by distance $R$. In this case, the mutual information is
\be
\label{MI}
I(A:B) = S(A) + S(B) - S(A \cup B)
\ee
where from (\ref{S2D}), we have
\be
S(A) = S(B) = {c \over 3} \log {L \over \epsilon} \; .
\ee
To calculated $S(A \cup B)$ holographically, we actually need to consider two different possible extremal surfaces with boundary $\partial (A \cup B)$, both a disconnected union of two parts, as shown in figure \ref{mutual}. For $R > (\sqrt{2} - 1)L$, the minimal-area extremal surface is simply the union of surfaces $\tilde{A}$ and $\tilde{B}$ computing $S(A)$ and $S(B)$, so the mutual information vanishes (at leading order in $1/N$). For $R < (\sqrt{2} - 1) L$, the alternative (red) extremal surface has less area. Thus, we have
\be
S(A \cup B) = \ba{ll} {c \over 3} \log {R + 2 L \over \epsilon} + {c \over 3}\log {R \over \epsilon} & \qquad \qquad  R < (\sqrt{2} - 1) L \cr
{c \over 3} \log {L \over \epsilon}  + {c \over 3} \log {L \over \epsilon} & \qquad \qquad R > (\sqrt{2} - 1) L \ea \; .
\ee
Combining our results as in (\ref{MI}), we obtain for the leading large $N$ behavior of the mutual information
\be
I(A:B) = \ba{ll} {c \over 3} \log {R(R + 2 L) \over L^2} & \qquad \qquad  R < (\sqrt{2} - 1) L \cr
0 & \qquad \qquad R > (\sqrt{2} - 1) L \ea
\ee
which is finite, as promised, and has an interesting first order phase transition as the separation between the intervals is increased.
\item
We can also consider the entanglement entropy for excited states, subtracting off the vacuum entanglement entropy for the same region. For example, consider the thermal state of the CFT with temperature $\beta$. The entanglement entropy in this state can be calculated \cite{Calabrese:2004eu} using a similar path-integral approach as for the vacuum calculation (described in appendix C), or for holographic theories, using the Ryu-Takayanagi formula applied to the BTZ black hole geometry (\ref{BTZ}). As for the vacuum case, the result is universal, depending only on the central charge of the CFT.

The divergent part of the entanglement entropy is the same as for the vacuum case, and subtracting the two, we find for an interval of length $L$,
\be
S_\beta - S_{vac} = {c \over 3} \log \left( {\beta \over \pi L} \sinh {\pi L \over \beta} \right)
\ee
In cases where $L$ is significantly larger than $\beta$, This gives
\be
\label{Stherm2D}
S_\beta - S_{vac} \sim  {\pi c \over 3 \beta} L = s(\beta) L
\ee
where $s(\beta)$ is ordinary thermodynamic entropy density for the CFT at temperature $\beta^{-1}$. Thus, (\ref{Stherm2D}) gives exactly the thermodynamic entropy of the subsystem.
\item
Finally, we can simply take the derivative of the entanglement entropy with respect to the size of the system, which gives the finite result
\be
{d S \over d L} = {c \over 3 L} \;
\ee
In higher dimensions, there are various ways to deform the shape of an region, and various possibilities for differential quantities that are UV finite (see, for example \cite{Liu:2012eea} and \cite{Bhattacharya:2014vja}).
\end{enumerate}

\subsection{Evidence for Ryu-Takayanagi}

Here, we mention a few pieces of evidence for the correctness of the Ryu-Takayanagi proposal.

We have already seen that the formula is correct when applied to an interval in the vacuum state or a thermal state of a holographic 2D CFT. Similar agreement can be shown for the case of multiple intervals in the vacuum state of holographic 2D CFTs \cite{Hartman:2013mia,Faulkner:2013yia}, for an interval in the thermal state of 2D CFTs, for ball-shaped regions in the vacuum of higher-dimensional CFTs \cite{casini2011towards}, for ball-shaped regions in CFTs deformed by relevant operators \cite{Faulkner:2014jva}, and for certain shape-deformations of these ball-shaped regions (see e.g. \cite{Faulkner:2015csl}).

In general, direct calculational checks are limited by our ability to calculated entanglement entropy in strongly coupled CFTs. However, it is possible to provide a general argument for the validity of the proposal starting directly from the assumed  equivalence between the CFT and gravity partition functions that lies at the heart of the AdS/CFT correspondence. The argument makes use of the fact that the CFT entanglement entropy is a limiting case of the Renyi entropies (\ref{Renyi}). As explained in appendix C, these can be calculated by evaluating the CFT path integral on a multi-sheeted Euclidean space defined by gluing multiple copies of the original boundary space together across the regions for which the entropy is being evaluated. According to the AdS/CFT correspondence, this field theory path integral should equal the path integral for the gravity theory on asymptotically AdS spacetimes with boundary geometry equal to the multi-sheeted space appearing in the CFT path integral. In the classical limit, the gravity path integral is dominated by a single saddle-point geometry, which is the solution to the classical gravitational equations with these boundary conditions. The result is then $e^{-S_{grav}}$ evaluated for this solution. Thus, we have a way in principle to compute Renyi entropies and therefore entanglement entropies by solving a gravity problem, without assuming the Ryu-Takayanagi formula. In general, finding the required solutions to compute the Renyi entropy $S_n$ would be a difficult problem requiring numerics. However, in the limit $n \to 1$ needed to compute entanglement entropies, Lewkowycz and Maldacena have argued (in the case of static or time-reflection symmetric spacetimes) \cite{Lewkowycz:2013nqa} that the problem reduces to precisely the problem of computing extremal surface areas in the original geometry. The argument has recently been extended to the time-dependent case by \cite{Dong:2016hjy}.

\subsubsection*{Entanglement entropy for complementary subsystems}

The holographic entanglement entropy formula also obeys various consistency checks. We recall that for any pure state of a system with subsystems $A$ and $\bar{A}$, the entanglement entropy of $A$ matches with the entanglement entropy of $\bar{A}$:
\be
\label{purestate}
S(A) = S(\bar{A}) \qquad \qquad |\Psi \rangle_{A \cup \bar{A}} \; {\rm pure}
\ee
In the RT formula, the calculation of the entanglement entropy for a region $A$ involves finding the minimal area bulk extremal surface whose boundary is the same as the boundary of $A$. But the boundary of $\bar{A}$ is the same as the boundary of $A$. Thus, the minimal-area extremal surface is the same for both cases, and we will get the same entanglement entropy, {\it provided that such a surface is homologous to both $A$ and $\bar{A}$.} The latter condition requires that $A$ and $\bar{A}$ are homologous in the full geometry, which is equivalent to saying that the bulk geometry has no boundary components other than $A \cup \bar{A}$.

\begin{figure}
\centering
\includegraphics[width = 0.4 \textwidth]{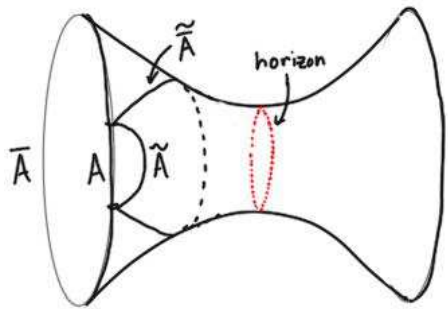}
\caption{Ryu-Takayanagi surfaces computing entanglement entropy for complementary regions $A$ and $\bar{A}$ in the thermal state on $S^d$ dual to a black hole. For small enough $A$, the minimal-area extremal surface corresponding to $\tilde{A}$ becomes the union of $\tilde{A}$ with the black hole horizon.}
\label{BHhomology}
\end{figure}

An example where $A$ and $\bar{A}$ can fail to be homologous is the case where the bulk geometry is a black hole. For example, in the maximally extended black hole geometry, $A$ will not be homologous to $\bar{A}$ since there is a second asymptotic region with its own boundary. In this case, the entanglement entropy for $A$ and $\bar{A}$ will be computed by separate surfaces (see figure \ref{BHhomology}) and will generally be different. But this is exactly what we expect, since the state of the CFT on $A \cup \bar{A}$ is no longer pure so (\ref{purestate}) doesn't apply.

An interesting example is provided by the case where $A$ is a small region and $\bar{A}$ makes up the remainder of the sphere, with the CFT on the sphere taken to be in some high-energy thermal state. In this case, the extremal surface corresponding to the region $A$ will remain close to the AdS boundary. For $\bar{A}$ even though the boundary of $\bar{A}$ is small, the extremal surface must ``wrap around'' the black hole (as shown in figure \ref{BHhomology}) in order to be homologous to $\bar{A}$. In this case, it turns out that the minimal area extremal surface is actually not  connected, but rather a disconnected surface made up of the union of $\tilde{A}$ with the black hole horizon. In this case we have the interesting relation that
\be
S(\bar{A}) = S(A) + S_{BH} \; .
\ee

\subsubsection*{Subadditivity and mutual information}

For general quantum systems with disjoint subsystems $A$ and $B$, it is straightforward to demonstrate that\footnote{For a more detailed discussion on the entanglement constraints in this and later sections, see \cite{nielsen2010quantum}.}
\be
\label{subadditivity}
I(A;B) \equiv S(A) + S(B) - S(A \cup B) \ge 0
\ee
This quantity, known as MUTUAL INFORMATION is a measure of entanglement and correlations between the two subsystems $A$ and $B$. The fact that it is always positive is known as the SUBADDITIVITY of entanglement entropies. Mutual information provides an upper bound for all correlations between the two subsystems; if ${\cal O}_A$ and ${\cal O}_B$ are two bounded operators acting on ${\cal H}_A$ and ${\cal H}_B$, it can be shown that \cite{Wolfetal}
\be
{(\langle {\cal O}_A {\cal O}_B \rangle - \langle {\cal O}_A \rangle \langle {\cal O}_B \rangle)^2 \over 2|\langle {\cal O}_A \rangle|^2 |\langle {\cal O}_B \rangle|^2} \le I(A;B) \; .
\ee

In a CFT, taking $A$ and $B$ to be disjoint spatial subsystems, the combination of entanglement entropies in \ref{subadditivity} is finite, so the positivity constraint is meaningful. Using the holographic  formula (\ref{HRT}) this translates to a condition on the areas of surfaces; as a consistency check, we should verify that this is satisfied. Given the extremal surfaces $\tilde{A}$ and $\tilde{B}$ for $A$ and $B$, the surface $\tilde{A} \cup \tilde{B}$ is an extremal surface with the same boundary as $A \cup B$. The surface that computes the entanglement entropy $S(A \cup B)$ is defined to be the minimal-area extremal surface with the same boundary as $A \cup B$. Thus, its area must be less than or equal to the area of of $\tilde{A} \cup \tilde{B}$, and the subadditivity relation (\ref{subadditivity}) follows immediately.

\subsubsection*{Strong subadditivity}

More constraints arise when we consider additional subsystems. For general quantum systems with disjoint subsystems $A$, $B$, and $C$ entanglement entropy obeys the relation
\be
\label{SSA}
S(A \cup B) + S(B \cup C) \ge S(B) + S(A \cup B \cup C) \ge 0 \; ,
\ee
known as STRONG SUBADDITIVITY. This can be reexpressed in terms of mutual information as
\be
I(A;B \cup C) \ge I(A;B) \; ,
\ee
i.e. that the mutual information between $A$ and the combined system $BC$ must be larger than the mutual information between $A$ and $B$. This sounds very plausible, but it turns out that the proof of strong subadditivity is rather difficult.

Again, we can ask whether the holographic formula for entanglement entropy respects this. For static geometries and regions $A$, $B$, and $C$ all on a preferred time slice, where we can use the original RT formula, the geometrical version of strong subadditivity may be easily demonstrated to hold for any geometry, as shown by Headrick and Takayanagi \cite{headrick2007holographic}. If we consider regions on more general time slices, or time-dependent geometries, the strong subadditivity relation is much more difficult to demonstrate, and in fact only holds when the geometry satisfies certain conditions. While the necessary conditions on the geometry are not known in general, it has been shown that the null energy condition is sufficient \cite{Wall:2012uf}.

\subsection{Generalizations of the Ryu-Takayanagi formula}

The Ryu-Takayanagi formula should be understood as holding in a particular limit where the gravitational theory is well-described by classical Einstein gravity coupled to matter (without curvature couplings). Of course, the entanglement entropy is a precisely defined quantity (up to the issue of divergences) for any CFT, so we might expect some version of the RT formula to hold even away from this limit of classical Einstein gravity.

For classical theories of gravity which are not Einstein gravity, there is a well-known generalization of the entropy-area connection for black holes due to Wald \cite{Wald:1993nt}. In this case, area is replaced by a more general covariant quantity calculated for the black hole horizon. The appropriate quantity can be determined directly from the form of the gravitational Lagrangian. It is natural to suppose that this more general quantity also replaces area in the holographic entanglement formula. It turns out that the full story is slightly more complicated, since there can be additional terms that make no contribution when evaluated on a black hole horizon. However, there are now precise proposals \cite{Camps:2013zua,Dong:2013qoa} for the correct quantity, based on the Lewkowycz-Maldacena derivation of the RT formula applied to these more general theories.

Away from large $N$, there will also be quantum effects in the bulk theory. The leading corrections to the classical limit are captured by ``semiclassical gravity,'' in which we include the effects of quantum fluctuations of the metric and other fields via a quantum field theory living on the spacetime geometry. At this level of approximation, we can think of CFT states $|\Psi \rangle$ as corresponding to some geometry $M$ plus some quantum field theory state $|\psi_{bulk} \rangle$ defined on $M$. On the CFT side, effects associated with these quantum field fluctuations correspond to $1/N$ corrections. For the entanglement entropy, this means that the entanglement entropy will have some leading asymptotic behavior in large $N$ plus subleading terms suppressed by powers of $1/N$. Faulkner, Leukowycz, and Maldacena have proposed \cite{Faulkner:2013ana} that the leading  $1/N$ correction to entanglement entropy in the field theory corresponds to the entanglement entropy of the bulk quantum fields across the extremal surface $\tilde{A}$
\be
\label{FLM}
S^{CFT}_A = {1 \over 4 G_N} {\rm Area}(\tilde{A}) + S^{bulk}_{\tilde{A}} \; .
\ee
For a continuum quantum field theory, the second term would have a divergence proportional to the area of $\tilde{A}$, but in the context of a quantum gravitational theory, we expect that there is a natural UV cutoff provided by the Planck scale that renders this term finite.

\subsection{Implications of the holographic entanglement entropy formula}

We conclude this section with some qualitative remarks on the plausible implications of the Ryu-Takayanagi formula.

\subsubsection*{Reconstructing the geometry}

An important implication of the holographic entanglement entropy formula is that much of the dual spacetime geometry for a holographic CFT state is encoded in the entanglement structure of the state. In principle, we can recover the dual spacetime geometry by calculating entanglement entropies for many different regions and then finding a geometry $M$ whose extremal surface areas match with the entropies. This should be a highly overconstrained problem, since the entanglement entropies give us some function on the space of subsets of the boundary spacetime, while geometries $M$ are specified by a handful of functions of a few coordinates (a much smaller space). Thus, for a general state in a general CFT, we should expect that no geometry will reproduce all the entanglement entropies.

Having a geometrical dual will therefore require a very special structure of entanglement for the CFT state. Given such a state, we can expect that the entanglement entropies fix the geometry uniquely, with some limitations. Most importantly, there can be regions of a spacetime, such as the region behind the horizon of a black hole, where no extremal surface penetrates. If a CFT state is dual to some geometry with such a region, known as the ENTANGLEMENT SHADOW (see e.g. \cite{Engelhardt:2013tra} for a discussion), we will clearly not be able to learn about this region by computing spatial entanglement entropies. It may be that the information about the geometry in these regions may be contained in more general types of entanglement (see e.g. \cite{Balasubramanian:2014sra,Lin:2016fqk}).\footnote{It is possible to associate an entanglement entropy to any subalgebra of the full algebra of observables for a quantum system. The entanglement entropies that we have discussed so far correspond to the full subalgebra of operators associated to spatial regions, but there are more general possibilities. See \cite{Radicevic:2016tlt} for some examples and the appendix of \cite{Harlow:2016vwg} for a general review.}

\subsubsection*{Spacetime from entanglement}

In the example of the maximally extended Schwarzschild black hole, the the two CFTs have nothing to do with one another except that their states are entangled via the thermofield double state. In that example, it is clear that this entanglement is a necessary condition for the two asymptotic regions of the dual spacetime to be connected. Without entanglement, we have a product state in two non-interacting systems, and the only possible interpretation would be two disconnected spacetimes.

Motivated by this, it is interesting to ask whether even in simpler spacetimes, dual to states of a single CFT, the connectedness of the geometry is related to the large amount of entanglement present in low-energy CFT states. To probe this, we can consider the following thought experiment \cite{VanRaamsdonk:2009ar, VanRaamsdonk:2010pw}. Starting with the vacuum state of a CFT on the sphere, imagine arbitrarily dividing the sphere into two hemispheres. Now, consider a one parameter family of CFT states for which the entanglement between the two hemispheres decreases from the initial value in the vacuum state. What happens to the dual spacetime?

\begin{figure}
\centering
\includegraphics[width = 0.8 \textwidth]{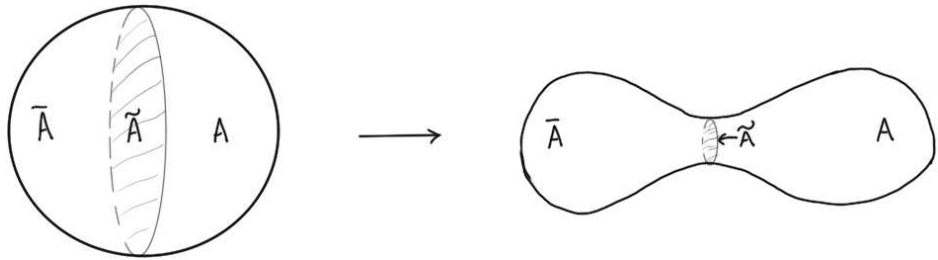}
\caption{Effect on the dual spatial geometry of disentangling the two hemispheres starting from the vacuum state of a CFT on $S^d$. Note that it is actually the bulk metric is changing, with the boundary metric remaining fixed.}
\label{Thoughtexpt}
\end{figure}

Using the Ryu-Takayanagi formula, we learn that as the entanglement between the two sides decreases, the area of the surface dividing the two sides of the bulk spacetime decreases. A separate argument suggests that the two sides also become further and further apart. Thus, the picture is that the spatial geometry of the corresponding spacetime stretches apart and pinches off in the middle, as shown in figure \ref{Thoughtexpt}. This picture can be checked by explicit calculations in some one parameter families of states \cite{Czech:2012be}. Thus, by removing the entanglement between the two halves of the CFT, it appears that we are able to pull apart the corresponding spacetime. We can go further, subdividing the boundary into more regions and removing the entanglement between these. As we will see in more detail below, the result is that the spacetime splits up into many small disconnected fragments. By removing all the entanglement (which, for the case of a continuum CFT, costs an infinite amount of energy), the dual spacetime disappears entirely!

These qualitative arguments suggest a rather remarkable conclusion: quantum entanglement between the underlying degrees of freedom is crucial for the existence of classical spacetime. We might even interpret the Ryu-Takayanagi formula as telling us that spacetime is a geometrical representation of the entanglement structure of a CFT state \cite{Swingle:2009bg,VanRaamsdonk:2009ar}.\footnote{There are many caveats to such a conclusion. It is an open question whether spacetime geometry in the entanglement shadow, e.g. behind black hole horizons, can be understood this way. Also, there are examples of holographic duality where the field theory is a simple matrix quantum mechanics. In this case, there are no spatial subsystems, so a connection between entanglement and geometry would have to be based on a more general type of entanglement within the system, such as those mentioned in the previous footnote.}

\section{The entanglement structure of the CFT vacuum}

In this section, we will begin to take a closer look at entanglement and other quantum-information theoretic quantities in quantum field theories in order to understand how they are related to features of the dual spacetime geometries and to gravitational physics in these spacetimes.

\subsubsection*{Domains of dependence}

Consider a local quantum field theory on some spacetime ${\cal B}$. To define a spatial subsystem, we first choose a spatial slice $\Sigma_{\cal B}$ and then let $A$ be a subset of this spatial slice. This subset can be connected or disconnected. The fields in $A$ represent a subset of the degrees of freedom of the quantum field theory. This is most clear in a lattice regularization, where we have degrees of freedom associated with specific points. There are subtleties in the case of gauge theories; for more discussion on these, see \cite{Donnelly:2011hn,Casini:2013rba}.

\begin{figure}
\centering
\includegraphics[width = 0.7 \textwidth]{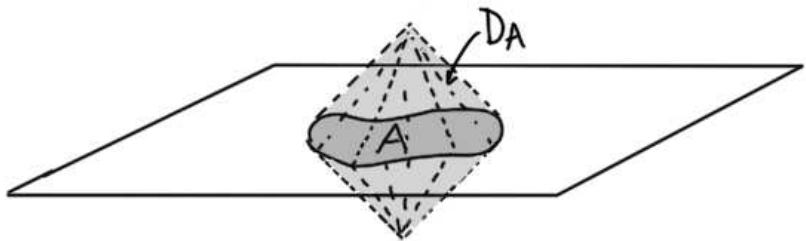}
\caption{Domain of dependence region $D_A$ (light shaded) for a subset $A$ (dark shaded) of a spatial slice.}
\label{DomD}
\end{figure}

In a relativistic quantum field theory, there is a spacetime region $D_A$ that is canonically associated with the spatial subsystem $A$. To define $D_A$, we note that for some spacetime points $p$ in $B$, all causal (i.e. timelike or lightlike) curves through $p$ necessarily pass through $A$. The region $D_A$ is the collection of all such points, known as the DOMAIN OF DEPENDENCE of $A$, or alternatively as the causal development region or causal diamond associated with $A$. This is illustrated in figure \ref{DomD}.

At the classical level, the domain of dependence is important because it is the region where the solution of any relativistic field equation is fully determined given initial data on $A$. Changes to the fields in $\bar{A} \subset S\Sigma_{\cal B}$ cannot affect this region. At the quantum level, any operator in $D_A$ can be expressed in terms of the field operators living on the region $A$, making use of the field theory Hamiltonian to evolve field operators backwards and forwards in time. Operators in $D_A$ commute with operators localized in $\bar{A}$ or $D_{\bar{A}}$.

\begin{figure}
\centering
\includegraphics[width = 0.3 \textwidth]{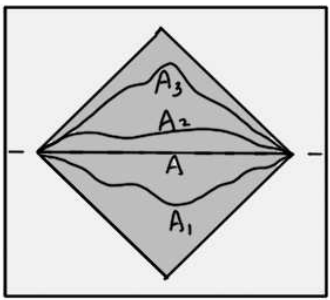}
\caption{Different spacelike regions with the same domain of dependence $D_A$.}
\label{DDAs}
\end{figure}

The domain of dependence $D_A$ associated with a region $A$ is also the domain of dependence for an infinite number of other spatial regions, essentially any spacelike surface in $D_A$ whose boundary is the same as the boundary of $A$, as in figure \ref{DDAs}. The density matrix associated with any of these other regions contains the same information as the density matrix associated with $\rho_A$, so it is useful to think of $\rho_A$ as being a density matrix associated with the whole spacetime region $D_A$.

The full set of operators with support in $A$ forms a closed algebra of observables for the field theory; in a sense, the field theory on $D_A$ represents a complete quantum system on its own. The full state of the CFT can then be understood as an entangled state between this system and a complementary system given by the field theory on $D_{\bar{A}}$,
\be
\label{entrep}
|\Psi \rangle = \sum_i \sqrt{p_i} |\Psi_i^A \rangle \otimes |\Psi_i^{\bar{A}} \rangle \; .
\ee
This expression is a bit heuristic, since the spectrum in this case is generally not discrete.

The representation (\ref{entrep}) suggests a natural decomposition of the quantum information associated with the full state as
\be
\label{psidecomp}
|\Psi \rangle \to \rho_A + \rho_{\bar{A}} + {\rm entanglement \; info} \; .
\ee
Here, it is important to note that the pair of density matrices $\rho_A,\rho_{\bar{A}}$ represent only a subset of the information contained in the state $|\Psi \rangle$. The rest of the information is contained in the details of how the two subsystems are entangled with one another in (\ref{entrep}).

In our discussion below, we will see that there is a natural spacetime analog to the decomposition (\ref{psidecomp}). This can be motivated by noting that the representation (\ref{entrep}) bears a formal similarity to the CFT description of the two-sided black hole, where (as we will argue below) the split (\ref{psidecomp}) corresponds to the natural spacetime decomposition into the outside-the-horizon regions on the two sides plus the black hole interior.

\subsubsection*{Density matrix for a half-space}

\begin{figure}
\centering
\includegraphics[width = 0.4 \textwidth]{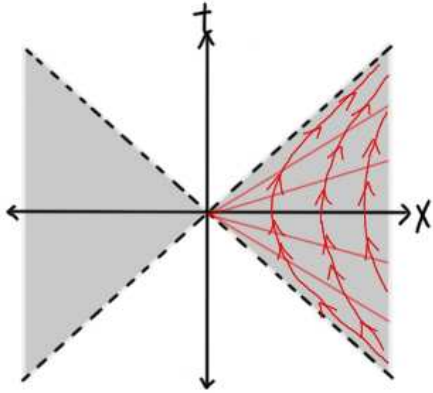}
\caption{Rindler wedges in Minkowski space. Lines of constant $\eta$ and the flow generated by $\partial_\eta$ are shown in red for the $x > 0$ wedge.}
\label{Rindler}
\end{figure}

In general, it is difficult to describe the density matrix for a spatial subsystem of a quantum field theory, even for the vacuum state. One special exception is the case of a half-space (e.g. $x > 0$, where $x$ is one of the spatial coordinates) of Minkowski space, whose domain of dependence is known as a RINDLER WEDGE. We can describe the Rindler wedge as $\{x > 0, |t| < x\}$ or alternatively as the region $\{r > 0\}$ in the metric
\be
ds^2 = dr^2 - r^2 d \eta^2 \; ,
\ee
where $(r,\eta)$ are related to the usual Minkowski coordinates by
\be
x = r \cosh(\eta) \qquad \qquad t = r \sinh(\eta) \; .
\ee
This wedge (and the complementary $x < 0$ wedge) are depicted in figure \ref{Rindler}.

A simple path-integral argument (see appendix \ref{AppPI}) shows that for the vacuum state of any Lorentz-invariant quantum field theory, the density matrix for the half space is a thermal density matrix
\be
\label{Rindlerrho}
\rho = {1 \over Z} e^{- 2 \pi H_\eta} \; .
\ee
with respect to the Hamiltonian $H_\eta$ that generates translations in the coordinate $\eta$. In the usual coordinates, these are Lorentz boosts, generated by the vector field\footnote{Recall that $\partial_i$ denotes the coordinate unit vector field in the $i$ direction.}
\be
\partial_\eta = x \partial_t + t \partial_{x} \equiv \eta^\mu \partial_\mu  \; .
\ee
We can write $H_\eta$, also known as the Rindler Hamiltonian, explicitly as
\be
\label{Rindham}
H_\eta = \int_{t=0, x > 0} d^{d-1} x \{ x T_{00}(x^\mu) \} \; .
\ee
More covariantly, we can write $H_\eta$ as
\be
H_\eta = \int_\Sigma \eta^\mu T_{\mu \nu} \epsilon^\nu
\ee
where $\Sigma$ is an arbitrary spacelike surface in the Rindler wedge with boundary $\{x=0,t=0\}$, and we have introduced the volume form
\be
\label{defepsilon}
\epsilon^\mu = \epsilon^\mu {}_{\mu_2 \cdots \mu_{d}} dx^{\mu_2} \wedge \cdots \wedge dx^{\mu_d} \; .
\ee
The fact that the density matrix is thermal with respect to this Hamiltonian which generates evolution on accelerated trajectories is related to the result of Unruh that an accelerated observer observing the vacuum state will see a thermal spectrum of particles.

We can also consider the complementary Rindler wedge, associated with the spatial region $x < 0$. The density matrix for this region is also thermal, with respect to the Hamiltonian $H_{\eta'}$ generating boosts in this complementary wedge. The full vacuum state of the CFT can be understood as an entangled state between the field theories defined on the two Rindler wedges. In fact, it is exactly the thermofield double state
\be
\label{QFTTFD}
|vac \rangle = \sum_i e^{- \beta E_i / 2} | E_i \rangle \otimes |E'_i \rangle \; ,
\ee
where $|E_i \rangle$ and $|E'_i \rangle$ represent the energy eigenstates for the Hamiltonians $H_\eta$ and $H_{\eta'}$.

In the case of a free field theory, we can define modes of the fields living in each of the wedges, and build the energy eigenstates of $H_\eta$ and $H_{\eta'}$ on the respective sides by occupying these modes. In this case, we find that in the vacuum state (\ref{QFTTFD}), each mode in one wedge is entangled with the corresponding mode in the other wedge. We can also consider states in which this entanglement is removed, i.e. a product state between the two wedges. For these states, we find that the stress tensor is singular at $x=t=0$ and on the boundaries of the two Rindler wedges. Thus, entanglement of the field theory modes on either side of the $x=0$ surface (or any other surface) is crucial to obtain a well-behaved field theory state.

\subsubsection*{Density matrix for a ball-shaped region in a CFT}

A Rindler wedge can be related by a conformal transformation to the domain of dependence of a ball-shaped region of Minkowski space.\footnote{See appendix \ref{ConfTrans} for a review of conformal transformations.} In a conformal field theory, such a transformation is associated with a mapping of states that leaves the vacuum state invariant. Thus, the vacuum density matrix for a ball-shaped region is the image under the conformal transformation of the vacuum density matrix (\ref{Rindlerrho}) for the Rindler wedge. The Rindler density matrix is expressed directly in terms of a symmetry generator $H_\eta$; this maps under the conformal transformation to the generator of a symmetry that acts within the domain of dependence of the ball:
\be
\label{DMball}
\rho_B = U \rho_{Rindler} U^\dagger = {1 \over Z} e^{- 2 \pi U H_\eta U^\dagger} \equiv {1 \over Z} e^{- H_\zeta} \; .
\ee
Note that we have absorbed the factor of $2 \pi$ in (\ref{Rindlerrho}) into the definition of $H_\zeta$.

As an example, for the ball $B$ of radius $R$ centered at the origin of Minkowski space, this symmetry is a spacetime transformation generated by the vector field
\be\label{defzeta}
\zeta = \frac{\pi}{R} \left\{ (R^2 - t^2 - |\vec{x}|^2) \partial_t - 2 t x^i \partial_i \right\}
\ee
which is a conformal Killing vector of Minkowski space. The associated Hamiltonian can be written as \cite{casini2011towards}
\be
\label{modHball}
H_\zeta = \frac{\pi}{R} \int_B d^{d-1} x (R^2 - |\vec{x}|^2) T_{00}(x) \; ,
\ee
or more covariantly as
\be
H_\zeta = \int_{B'} \zeta^\mu T_{\mu \nu} \epsilon^\nu
\ee
where $\epsilon$ is defined in (\ref{defepsilon}) and $B'$ is any spacelike surface in $D_B$ with the same boundary as $B$.

\subsubsection*{Pure AdS is a maximally extended black hole}

\begin{figure}
\centering
\includegraphics[width = 0.8 \textwidth]{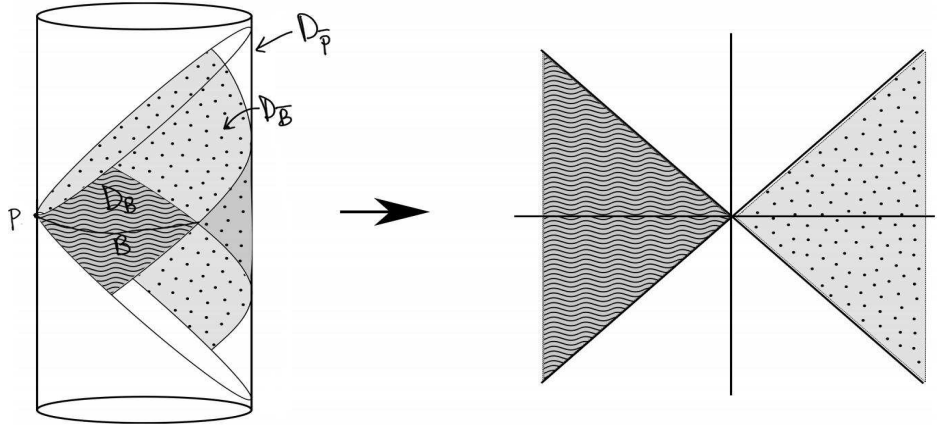}
\caption{A Conformal transformation maps the domains of dependence for complementary balls $B$ and $\bar{B}$ on the sphere to complementary Rindler wedges in Minkowski space, the conformal image of the domain of dependence $D_{\bar{p}}$ of the sphere minus the point $p$.}
\label{DptoRind}
\end{figure}

Our results for Minkowski space translate directly to results for the vacuum state of a single CFT on $S^d \times R$. On the spatial sphere $S^d_t$ at some time, consider any ball $B$ (e.g. a region $\theta < \theta_0$ in polar coordinates) and the complementary ball $\bar{B}$. Choosing some point $P$ on the boundary of $B$, let $D_{\bar{p}}$ be the domain of dependence region of $S^d_t - P$ (i.e. the complement of $p$ on $S^d_t$). Then there is a conformal transformation that maps $D_{\bar{p}}$ to Minkowski space, taking $D_B$ and $D_{\bar{B}}$ to complementary Rindler wedges (defined as the domain of dependence of a half space), as shown in figure \ref{DptoRind}.

This conformal transformation maps the vacuum state on the sphere to the vacuum state on Minkowski space. Thus, as in the previous section, the vacuum density matrices for the balls $B$ and $\bar{B}$  are related by the conformal transformation to the vacuum density matrix of the half space. Again, we have symmetry generators $H_B$ and $H_{\bar{B}}$ defining a notion of time evolution in $D_B$ and $D_{\bar{B}}$ such that the density matrices for the two regions are $\rho_B = \exp(-H_B)/Z$ and $\rho_{\bar{B}} = \exp(-H_{\bar{B}}/Z)$. Furthermore, using the result (\ref{QFTTFD}), we can say that the vacuum state of the CFT on the sphere may be written as the thermofield double state
\be
|vac \rangle_{S^d} = \sum_i e^{- \beta E_i /2} | E^{B}_i \rangle \otimes |E^{\bar{B}}_i \rangle
\ee
where $|E^{B}_i \rangle$ and $|E^{\bar{B}}_i \rangle$ represent energy eigenstates with respect to the Hamiltonians $H_B$ and $H_{\bar{B}}$.

This description of the CFT vacuum has a formal similarity to the CFT description (\ref{TFD}) of the maximally extended Schwarzschild black hole in AdS. We can make an even closer analogy by noting that the regions $D_B$ and $D_{\bar{B}}$ can each be mapped by another conformal transformation (see appendix \ref{ConfTrans}) to the static spacetime $H^d \times R$, i.e. negatively curved hyperbolic space times time, which we can represent as
\be
\label{hyperbolic_metric}
ds^2 = -dt^2 + dr^2 + R_H^2 \sinh^2(r/R_H) d \Omega^2_{d-2} \; .
\ee
Here, $R_H$ is length scale that sets the curvature of the hyperbolic space. Under this transformation, the Hamiltonian $H_B$ maps to the standard CFT Hamiltonian generating evolution in the $t$ coordinate. Thus, any state of the CFT on $S^d$ maps to an entangled state of a pair of CFTs on $H^d$. Specifically, the vacuum state maps to the thermofield double state with temperature $T = 1/(2 \pi R_H)$, with each of the CFTs in a thermal state.

Our experience with thermal states of a holographic CFT on Minkowski space or on a sphere might suggest that these hyperbolic space thermal states have something to do with black hole dual geometries. On the other hand, the hyperbolic space thermofield double state we have arrived at is simply an alternative description of the CFT vacuum state on a sphere, which is dual to pure global AdS spacetime. This apparent tension can be resolved by noting that there is a way to think of pure global AdS spacetime as a black hole.

To understand this, we note that any thermal state of a CFT on hyperbolic space is indeed dual to an AdS-Schwarzshild black hole, but this time with a non-compact horizon with the geometry of hyperbolic space. This can be represented in the standard Schwarzschild coordinates as
\be
\label{HypBHmetric}
ds^2 = -f_\beta(r) dt^2 + {dr^2 \over f_\beta(r)} + {r^2 \over \ell^2} dH^2_{d-1}
\ee
where $f_\beta$ is a temperature-dependent function of $r$ that vanishes at the black hole horizon.
However, for the special case $\beta = 2 \pi R_H$, we have $f_\beta(r) = r^2/\ell^2 - 1$, and it turns out that the geometry (\ref{HypBHmetric}) describes a wedge $R_B$ of global AdS equal to the intersection of the causal future of $D_B$ and the causal past of $D_B$. This is a Rindler wedge of AdS in the sense that it is the bulk domain of dependence of a half-space bounded by the extremal surface $\tilde{B}$.

We now see that pure global AdS can be thought of as a maximally extended black hole geometry.
The wedge $R_B$ represents the exterior of the hyperbolic black hole, the complementary wedge $R_{\bar{B}}$ represents the exterior of a black hole in a second asymptotic region, and the remaining regions to the past and future of these wedges represent the black hole interior.

\subsubsection*{The gravity dual of a density matrix}

We have seen that given a ball-shaped region $B$ on the sphere, we can represent the CFT vacuum state as an entangled state between $B$ and $\bar{B}$, where the density matrices $\rho_B$ and $\rho_{\bar{B}}$ are thermal density matrices for a particular temperature, and the full state is the thermofield double state. On the bulk side, this decomposition of the quantum state naturally corresponds to a decomposition of global AdS into two complementary Rindler wedges plus a behind-the-horizon region; we will now explain this in more detail.

To motivate this correspondence, let us consider a somewhat more general question \cite{Bousso:2012sj,Czech:2012bh,Hubeny:2012wa}: given a state $|\Psi \rangle$ dual to some geometry $M$, and given the reduced density matrix $\rho_A$ calculated from this state for a subsystem $A$, what information about the dual spacetime is contained in $A$?

Since AdS/CFT is a non-local duality, there is no reason a priori that $\rho_A$ should tell us about any particular subset of the dual geometry. It could be that the information about any local region in the dual spacetime is completely spread throughout the CFT degrees of freedom. In this case, no spatial CFT subsystem would contain all the information about the bulk region. However, we will now argue that $\rho_A$ does encode the information about the bulk physics in a particular subset of the dual geometry naturally associated with the domain of dependence region $D_A$.

We note first that given the density matrix $\rho_A$, we can compute the expectation value of any local or nonlocal CFT observable contained within $D_A$. We can also calculate more general quantities, such as the entanglement entropies of spatial subsystems within $D_A$. Through the AdS/CFT dictionary, these provide information about the dual geometry.

\subsubsection*{The causal wedge}

As an example, we can consider response functions in the CFT, i.e. two point functions at timelike separated points contained within $D_A$. We can think of these as perturbing the system at the location of the earlier operator and then measuring the response at the location of the later operator. In the bulk picture, we can think of this operation as making a perturbation at the boundary of AdS that propagates causally into the bulk and then measuring the response of this perturbation at the AdS boundary at some later time. The subset of the bulk geometry sensitive to this kind of measurement is the intersection of the causal future of $D_A$ (the region to which we can send a signal) and the causal past of $D_A$ (the region from which we can receive a signal). This is known as the CAUSAL WEDGE associated with the subsystem $A$.

By measuring response functions and other CFT observables, it is very plausible that we can recover the physics in the causal wedge. For example, we can imagine inserting an arbitrarily small mirror anywhere in the causal wedge; this would have a significant effect on certain response functions, since now light signals from certain boundary points would bounce directly back to other boundary points creating a very different response. Perturbatively around AdS, we can even write explicit CFT expressions corresponding to bulk operators in this causal wedge. For a ball-shaped region on the boundary of pure AdS, the causal wedge is precisely the Rindler wedge $R_B$ discussed above.

\begin{figure}
\centering
\includegraphics[width = 0.3 \textwidth]{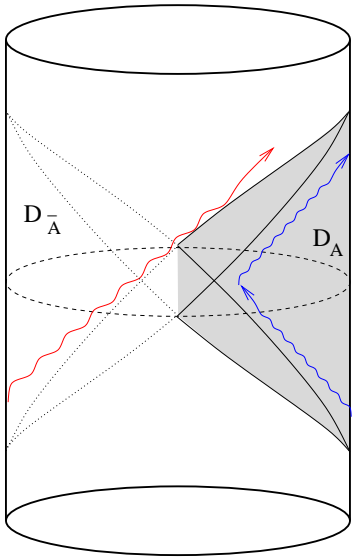}
\caption{The causal wedge associated with $D_A$ (shaded) is the region for which an observer restricted to the boundary region $D_A$ can send a signal to and receive a signal back. Regions of spacetime causally connected to $D_{\bar{A}}$ cannot be described by density matrix $\rho_A$ since we can have perturbations that do not affect $\rho_A$ altering the spacetime anywhere in this region (e.g. via the light wave denoted in red).}
\label{Causal}
\end{figure}

\subsubsection*{An upper bound on the size of the region dual to $\rho_A$}

We can place a useful upper bound on the size of the region encoded in $\rho_A$ as follows. Consider a perturbation to the field theory state which takes the form $|\Psi \rangle \to U_{\bar{A}} \otimes \identity_A |\Psi \rangle$. For the perturbed state, the density matrix $\rho_A$ is the same as for the unperturbed state, while the density matrix $\rho_{\bar{A}}$ is generally changed. As an example, consider local perturbations contained within $D_{\bar{A}}$. These correspond to bulk perturbations near the boundary of AdS which can propagate causally into the geometry, as for the red signal coming from the left in figure \ref{Causal}. In general, the region of the bulk affected by such perturbations is the union of the causal future and the causal past of $D_{\bar{A}}$, $J^+(D_{\bar{A}}) \cup J^-(D_{\bar{A}})$. Since we can alter this region without altering $\rho_A$, we concluded that the region dual to $\rho_A$ must be contained in the complement of $J^+(D_{\bar{A}}) \cup J^-(D_{\bar{A}})$.\footnote{We can enlarge this by replacing $D_{\bar{A}}$ with any bulk region that can be affected by changes in $\rho_{\bar{A}}$ that do not affect $\rho_A$.}

In the case of pure AdS and a region $B$ on the boundary, the complement of $J^+(D_{\bar{B}}) \cup J^-(D_{\bar{B}})$ is again the Rindler wedge $R_B$, so in this case, it seems that the information contained in $\rho_B$ corresponds precisely to bulk information about the physics inside the Rindler wedge $R_B$.

For a generic spacetime, the causal wedge of a region $A$ is generally smaller than the complement of $J^+(D_{\bar{A}}) \cup J^-(D_{\bar{A}})$. Thus it is possible, and, as we will now see, plausible, that the region dual to $\rho_A$ is actually larger than the causal wedge.

\subsubsection*{The entanglement wedge}

Another type of quantity that we can compute using the density matrix $\rho_A$ is the set of entanglement entropies for spatial regions contained within $D_A$. According to the RT/HRT formulas, these tell us about the areas of various extremal surfaces in the bulk geometry. For larger boundary regions, the extremal surfaces generally penetrate deeper into the bulk.\footnote{In fact, assuming the bulk geometry satisfies the null energy condition, it has been proven that the surface moves outward spatially as the size of the region is increased \cite{Wall:2012uf}.} The deepest such surface is the surface $\tilde{A}$ that calculates the entanglement entropy of the whole region $A$. For a generic asymptotically AdS spacetime, it turns out that this surface (and many other surfaces corresponding to large subsystems in $D_A$) lies outside the causal wedge. Since the density matrix contains information about the areas of all of these surfaces, it seems likely that the density matrix $\rho_A$ is dual to a larger region than just the causal wedge. Motivated by these observations, it has been proposed \cite{Czech:2012bh,Wall:2012uf,Entwedge} that the relevant region is the ENTANGLEMENT WEDGE, defined to be the domain of dependence of any spatial bulk region which has $\tilde{A}$ as one boundary component and the remaining boundary component some Cauchy surface in $D_A$. There have been various arguments supporting the validity of this claim.\footnote{See \cite{Dong:2016eik} for a recent discussion.}

One nice feature of the entanglement wedge proposal is that for a pure state of the boundary theory, for which the regions $A$ and $\bar{A}$ have the same surface $\tilde{A}$ computing their entanglement entropies, the entanglement wedges for these complementary regions meet at the surface $\tilde{A}$. Thus, the spacetime naturally divides up into the two wedges plus regions in the future and past of the wedges, as for the case of the maximally extended black hole geometry. Note that the common information between the two density matrices $\rho_A$ and $\rho_{\bar{A}}$ is the spectrum of non-zero eigenvalues, which allows us to compute the entanglement entropy common to the two density matrices. It is thus natural that the common information between the two wedges is the surface $\tilde{A}$, whose geometry allows us to compute the entanglement entropy holographically.\footnote{It is interesting to ask what other covariant geometrical quantities associated to this surface correspond to in the field theory.}

Note that for the case of pure AdS, the causal wedge and the entanglement wedge coincide, but generically, the entanglement wedge is larger.\footnote{This is implied by a theorem of Wald and Gao that light rays generically take longer to pass through the bulk than to go along the boundary.}

\subsubsection*{Summary}

For a given spatial region $A$ and its complement $\bar{A}$, the information contained in the state of a field theory on $A \cup \bar{A}$ decomposes naturally as
\be
|\Psi \rangle \to \rho_A + \rho_{\bar{A}} + {\rm entanglement \; info} \; .
\ee
We have seen that the two density matrices likely encode the physics in the entanglement wedges associated with $A$ and $\bar{A}$. In this case, the remaining region of the spacetime, which lies within the light cone of the bulk surface $\tilde{A}$, should correspond to the information about how the two field theory subsystems are entangled with one another.

To highlight this point, we can consider field theory states for which simple observables contained within the regions $A$ and $\bar{A}$ are almost unchanged, but for which the entanglement between the two subsystems is reduced. For example, considering the vacuum state of the CFT on a sphere with ball-shaped regions $B$ and $\bar{B}$, we recall that $\rho_B$ and $\rho_{\bar{B}}$ represent thermal ensembles with respect to the Hamiltonians $H_B$ and $H_{\bar{B}}$. Choosing typical pure states $|E_B \rangle$ and $|E_{\bar{B}} \rangle$ from these ensembles, we can consider the product state $|\Psi' \rangle = |E_B \rangle \otimes |E_{\bar{B}} \rangle$.\footnote{To make this precise, we should consider a regulated version of the field theory.}

\begin{figure}
\centering
\includegraphics[width = 0.7 \textwidth]{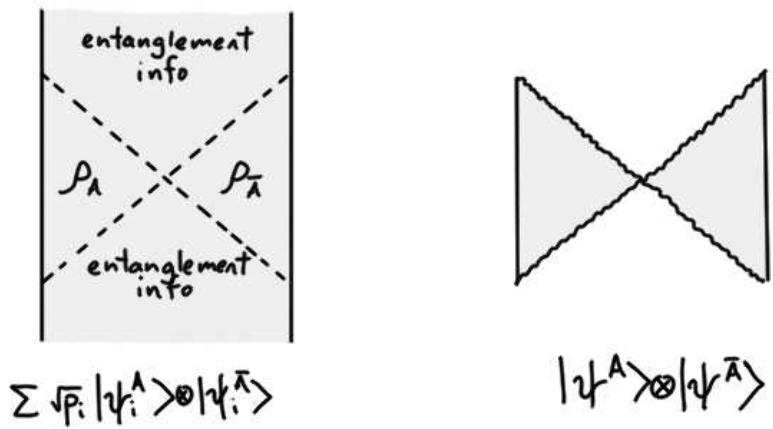}
\caption{Spacetime interpretation for the decomposition of quantum information into the density matrices for complementary subsystems plus entanglement information. The subsystem density matrices encode information about the corresponding entanglement wedges in the bulk, the remaining geometry is encoded in the details of how the two subsystems are entangled. Replacing the state with a product state $|\psi^A \rangle \otimes |\psi^{\bar{A}} \rangle$ where $|\psi^A \rangle$ and $|\psi^{\bar{A}} \rangle$ are typical states in the ensembles $\rho_A$ and $\rho_{\bar{A}}$, the resulting geometry corresponds to two disconnected wedges with singular boundaries.}
\label{disentangle}
\end{figure}

In this new state, almost any simple observable localized within $B$ or $\bar{B}$ should have an expectation value that agrees with the expectation value in the state $|\Psi \rangle$; this is the basic expectation of thermodynamics. In particular, the observables such as response functions used to deduce the spacetime physics dual to $\rho_B$ will give nearly the same results as before, so we can say that interiors of the wedges dual to $\rho_B$ and $\rho_{\bar{B}}$ will be essentially unchanged. On the other hand, the entanglement between the two subsystems in the new state has gone to zero. Taking the RT formula literally in this case, we would conclude that the area of the minimal surface in the bulk spacetime separating $B$ and $\bar{B}$ has gone to zero.\footnote{Alternatively, it may be that the RT formula breaks down somehow, which presumably also indicates a failure of classical geometry.} This suggests some kind of singularity at the boundary of the bulk wedges associated with $B$ and $\bar{B}$. Thus, in the state with entanglement between $B$ and $\bar{B}$ removed, we expect that the dual spacetime still includes two regions that are essentially the same as the AdS Rindler wedges $R_B$ and $R_{\bar{B}}$, but instead of joining smoothly to produce global AdS spacetime, we just have two disconnected wedges each ending at some type of bulk singularity. These observations are summarized in figure \ref{disentangle}.

These observations add weight to the suggestion that classical spacetime is connected via entanglement in the CFT. Indeed, if we consider {\it any} point $p$ in AdS and {\it any} spacelike codimension-two plane $S_p$ in the tangent space to that point, there will be precisely one extremal surface through $p$ tangent to $S_p$ and ending on the boundary of a ball-shaped region $B$ on the AdS boundary. Removing the entanglement between $B$ and $\bar{B}$ in the CFT as in the previous paragraph, we get two disconnected Rindler wedges in the bulk (as in figure \ref{disentangle}), each ending at the point $p$. Thus, the fact that the space on the two sides of the plane $S_p$ is connected smoothly at the point $p$ is directly related to the CFT entanglement between $B$ and $\bar{B}$.

\subsubsection*{Aside: Black holes and firewalls}

In 2012 and subsequently, there was a great deal of discussion following a paper \cite{AMPS} of Almheiri, Marolf Polchinski, and Sully (AMPS) on the issue of whether black hole horizons are smooth as is expected from classical general relativity, or whether some quantum effects lead to singular behavior there.

The argument in (\cite{AMPS}) {\it very roughly} goes as follows; see the paper for details. In order to have smooth spacetime at a black hole horizon, we must have entanglement between the quantum field theory modes inside and outside the horizon as in our discussion of Rindler space above. As a black hole evaporates, it becomes entangled with its outgoing Hawking radiation. For a sufficiently old black hole, this entanglement becomes maximal. The inside-the-horizon modes must then be entangled with some radiation subsystem and therefore (due to basic ``monogamy'' constraints on entanglement) cannot also have the entanglement with the outside-the-horizon modes required to ensure a nonsingular space. Thus, the horizon of an old black hole must be singular i.e. a firewall.

Many arguments for and against firewalls appeared in the literature in the months and years following the AMPS paper. Summarizing these would be beyond the scope of the current notes; however, I'll mention a couple of interesting connections to the discussion in this section.

First, there is a possible way to avoid the firewall conclusion proposed by Maldacena and Susskind \cite{Maldacena:2013xja} using some of the ideas discussed above. As we have discussed, in the example of the maximally-extended AdS-Schwarzschild black hole, the spacetime behind the horizon is encoded in the structure of entanglement between the two CFTs. From the bulk point of view, we can interpret this entanglement as being between the black holes in the two separate asymptotically AdS spacetimes. Thus, the existence of smooth spacetime behind the horizon actually {\it requires} the entanglement of the black hole with some other system. If something like this is also true in the AMPS example of an evaporating black hole, it might be that the entanglement of the black hole with its outgoing radiation does not forbid a smooth spacetime behind the horizon, but rather is somehow the origin of it!\footnote{We have argued that entanglement between fundemental degrees of freedom underlies the connectivity of spacetime. Maldacena and Susskind's suggestion in \cite{Maldacena:2013xja} is that not only is this entanglement a necessary condition for connectedness, it is also sufficient. That is, they propose that any time two subsystems are entangled (e.g. the black hole with its Hawking radiation), there is some geometrical connection (they envision a ``wormhole'' connecting the black hole exterior with the radiation system). This is known as the ``ER=EPR'' (Einstein-Rosen equals Einstein-Podolsky-Rosen) proposal.}

The best chance for a compelling argument for or against the firewall proposal is within the context of a complete quantum theory of gravity. Thus, it is interesting to ask whether black holes as described in the AdS/CFT correspondence have firewalls.\footnote{In the case of large AdS black holes, the black holes do not evaporate, but rather are in thermal equilibrium with their Hawking radiation.} The following argument \cite{VanRaamsdonk:2013sza} (see also \cite{Marolf:2013dba} for a closely related argument) suggests that either these black holes have firewalls, or that there is no state-independent observable in the CFT that can be used to learn about the geometry behind the horizon.

Consider a high-temperature canonical ensemble $\rho_{thermal}$ of states in a holographic CFT. Typical states in this ensemble correspond to large black hole microstates in AdS. Outside their horizons, we expect that the geometry for each of these states is well-approximated by the AdS-Schwarzschild black hole. We would like to ask whether the black holes corresponding to these microstates have smooth spacetime behind their horizon. Suppose that 1) typical states have smooth spacetime behind their horizon and 2) there is some bounded CFT operator ${\cal O}$ that we can use to verify this (e.g. $\langle {\cal O} \rangle = 1$ if an observer on some particular trajectory encounters some smooth spacetime behind the black hole horizon and less than one otherwise).

\begin{figure}
\centering
\includegraphics[width = 0.7 \textwidth]{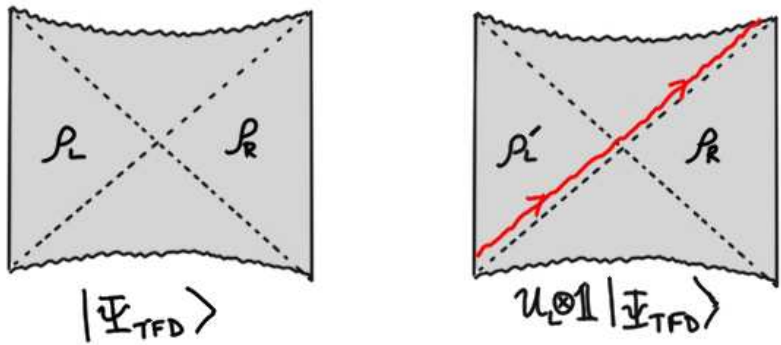}
\caption{Two different purifications of a high-temperature thermal state $\rho_R$ for a CFT on $S^d$. In the thermofield double, the geometry behind the black hole horizon on the right is smooth, while in the perturbed state on the right, there will be a shockwave behind the horizon. In either case, the density matrix $\rho_R$ represents the ensemble of single-sided black hole microstates. Thus, there cannot be a single operator}
\label{firewall}
\end{figure}

In this case, we must have $\tr(\rho_{thermal} {\cal O})= 1$: the expectation value in the canonical ensemble should match with the expectation value in typical states since the canonical ensemble is dominated by these states. The same result will hold for any purification of $\rho_{thermal}$ to some state of a larger system. Thus, we would conclude that the full spacetime dual to any such purification would also have smooth spacetime behind the horizon. This is certainly expected to be true for the thermofield double state, which is dual to the maximally extended black hole geometry. However, it will not be true for other purifications; for example, various states of the form $U_L \otimes \identity_R |\Psi \rangle_{TFD}$ will correspond to perturbations to the maximally extended black hole geometry in which a perturbation falling into the black hole on the left will be detected as a shockwave immediately behind the black hole horizon on the right, as shown in figure \ref{firewall}. Thus, there will be states for which $\tr(\rho_{thermal} {\cal O})= 1$ but the spacetime behind the black hole horizon is not smooth, contradicting our assumptions.

We conclude that {\it either} the typical black hole microstates do not all have smooth spacetime behind their horizon, {\it or} the CFT operators used to probe behind-the-horizon physics are state-dependent, i.e. there is no single operator ${\cal O}$ that we can use for all CFT states to find out if there is smooth geometry.\footnote{Another possibility is the existence of ``superselection sectors'' \cite{Marolf:2012xe}, which is roughly the idea that additional information beyond the CFT state is necessary to determine the dual spacetime.} For more discussion on the issue of state dependence, see \cite{Papadodimas:2015jra} and references therein.

\section{Spacetime physics from entanglement constraints}

We have seen that the entanglement structure of CFT states is directly linked to the geometrical structure of the dual spacetime. It is interesting to ask how much of spacetime dynamics, i.e.  gravitation, can be understood from the physics of entanglement.

It is not hard to see that geometries which capture the entanglement entropies of CFTs states via the RT/HRT formulae must obey certain constraints. We have already seen that entanglement entropies in general quantum systems are constrained, for example by the subadditivity (\ref{subadditivity}) and strong subadditivity (\ref{SSA}) inequalities. For a CFT on some spacetime ${\cal B}$, these constraints govern which functions $S(A)$ from subsets of ${\cal B}$ to real numbers can represent the entanglement entropy of a consistent state.

Starting from an arbitrary asymptotically $AdS$ geometry ${\cal M}$ with boundary ${\cal B}$, the HRT formula also gives us a map $S(A)$ from subsets of ${\cal B}$ to real numbers. In cases where this map does not satisfy the restrictions arising from entanglement constraints, we can conclude that the geometry $M$ cannot correspond to a consistent CFT state in a theory for which the HRT formula is valid. If it is true that any UV complete theory of Einstein gravity coupled to matter has a CFT dual in which the HRT formula holds, then we must conclude that the spacetime ${\cal M}$ is unphysical.

\begin{figure}
\centering
\includegraphics[width=\textwidth]{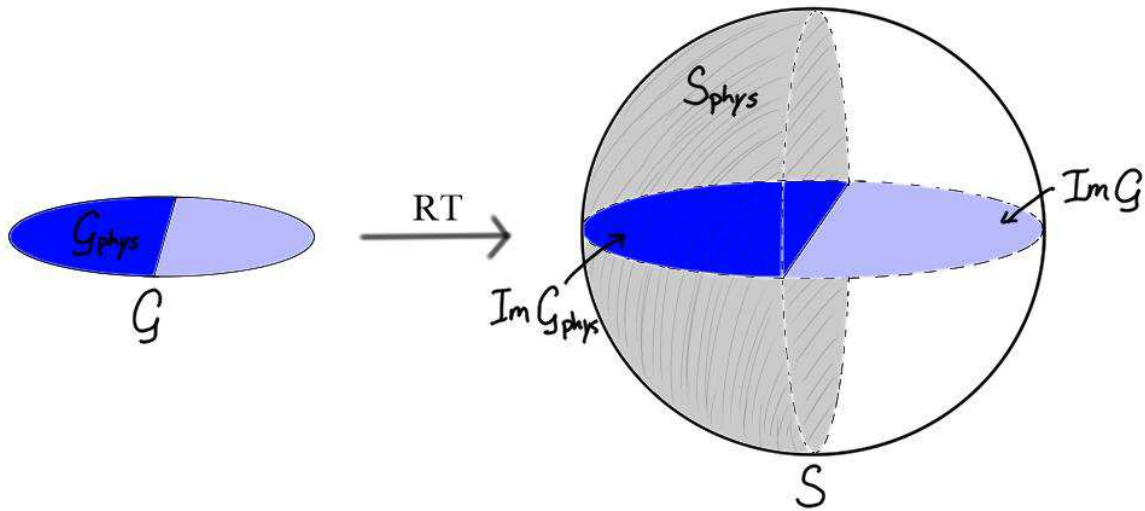}
\caption{Ryu-Takayanagi formula as a map from the space ${\bf \cal G}$ of geometries with boundary ${\cal B}$ to the space ${\bf \cal S}$ of mappings from subsets of ${\cal B}$ to real numbers. Mappings in region  ${\bf \cal S}_{phys}$ (shaded) correspond to physically allowed entanglement entropies. Geometries in region ${\bf \cal G}_{phys}$ map into ${\bf \cal S}_{phys}$ while the remaining geometries are unphysical in any consistent theory for which the Ryu-Takayanagi formula holds (plausibly equal to the set of gravity theories with Einstein gravity coupled to matter in the classical limit).}
\label{intropic}
\end{figure}

In this section, we first describe a variety of basic constraints on entanglement structure, and then understand the implications of these constraints for geometries dual to CFT states.

\subsection{Entanglement constraints}

In this subsection, we review a number of basic constraints on entanglement structure in quantum mechanical theories. We discuss the constraints as they apply to general quantum mechanical systems and then mention the specific results for conformal field theories.  In the next subsection, we will understand how these constraints applied to holographic CFTs translate to gravitational constraints.

\subsubsection*{The first law of entanglement}

We begin with a very useful result for the first order change in entanglement entropy upon variation in the state.

Consider a one-parameter family of states $|\Psi(\lambda) \rangle$ of a general quantum system with subsystem $A$. Then from the definition (\ref{entropy}) of entanglement entropy, the first order variation in the entanglement entropy for the subsystem $A$ is given by\footnote{Here, we should be worried about operator ordering in our manipulations, but using the cyclicity property of the trace, it follows that $d/d \lambda\tr(f(\rho)) = \tr(f'(\rho)d \rho / d \lambda)$.}
\be
\label{firstlaw1}
{d \over d \lambda} S_A = -\tr \left( \log \rho_A {d \over d \lambda} \rho_A  \right)  -\tr \left( {d \over d \lambda} \rho_A \right)
\ee
Since the density matrix $\rho_A$ for any state will have unit trace, the last term vanishes. To simplify the expression further, we can define $H_A = - \log \rho_A(\lambda = 0)$, usually known as the MODULAR HAMILTONIAN associated with the unperturbed density matrix $\rho_A(\lambda = 0)$. With this definition, we can rewrite (\ref{firstlaw1}) as \cite{blanco2013relative}
\be
\label{firstlaw}
{d \over d \lambda} S_A = {d \over d \lambda} \langle H_A \rangle \equiv {d \over d \lambda} \tr( H_A \rho_A) \; .
\ee
We emphasize that the modular Hamiltonian $H_A$ is defined in terms of the unperturbed density matrix and is not a function of $\lambda$; thus, the derivative on the right side acts only on the state $\rho_A$.

The result (\ref{firstlaw}) has become known as the FIRST LAW OF ENTANGLEMENT, since it represents a quantum generalization of the first law of thermodynamics. To see this, consider the special case where the unperturbed density matrix is thermal with respect to some Hamiltonian $H$,
\be
\rho_A = {1 \over Z} e^{-H/T} \; .
\ee
Then (\ref{firstlaw}) gives
\be
\label{firstlaw_thermo}
{d \over d \lambda} S_A = {1 \over T} {d \over d \lambda} \langle H \rangle \; ,
\ee
or more simply, $dE = T dS$. This is the usual first law of thermodynamics, though in the form (\ref{firstlaw_thermo}), we are allowed to apply it to any perturbation of the state, not just perturbations to some nearby equilibrium state.

The entanglement first law (\ref{firstlaw}) is most useful in cases where we have an explicit expression for the modular Hamiltonian, as when the unperturbed density matrix is thermal. For a conformal field theory on Minkowski space, a useful example is provided by the case where the full state is the vacuum state and the subsystem is taken to be a ball-shaped region. In this case, we can use our result (\ref{DMball}),(\ref{modHball}) for the density matrix of the ball to rewrite (\ref{firstlaw}) as \cite{blanco2013relative}
\be
\label{firstlaw_CFT}
{d \over d \lambda} S_B = {\pi \over R} \int_B d^{d-1} x (R^2 - |\vec{x}|^2) {d \over d \lambda} \langle T_{00}(x) \rangle \; .
\ee

\subsubsection*{Positivity and Monotonicity of Relative Entropy}

The entanglement first law (\ref{firstlaw}) has a natural generalization to finite perturbations. In this case, the statement is an inequality, that
\be
\label{RE1}
 \Delta \langle H_A \rangle - \Delta S_A \ge 0 \; .
\ee
To understand the origin of this constraint, we need to introduce a new quantum information theoretic quantity, the RELATIVE ENTROPY of two states.

Relative entropy is a measure of distinguishibility between a (mixed) state $\rho$ and a reference state $\sigma$ associated with the same Hilbert space. It is defined as
\be
\label{REdef}
S(\rho || \sigma) = \tr(\rho \log \rho) - \tr(\rho \log \sigma) \; .
\ee
Relative entropy vanishes if and only if $\rho = \sigma$, and is otherwise a positive quantity. Further, if $A \subset B$ are two subsystems of a general quantum system, it follows that
\be
S(\rho_A || \sigma_A) \le S(\rho_B || \sigma_B) \; .
\ee
Roughly, this says that having access to a larger subsystem allows us to more easily distinguish two states of the full system.

In terms of the modular Hamiltonian $H_\sigma = -\log \sigma$ for the reference state, we can rewrite (\ref{REdef}) as
\bea
S(\rho || \sigma) &=&   [\tr(\rho (-\log \sigma)) - \tr(\sigma (-\log \sigma))] - [\tr(-\sigma \log \sigma) -   \tr(-\rho \log \rho)] \cr
&=& \Delta \langle H_\sigma \rangle - \Delta S
\label{RE2}
\eea
where $\Delta$ indicates the difference between the values for the state $\rho$ and the reference state $\sigma$. Using this expression, positivity of relative entropy immediately gives the finite generalization (\ref{RE1}) of the entanglement first law. Further, monotonicity implies that the difference on the right side of (\ref{RE2}) is non-decreasing as we move to a larger system size.

For the case of a conformal field theory, we have an explicit expression for the modular Hamiltonian when the reference state is taken to be the vacuum state and we consider a ball-shaped region. In this case, we have the explicit constraints that
\be
\label{RE_CFT}
{\pi \over R} \int_B d^{d-1} x (R^2 - |\vec{x}|^2) \langle T_{00}(x) \rangle - \Delta S_B \ge 0.
\ee
and that the left side increases as we deform $B$ to any larger ball (or, more generally to any ball  whose domain of dependence contains the original ball). These generalize the first law constraint (\ref{firstlaw_CFT}) to finite perturbations.

\subsubsection*{Quantum Fisher Information}

For a given reference state $\sigma$, the positivity of relative entropy is equivalent to the statement that $S(\rho||\sigma)$ has a global minimum at $\rho = \sigma$ on the space of density matrices $\rho_A$. Thus, the first order variation away from $\rho= \sigma$ must vanish, while the second order variation must be positive. The first order statement is
\be
\label{firstlawA}
\delta \langle H_\sigma \rangle - \delta S_A \ge 0 = 0 \; ,
\ee
which is exactly the first law of entanglement that we derived directly above. At second order, we have
\beas
\label{deffisher}
S(\sigma + \delta \rho||\sigma) &=& \tr \left(\delta \rho \frac{d}{d\lambda}\log(\sigma+\lambda \delta \rho)\Big |_{\lambda=0} \right) + \tr \left( \sigma {1 \over 2} \frac{d^2}{d\lambda^2}\log(\sigma+\lambda \delta \rho)\Big |_{\lambda=0} \right) \cr
&=& \frac{1}{2}\tr \left(\delta \rho \frac{d}{d\lambda}\log(\sigma+\lambda \delta \rho)\Big |_{\lambda=0} \right) \cr
&\equiv& \langle \delta \rho, \delta \rho \rangle_\sigma \; .
\eeas
To obtain the second line, we have used that
\be\label{tracelesseq}
\tr \left(\delta \rho \frac{d}{d\lambda}\log(\sigma+\lambda \delta \rho)\Big |_{\lambda=0} \right) + \tr \left( \sigma \frac{d^2}{d\lambda^2}\log(\sigma+\lambda \delta \rho)\Big |_{\lambda=0} \right)=0 \; ,
\ee
which follows by taking a  $\lambda$ derivative of  $\tr(\rho(\lambda)\partial_\lambda \log\rho(\lambda))=\tr(\delta\rho)=0$.

The quantity $\langle \delta \rho, \delta \rho \rangle_\sigma$, a quadratic function of the first order perturbations, is known as {\it quantum Fisher information}. It can be promoted to an inner product on the tangent space to the manifold of states at $\sigma$ via
\be
\label{Fishermetric}
\langle \delta \rho_1, \delta \rho_2 \rangle_{\sigma} \equiv {1 \over 2} (\langle \delta \rho_1 + \delta \rho_2, \delta \rho_1 + \delta \rho_2 \rangle - \langle \delta \rho_1 , \delta \rho_1 \rangle - \langle \delta \rho_2, \delta \rho_2 \rangle) \; .
\ee
By the positivity of relative entropy, the quantum Fisher information metric is non-degenerate, non-negative and can be thought of as defining a Riemannian metric on the space of states.

\subsubsection*{A minimal set of strong subadditivity constraints}

We have previously mentioned the constraints of subadditivity (\ref{subadditivity})
\be
\label{SA2}
S(A) + S(B) \ge S(A \cup B)
\ee
and strong subadditivity
\be
\label{SSA2}
S(A \cup B) + S(B \cup C) \ge S(B) + S(A \cup B \cup C) ; ,
\ee
where $A$, $B$, and $C$ are disjoint subsystems of a quantum system. For a quantum field theory on fixed spacetime ${\cal B}$, we consider any Cauchy slice of ${\cal B}$. Then the strong subadditivity constraint must apply for the case where $A$, $B$, and $C$ are any three disjoint spatial regions of ${\cal B}$.

In our discussions below, it will be useful to note \cite{Bhattacharya:2014vja, Lashkari:2014kda} that the full set of strong subadditivity constraints is implied by a smaller set, where $B$ is taken to be arbitrary while $A$ and $C$ are taken to be infinitesimal. To see this, consider regions $A$, $B$, $C_1$ and $C_2$. Then the strong subadditivity constraints applied to triples $\{A,B,C_1 \}$ and $\{A,B \cup C_1\, C_2 \}$ give
\beas
S(A \cup B) + S(B \cup C_1) &\ge& S(A \cup B \cup C_1) + S(B) \cr
S(A \cup B \cup C_1) + S(B \cup C_1 \cup C_2) &\ge& S(A \cup B \cup C_1 \cup C_2) + S(B \cup C_1) \; .
\eeas
Adding these, we find
\[
S(A \cup B) + S(B \cup \left\{C_1 \cup C_2 \right\}) \ge S(A \cup B \cup \left\{C_1 \cup C_2 \right\}) + S(B) \; .
\]
Thus, if the strong subadditivity constraint applies for regions $\{A,B,C_1\}$ and regions $\{A,B,C_2\}$, it applies for regions $\{A,B,C_1 \cup C_2\}$. Given general regions $\{A,B,C\}$, we can then divide up $A$ and $C$ into infinitesimal parts $A_i$, and $C_i$. Starting from the strong subadditivity constraint for $\{A_1,B,C_1\}$, we can then use the results above to keep appending infinitesimal parts $A_i$ and $C_i$, eventually building up to the full constraint for $\{A,B,C\} = \{\cup A_i,B, \cup C_i\}$.

\subsubsection*{Two dimensions}

As an example, consider the strong subadditivity constraints in the case of a field theory in 1+1 dimensions.

\begin{figure}
\centering
\includegraphics[width=0.4\textwidth]{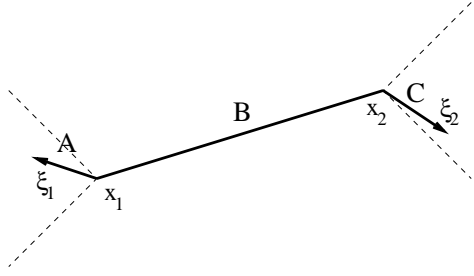}
\caption{Spacelike intervals for strong subadditivity.}
\label{figSSA}
\end{figure}

Let $B$ be the interval $[x_1,x_2]$ while $A$ and $C$ are the intervals $[x_1+\epsilon \xi_1,x_1]$ and $[x_2,x_2 + \delta \xi_2]$, as shown in figure \ref{figSSA}. In this case, the constraint (\ref{SSA}) gives
\beas
&& S([x_1+\epsilon \xi_1,x_1] \cup [x_1,x_2]) + S([x_1,x_2] \cup [x_2 + \delta \xi_2]) \cr && \hskip 2 in \ge S([x_1+\epsilon \xi_1,x_1] \cup [x_1,x_2] \cup [x_2 + \delta \xi_2]) + S([x_1,x_2]) \cr
\implies && S(x_1+\epsilon \xi_1,x_2) + S(x_1,x_2 + \delta \xi_2) - S(x_1+\epsilon \xi_1,x_2 + \delta \xi_2) - S(x_1,x_2) \ge 0
\eeas
Expanding to first order in both $\delta$ and $\epsilon$, this gives
\be
\label{ent_density}
\xi_1^\alpha \xi_2^\beta \partial^1_\alpha \partial^2_\beta S(x_1,x_2) \le 0 \; .
\ee
This second derivative of entanglement entropy has been dubbed the ``entanglement density'' \cite{Bhattacharya:2014vja}.

Since the constraint (\ref{ent_density}) is linear in the spacelike vectors $\xi_1$ and $\xi_2$, it is sufficient to require that the constraint be satisfied in the lightlike limit of $\xi_1$ and $\xi_2$, i.e. when  $\xi_1$ and $\xi_2$ lie along the dotted lines in figure \ref{figSSA}. For example, the constraint for a general spacelike $\xi_1$ is obtained as a linear combination of the constraints for $\xi_1$ in the future and past lightlike directions.

Thus, a minimal set of strong subadditivity constraints that imply all constraints for connected regions is
\be
\label{EDlightlike}
\partial^1_+ \partial^2_+ S(x_1,x_2) \le 0 \qquad
\partial^1_+ \partial^2_- S(x_1,x_2) \le 0 \qquad
\partial^1_- \partial^2_+ S(x_1,x_2) \le 0 \qquad
\partial^1_- \partial^2_- S(x_1,x_2) \le 0\; .
\ee
The latter two constraints are saturated in the case of the vacuum state, so these are likely to be more restrictive in general.

\subsubsection*{Aside: a one sentence proof of the c-theorem}

As an aside, we note that the basic strong subadditivity constraints (\ref{EDlightlike}) lead immediately to a beautiful proof of the c-theorem in general two-dimensional CFTs. In fact, we can write the proof, originally discovered by Casini and Huerta \cite{Casini:2004bw}, using a single sentence. There are many other interesting connections between entanglement and RG flows; see for example \cite{Myers:2010tj,Casini:2016fgb,Casini:2015woa}.
\\
\noindent
{\bf Theorem:} For any Lorentz-invariant local quantum field theory which flows from a UV fixed point described by a CFT with central charge $c_{UV}$ to an IR fixed point described by a CFT with central charge $c_{IR}$, we must have $c_{IR} \le c_{UV}$.
\\
\noindent
{\bf Proof:} Defining $S_{vac}(R)$ to be the vacuum entanglement entropy for an interval of proper length $R$, and using
\be
S([x_1,x_2]) = S_{vac}\left(\sqrt{(x_2 - x_1)^\mu (x_2 - x_1)_\mu}\right) \;
\ee
by Lorentz invariance, we find that the third or fourth strong subadditivity constraints in (\ref{EDlightlike}) applied to an interval of proper length $R$ gives
\be
\label{defC}
c'(R) \le 0 \qquad \qquad c(R) \equiv R {d S \over d R} \; ,
\ee
which proves the c-theorem, since in the limits of small and large $R$ where the result is governed by the UV and IR CFTs respectively, (\ref{S2D}) shows that the monotonically decreasing function $c(R)$ reduces to $c_{UV}$ and $c_{IR}$.

\subsection{Gravitational interpretation of the constraints for holographic CFTs}.

We would now like to understand the gravitational interpretation of the various constraints discussed above applied to subsystems of a holographic CFT.

\subsubsection{The entanglement first law in holographic CFTs: deriving the Einstein Equations for small perturbations to AdS}

First consider the entanglement first law (\ref{firstlaw_CFT}), applied to a one parameter family of states $|\Psi(\lambda) \rangle$ for which there are asymptotically AdS spacetimes ${\cal M}(\lambda)$ which compute the spatial entanglement entropies for the states $|\Psi(\lambda) \rangle$ via the HRT formula. While we have in mind that these states are part of some holographic CFT, we will be able to derive the constraints on ${\cal M}$ without assuming anything but the validity of the HRT formula.

Since the vacuum state $|\Psi(0) \rangle$ corresponds to pure AdS spacetime, small perturbations to this state should be represented by spacetimes that are close to pure AdS. Using Fefferman-Graham coordinates (\ref{FG}) we can represent these by
\be
\label{FGpert}
ds^2 = { \ell^2 \over z^2} \left( dz^2 + dx_\mu dx^\mu + z^d H_{\mu \nu}(z,x) dx^\mu dx^\nu \right)
\ee
where $H$ vanishes in the limit $\lambda \to 0$. We would like to understand what the first law (\ref{firstlaw_CFT}) tells us about the perturbation $H$.

\subsubsection*{First law for infinitesimal balls}

It is useful to first consider the case where the ball-shaped region is an infinitesimal ball around some point $x_0$ \cite{faulkner2014gravitation}. In this case, the right-hand side of (\ref{firstlaw_CFT}) depends only on the expectation value of the energy density at the point $x_0$. Using the HRT formula, the left side will be determined by the area of the extremal bulk surface $\tilde{B}$ with the same boundary as the infinitesimal ball. This surface is localized near the boundary of AdS, so its area depends only on the asymptotic behaviour of the metric function $H$ in (\ref{FGpert}). A direct calculation shows that the first law translates to
\be
\label{holoT1}
\frac{d \ell^{d-3}}{16 \pi G_N}\,  {d \over d \lambda} H^{i}{}_{i} (x_0,z=0) =  {d \over d \lambda} \langle T_{00}(x_0) \rangle \ .
\ee
Thus, the entanglement first law together with the HRT formula implies that the asymptotic metric of ${\cal M}$ must be related to the CFT stress tensor expectation value. This is a standard result in AdS/CFT, but we have seen that it follows directly from the HRT formula.

We can obtain a more covariant version of (\ref{holoT1}) by noting that we have worked so far with balls in the $t=0$ slice, i.e. in the frame of reference of an observer with four-velocity $u^\mu = (1,\vec{0})$. In term of this vector, we have found that
\be
\label{holoT2}
 {d \over d \lambda} u^\mu u^\nu \langle T_{\mu \nu} \rangle = \frac{d \ell^{d-3}}{16 \pi G_N}\,  {d \over d \lambda} u^\mu u^\nu \left(H_{\mu \nu} - \eta_{\mu \nu} H^\alpha {}_\alpha \right)|_{z=0}  \ .
\ee
But the same calculation goes through in any frame of reference, so the result (\ref{holoT1}) must hold for any timelike $u$. This is possible if and only if
\be
\label{holoT3}
 {d \over d \lambda} \langle T_{\mu \nu} \rangle = \frac{d \ell^{d-3}}{16 \pi G_N}\,  {d \over d \lambda} \left(H_{\mu \nu} - \eta_{\mu \nu} H^\alpha {}_\alpha \right)|_{z=0}  \ .
\ee
For any conformal field theory, translation invariance and scaling symmetry imply that the stress tensor expectation value must be conserved and traceless,
\be
\partial_\mu \langle T^{\mu \nu} \rangle = \langle T^{\alpha} {}_\alpha \rangle = 0 \; .
\ee
From (\ref{holoT2}), the dual geometry must then satisfy the constraints that
\be
\label{asymptconstr}
\partial_\mu H^{\mu \nu}|_{z=0} = H^\alpha {}_\alpha|_{z=0} = 0 \; .
\ee
These asymptotic constraints give our first restrictions on the geometry of a spacetime dual to a CFT state. Taking these into account, we can simplify (\ref{holoT2}) to
\be
\label{holoT}
 {d \over d \lambda} \langle T_{\mu \nu} \rangle = \frac{d \ell^{d-3}}{16 \pi G_N}\,  {d \over d \lambda} H_{\mu \nu} |_{z=0}  \ .
\ee

\subsubsection*{First law for general balls}

Now consider the first law (\ref{firstlaw_CFT}) applied to general balls \cite{Lashkari:2013koa,faulkner2014gravitation}. Using the result $(\ref{holoT})$ and the HRT formula, we can translate the first law constraint directly to a gravitational constraint as
\be
\label{firstlaw_grav1}
{d \over d \lambda} {\rm Area}(\tilde{B}) = {d \ell^{d-3} \over 4 R} \int_B d^{d-1} x (R^2 - |\vec{x}|^2) {d \over d \lambda} H_{00}(x,z=0)
\ee
The left hand side can be written more explicitly starting from the area functional (\ref{Area_explicit}). To calculate the variation in area, we can use the fact that the original surface extremizes the area functional for the original metric. Thinking of the area as a functional ${\rm Area}(G,X)$ depending on both the spacetime metric $G$ and the embedding function $X$, we can write the first order variation schematically as
\be
\delta {\rm Area} = {\delta {\rm Area} \over \delta G} \delta G + {\delta {\rm Area} \over \delta X} {\delta X \over \delta G} \delta G \; .
\ee
However, the surface $X$ being extremal means that the first variation of the area with respect to variations in $X$ vanishes, so only the first term here is nonzero, i.e. we need only compute the variation of the action with respect to $G$ while keeping the surface $X$ fixed.

The basic relation $\det(M) = \exp(\tr(\ln(M)))$ can be used to show that
\be
{d \over d \lambda} \sqrt{\det{g_{ab}}} = {1 \over 2}  \sqrt{\det{g_{ab}}} g^{cd} {d \over d \lambda} g_{cd}
\ee
where $g_{ab}$ is the induced metric (\ref{Induced}). Using the spatial coordinates $x^i$ to parameterize the surface, and the explicit form $Z(x) = \sqrt{R^2-\vec{x}^2}$ for the surface, we find for the ball of radius $R$ centered at the origin that
\bea
{d \over d \lambda} {\rm Area}(\tilde{B}) &=& {1 \over 2} \int_{\tilde{B}}  \sqrt{\det{g_{ab}}} g^{cd} {d \over d \lambda} g_{cd} \cr
&=& {\ell^{d-3} \over 2 R} \int_{\tilde{B}} d^{d-1} x {d \over d \lambda}(R^2 H_{ii} - x^i x^j H_{ij})
\eea
Thus we have finally an explicit form of the constraint
\be
\label{firstlaw_grav}
\int_{\tilde{B}} d^{d-1} x (R^2 \delta H_{ii} - x^i x^j \delta H_{ij}) = {d  \over 2 } \int_B d^{d-1} x (R^2 - |\vec{x}|^2) \delta H_{ii}(x,z=0)
\ee
where $\delta H$ represents the metric perturbation at first order in $\lambda$, and we have used the tracelessness condition (\ref{asymptconstr}). Below, we will refer to the left and right sides of this expression as $\delta S^{grav}_B$ and $\delta E^{grav}_B$ respectively.

This result relates the metric perturbation in the interior of the geometry to asymptotic metric perturbation. It will only be satisfied for some special choices of $H$. We actually have an infinite number of such constraints, one for each pair $(x_0^i,R)$ labeling the center and radius of the ball. Associating each such constraint to a bulk point $(x = x_0, z = R)$ at the tip of the extremal surface, we note that there is one constraint for each bulk point. This motivates the possibility that this infinite family of non-local constraints is actually equivalent to a single local equation. We will now see that this is the case.

\subsubsection*{Converting the nonlocal constraints into a local equation}

\begin{figure}
\centering
\includegraphics[width=0.25\textwidth]{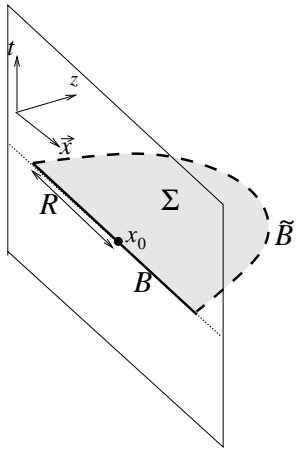}\\
\caption{Notation for regions in AdS$_{d+1}$, with radial coordinate $z$ and boundary space coordinate $\vec{x}$. $B$ is a $(d-1)$-dimensional ball on the $z=0$ boundary of radius $R$ centered at some position $\vec{x}_0$ on the spatial slice at time $t_0$. $\tilde{B}$ is the $(d-1)$-dimensional hemispherical surface in AdS ending on $\partial B$, and $\Sigma$ is the enclosed $d$-dimensional spatial region.}
\label{notation2}
\end{figure}

To obtain a local constraint on $H$ from (\ref{firstlaw_grav}), the approach is similar to the steps used to convert the integral form of Maxwell's equations to the differential form. Essentially, we use Stokes theorem. In detail, it turns out that for each choice of $B$, there exists a differential form $\chi_B$ with the following three properties:\footnote{If you must know, the explicit form of $\chi$ for the ball of radius $R$ centered at $x_0$ is
\bea
  \chi|_\Sigma &=&  {z^d \over 16\pi G_N}\left\{   \epsilon^t {}_z \left[\left({2\pi z \over R} + {d \over z} \xi^t + \xi^t \partial_z\right) H^i{}_i\right]+ \right. \\
  & & \left.\quad\quad +  \epsilon^t {}_i \left[ \left({2\pi (x^i-x_0^i) \over R} + \xi^t \partial^i \right) H^j{}_{j} - \left({2\pi (x^j-x_0^j) \over R} + \xi^t \partial^j \right) H^i{}_{j} \right]\right\} \notag
\eea
where $\xi^t = {\pi \over R} (R^2 - z^2 -|\vec{x}-\vec{x}_0|^2)$ and $
\epsilon_{ab} = \sqrt{-g} \epsilon_{a b c_1 \cdots c_{d-2}} dx^{c_1} \wedge \cdots \wedge dx^{c_{d-2}}$. We will understand the origin of this expression in the section ``What is the form $\chi$'' below.}
\begin{itemize}
\item
\be
\label{prop1}
\int_B \chi = \delta E^{grav}_B
\ee
\item
\be
\label{prop2}
\int_{\tilde{B}} \chi = \delta S^{grav}_B
\ee
\item
\be
\label{prop3}
d \chi = 2 \xi_B^0(x) \delta E_{00}(x) {vol}_\Sigma \;   .
\ee
\end{itemize}
In the last line, $\xi_B^0$, defined in (\ref{defxi}) below is a positive function in the spatial region $\Sigma$ between $B$ and $\tilde{B}$ (see figure \ref{notation2}), ${vol}_\Sigma$ represents the volume form on this region, and
\be
\delta E_{00}(x) \propto z^d \left(\partial_z^2 H^i{}_{i} + {d+1 \over z} \partial_z H^i{}_{i} +  \partial_j \partial^j H^i{}_{i} -  \partial^i \partial^j H_{ij}\right)
\ee
is (up to a constant) the time-time component of the Einstein equations linearized about AdS.

With these properties, it is straightforward to convert the nonlocal constraints (\ref{firstlaw_grav}) into a local equation. We have
\bea
&& \delta E^{grav}_B = \delta S^{grav}_B \cr
&\Leftrightarrow& \int_B \chi = \int_{\tilde{B}} \chi \cr
&\Leftrightarrow& \int_{\partial \Sigma} \chi = 0 \cr
&\Leftrightarrow& \int_\Sigma d \chi = 0 \cr
&\Leftrightarrow& \int_\Sigma \zeta_B^0(x) \delta E_{00}(x) {vol}_\Sigma = 0
\eea
This must be true for all possible half-ball regions $\Sigma$ defined by $\{t = t_0, z < \sqrt{R^2 - (\vec{x} - \vec{x}_0)^2}\}$. It is not hard to show \cite{faulkner2014gravitation} that the vanishing of all of these integrals will hold if and only if
\be
\delta E_{00}(x) = 0 \; .
\ee
Thus, the entanglement first law constraints for ball-shaped regions in slices of constant $t$ implies the time-time component of the linearized bulk Einstein equations. To obtain the remaining components, we first repeat the argument for ball-shaped regions in general frames of reference. Using precisely the same reasoning that we used leading up to $(\ref{holoT2})$, we obtain
\be
\delta E_{\mu \nu} = 0 \; ,
\ee
where $\mu$ and $\nu$ correspond to the field theory directions. The remaining equations,
\be
\delta E_{z \nu} = \delta E_{z z} = 0
\ee
are constraint equations. They hold everywhere if they hold at $z=0$, and it turns out that at $z=0$ they are precisely equivalent to the asymptotic constraints (\ref{asymptconstr}) that we obtained by considering small balls.

We have therefore shown that the first law of entanglement for ball-shaped regions in the CFTs is precisely equivalent to the linearized Einstein equations in the gravity dual theory \cite{Lashkari:2013koa,faulkner2014gravitation}.

\subsubsection*{Connection to black hole thermodynamics}

The equivalence of the entanglement first law with the linearized Einstein's equations is closely related to the the first law of black hole thermodynamics that served as one of the initial motivations for the connection between entropy and geometry. As we described above, a Rindler wedge of pure AdS is equivalent to a Schwarzschild-AdS black hole with non-compact hyperbolic horizon geometry, at a special value of the temperature. While this spacetime bears little resemblance to our standard notion of a black hole, it turns out that a general form of the first law of black hole thermodynamics applies to this spacetime. The result, proved by Wald and Iyer \cite{Iyer:1994ys} (we review this below), says that for any perturbation satisfying the Einstein's equations linearized about this background (and more generally, any black hole background with a ``bifurcate Killing horizon''), the change in area of the horizon is equal to the change in a certain energy defined based on the asymptotic metric. It turns out that this energy is exactly the quantity $\delta E^{grav}_B$ coming in on the right hand side of (\ref{firstlaw_grav}). Thus, the black hole first law is equivalent to the assertion that the linearized Einstein equation implies the gravitational version of the entanglement first law. The result that the linearized Einstein equation follows from the first law (applied to all boundary balls in all Lorentz frames) thus provides a converse to the theorem of Wald and Iyer.

\subsubsection*{Generalizations}

There are various generalizations of these first-order results. First, in higher-derivative theories of gravity, it is known that the correct gravitational expressions for black hole entropy (defined so that the laws of black hole thermodynamics work out correctly) involve more general covariant functionals rather than area. It is possible to show that starting with these more general functionals the entanglement first law implies the linearized gravitational equations corresponding to the associated higher-derivative Lagrangian \cite{faulkner2014gravitation}. In this derivation, possible terms in the entanglement entropy functional involving extrinsic curvatures do not contribute.

We can also take into account the leading quantum effects in the gravitational theory by including the quantum corrections (\ref{HRT}) in the HRT formula. For a ball-shaped region, the bulk entanglement term measures the entanglement of perturbative bulk fields inside the associated Rindler wedge with fields outside the wedge. This entanglement can be related directly to the expectation value of the bulk stress-energy tensor via a bulk version of the entanglement first law (\ref{firstlaw}). For this calculation, we require the modular Hamiltonian for a Rindler wedge of AdS, for general quantum field theory in its vacuum state. Fortunately, essentially the same path integral calculation (from appendix \ref{AppPI}) that gives the density matrix for a Rindler wedge of Minkowski space also works in AdS, yielding an expression analogous to (\ref{Rindham}) for the modular Hamiltonian. Using that expression in the first law, we find that the extra bulk entanglement term in the entanglement first law results in a source term for the linearized Einstein equations \cite{swingle2014universality}
\be
\delta E_{ab} = 8 \pi G_N \langle T_{ab} \rangle \; .
\ee
In this derivation, the universality of gravity, that all fields act as sources for gravitation, arises from the universality of entanglement, that all fields contribute to the bulk spatial entanglement entropy.\footnote{Various authors have argued that the linearized gravitational equations plus the assumption that the stress-energy tensor acts as a source for these equations requires that the complete theory is a generally-covariant theory of gravity. With the additional assumption that the black hole entropy functional is simply area (which follows from our assumption of HRT), the theory is further restricted to Einstein gravity coupled to matter. Thus, we have at least an indirect argument for the full non-linear gravitational equations starting from entanglement physics.}

\subsubsection*{What is the form $\chi$?}

For a deeper understanding of the relation between the entanglement first law and the linearized Einstein equations, and to understand the meaning of the form $\chi$ that played a crucial role in the derivation, it is helpful to review some formalism originally introduced by Wald and collaborators (see e.g. \cite{Iyer:1994ys,hollands2013stability}). This formalism is also the basis for the generalization to arbitrary covariant theories of gravity. This section is not essential for understanding the remaining sections, so the reader should not worry about fully absorbing the content here before continuing.

Consider a general classical theory of gravity with covariant Lagrangian ${\cal L}(g)$, where $g$ represents the metric as well as any matter fields. For this theory, the action can be written as the integral over spacetime of a differential form
\be
L = {\cal L} \epsilon
\ee
where
\be
\epsilon = \sqrt{g} \epsilon_{a_1 \cdots a_{d+1}} dx^{a_1} \wedge \dots \wedge dx^{a_{d+1}} \; .
\ee
When we vary the action to determine the equations of motion for the theory, the usual procedure is to make the variation and reorganize the result (via integration by parts) as a total derivative plus a term proportional to the variation $\delta g$ without derivatives on $\delta g$. Requiring that this variation vanishes for any $\delta g$ means that the term multiplying $\delta g$ must vanish. In terms of the differential form $L$, this means that we can write its variation under a variation of the metric as
\be
\delta L = E \cdot \delta g + d \theta
\ee
where $E$ represent quantities defined such that $E=0$ are the equations of motion, and $\cdot$ represents the various index contractions for the different fields (e.g. $E^{ab} \delta g_{ab}$ for the term proportional to the metric variation). The term $d \theta$ is the exterior derivative of a $d$-form $\theta$; this is the total derivative term that we separate off when integrating by parts.

The quantity $\theta$ plays an important role in the phase space formulation of our gravitational theory. We recall that in general classical mechanics systems, dynamics can be understood as evolution in phase space governed by Hamilton's equations $\dot{X} = \xi_H$ where $\xi_H$ is a vector field on phase space defined in terms of a symplectic two-form $W$ and a Hamiltonian function $H$ by
\be
\label{Ham}
\xi_H \cdot W = d H \, .
\ee
For our gravitational system, the symplectic form acts on a pair $(\delta g_1 \delta g_2)$ of deformations to the metric, and is given by
\be
W(g, \delta g_1, \delta g_2) = \int_\Sigma \omega(g, \delta g_1, \delta g_2)
\ee
where $\omega$ is defined in terms of $\theta$ as
\be
\omega(g, \delta g_1, \delta g_2) = \delta_1 \theta(g, \delta g_2) - \delta_2 \theta(g, \delta g_1) \; .
\ee
Now, consider the metric transformation $g \to g + {\cal L}_X g$ induced by a vector field $X$, where $({\cal L}_X g)_{ab} = \nabla_a X_b + \nabla_b X_a$ is the Lie derivative. This induces a flow on the gravitational phase space. Using the definition (\ref{Ham}), we can associate a phase space Hamiltonian $H_X$ to this flow if for any other perturbation $\delta g$ (viewed as second vector field on phase space to which the one forms on the left and right side of (\ref{Ham}) are applied) we have
\be\label{vh}
\delta H_X = W(\delta g, {\cal L}_X g) \, .
\ee
As we will see below, certain very useful notions of energy in general relativity can be defined in this way as the phase space Hamiltonian corresponding to specific vector fields.

The form $\theta$ also appears in the definition of conserved quantities associated with symmetries of the gravitational theory. Since the theory is diffeomorphism invariant, the coordinate transformation $\delta x = X$ for any vector field $X$ represents a symmetry of the theory. By the usual Noether procedure, we can find an associated conserved current, which turns out to be
\be
J_X = \theta(\delta_X g) - X \cdot L \ .
\ee
This has been written as a differential form, so that the equation of current conservation is simply
\be
d J_X = 0 \; .
\ee
Because this holds for all vector fields $X$, it follows \cite{Iyer:1994ys}, that we can find a $(d-1)$-form $Q$ such that
\be
\label{defJ}
J_X = d Q_X + X^a C_a
\ee
where $C_a$ are quantities proportional to the equation of motion that vanish on shell.

In terms of these definitions, the form $\chi$ used in the previous section is defined in terms of $X$ and an arbitrary variation of the fields as:
\be
\label{defchi}
\chi(g, \delta g) = \delta Q_X(g) - X \cdot \theta(\delta g)
\ee
The key property of $\chi$ is that it obeys an identity
\be
\label{fundid}
d \chi = \omega(g, \delta g, {\cal L}_X g) - X \cdot (E \delta g) - X^a \delta C_a \; .
\ee
In the case where the background metric $g$ satisfies the equations of motion $E=0$ and admits a Killing vector $X = \xi$ so that ${\cal L}_\xi g = 0$, this simplifies to
\be
\label{prop3a}
d \chi = - \xi^a \delta C_a \; .
\ee
Recalling that $C_a$ vanishes when the equations of motion are satisfied, we have that $\delta C_a$ is a quantity that vanishes when the variation $\delta g$ satisfies the equations of motion linearized about the background $g$.

This result leads directly to Iyer and Wald's derivation of the first law of black hole mechanics. In this case, we consider a black hole with a Killing horizon (i.e. a horizon where some Killing vector in the black hole exterior vanishes), and take $\xi$ to be the Killing vector vanishing at the horizon. Then for any on-shell variation of the metric, we have $d \chi = 0$. Integrating this from the spacetime boundary $B$ to the horizon $\tilde{B}$, we get $\int_B \chi = \int_{\tilde{B}} \delta Q_\xi$. For Einstein gravity coupled to matter, it turns out that the expression on the right is proportional to the variation in the area of the black hole horizon, while the expression on the left has the interpretation of energy, and we get the first law $dE = T dS$ once we identify $S = Area / 4 G$. For more general theories of gravity, everything still works, but $Q_\xi$ is no longer the area. So we have this more general quantity (usually called the Wald entropy functional) that describes the black hole entropy.

For our application to the previous section, we take $g$ to be pure AdS, and $\xi$ to be a Killing vector (described below around equation (\ref{defchi})) that vanishes at the extremal surface $\tilde{B}$. Then (\ref{prop3a}) is exactly the condition (\ref{prop3}) that we needed above. As in Wald's derivation, the integral of $\chi$ over $\tilde{B}$ is the variation in the area of $\tilde{B}$ under the metric perturbation, while the integral of $\chi$ over the boundary region $B$ has an interpretation as a variation in energy. These turn out to be exactly the conditions (\ref{prop1}) and (\ref{prop2}) that we needed above.

\subsubsection{Gravitational consequences of relative entropy inequalities}

For ball-shaped regions in a holographic CFT, the expression for relative entropy (\ref{RE_CFT}) can be translated directly to a gravitational quantity using the HRT formula and the result (\ref{holoT}) for the gravitational interpretation of the stress tensor expectation value. The resulting quantity, which we denote as $\Delta E^{grav} - \Delta S^{grav}$, must be positive and monotonic with increasing system size.\footnote{The application of relative entropy in holography was initiated in \cite{blanco2013relative}.} We will now see that this quantity has the interpretation as an energy associated to a bulk region that lies between the boundary domain of dependence region $D_B$ and the extremal surface $\tilde{B}$.

\subsubsection*{Positivity of Fisher Information}

We start by considering the interpretation of relative entropy for states that are small perturbations to the vacuum. We recall that the leading contributions to relative entropy come at second order and define the Fisher information (\ref{deffisher}).

\begin{figure}
\centering
\includegraphics[width=0.25\textwidth]{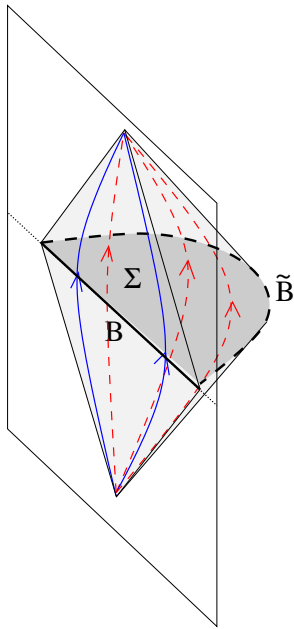}\\
\caption{AdS-Rindler patch associated with a ball $B(R, x_0)$ on a spatial slice of the boundary. Solid blue paths indicate the boundary flow associated with $H_B$ and the conformal Killing vector $\zeta$. Dashed red paths indicate the action of the Killing vector $\xi$.}
\label{hyperbolic2}
\end{figure}

On the gravity side, the positivity of Fisher information implies the positivity of a certain perturbative gravitational energy \cite{Lashkari:2015hha}. To define this, consider the Rindler wedge of the unperturbed AdS spacetime associated with the boundary ball $B$. Associated to this is a bulk Killing vector $\xi_B$ that vanishes on the Rindler horizon (i.e. the extremal surface $\tilde{B}$ in the unperturbed spacetime) and asymptotes to the boundary conformal Killing vector (\ref{defzeta}) near the AdS boundary, as shown in figure \ref{hyperbolic2}. Explicitly, in the Rindler wedge associated with a ball of radius $R$ situated at the origin on the boundary Minkowski space, the Killing vector takes the form
\be
\label{defxi}
\xi_B  = - \frac{2\pi}{ R}  t [z \partial_z + x^i \partial_i ] + \frac{\pi}{R} [R^2 - z^2 - t^2 - \vec{x}^2] \, \partial_t
\ee
We can verify that this satisfies ${\cal L}_\xi g_{AdS} = 0$, vanishes at $\vec{x}^2 + z^2 = R^2$ and reduces to (\ref{defzeta}) in the region $D_B$ at the boundary of AdS.  For the quantum field theory describing perturbative bulk fields on the pure AdS background, this timelike Killing vector defines to a symmetry, so there is an associated conserved energy, which includes contributions from all fields including the metric perturbation. This takes the form
\be
\label{defCE}
\int_\Sigma \xi_B^a (T^{matter}_{ab} + T^{grav}_{ab}) \epsilon^b
\ee
where $\Sigma$ is a Cauchy surface for the Rindler wedge, and the gravitational stress tensor is defined (up to a boundary term) as the expression quadratic in the first order metric perturbation that acts as a source for the second order metric perturbation in the perturbative Einstein equations.\footnote{To derive this result, we can use the results (\ref{prop1}) and (\ref{prop2}) that $\delta S^{grav}$ and $\delta E^{grav}$ are related to the integral of the form $\chi$ over $\tilde{B}$ and $B$ respectively. Assuming the background and perturbations satisfy the equations of motion, we can then use the identity (\ref{fundid}) to obtain
\be
{d \over d \lambda} (\delta S^{grav} - \delta E^{grav}) = \int \omega(g, {d \over d \lambda} g, {\cal L}_X g) \; ,
\ee
which can be shown to hold at any point on a one-parameter family of solutions. Taking another derivative and setting $\lambda = 0$ so that $g = g_{AdS}$, we get
\be
{d^2 \over d \lambda^2} (\delta S^{grav} - \delta E^{grav})|_{\lambda = 0} = \int \omega(g, {d \over d \lambda} g, {\cal L}_X {d \over d \lambda} g) \; .
\ee
The expression on the right is defined by Hollands and Wald \cite{hollands2013stability} to be the canonical energy associated with $\xi$, and shown to be equivalent to (\ref{defCE}).
}
This quantity is known as the canonical energy associated with the Killing vector $\xi_B$. Thus, Fisher information in the CFT is dual to canonical energy on the gravity side, and the relative entropy inequalities imply that this energy must be positive for the Rindler wedges associated with all possible boundary balls and must increase as we increase the size of the ball.

\subsubsection*{Relative entropy constraints for general asymptotically AdS geometries}

The fact that relative entropy corresponds to a natural gravitational energy at second order in perturbations to pure AdS suggests that this quantity might be related to some notion of energy more generally. This is very interesting, as it is a notoriously difficult problem to come up with precise covariantly defined notions of energy for gravitational subsystems.

For a general asymptotically AdS spacetime, there will be no Killing vectors, and no preferred definition of time apart from the boundary time used to define the ADM mass of the whole spacetime. It is therefore unclear a priori how to define useful notions of energy for subsystems. However, the following definition turns out to be the precise dual of relative entropy associated with a ball-shaped region $B$ \cite{Lashkari:2016idm}. In place of the Killing vector (\ref{defxi}) defined for a Rindler wedge of AdS, we define a vector $\xi$ in the entanglement wedge associated with $B$ that has the same behavior as the Rindler wedge Killing vector near the extremal surface and near the AdS boundary. Specifically, we demand that $\xi$ reduces to the conformal Killing vector $\zeta_B$ at the AdS boundary region $D_B$, that $\xi$ vanishes at the extremal surface $\tilde{B}$, and that the Killing equation $({\cal L}_\xi g)_{ab}  = \nabla_{(a} \xi_{b)}= 0$ is satisfied at the boundary (up to some order in $z$) and at the extremal surface. We also require that the antisymmetrized derivative at the extremal surface has the same behavior as for the Rindler wedge Killing vector. It is always possible to satisfy these conditions, and in fact there is a great deal of freedom in choosing $\xi$ since it is essentially unconstrained away from the boundaries of the entanglement wedge.

Though our general asymptotically AdS spacetime $M$ has no Killing vectors, the vector $\xi$ generates a diffeomorphism, and this is a symmetry of the gravitational theory. Thus, we can define an associated symmetry current (the quantity $J_\xi$ defined in (\ref{defJ})) and integrate over a spatial slice of the entanglement wedge to define the corresponding charge $E_B$. Remarkably, this charge is independent of the choice of $\xi$, and it is exactly the quantity dual to relative entropy. Thus, if the spacetime $M$ does correspond to some consistent CFT state, the energy $E_B$ associated with the entanglement wedge of $B$ must be positive for every choice of $B$ and be larger for a larger ball $B'$ with $D_B \subset D_{B'}$.

This result can be understood as a prediction for a new positive energy theorem in general relativity. We have derived it making use of basic quantum mechanical constraints in the underlying CFT description for the theory. However, it is plausible that assuming some standard energy condition in GR would be sufficient to derive the result directly as a theorem in classical general relativity.

\subsubsection*{Example: spherically symmetric geometries}

As a simple example, we can consider the constraints on regular spherically-symmetric asymptotically AdS geometries in four dimensional gravity, for example, the geometry associated with a spherically symmetric distribution of matter obeying some equation of state. If this solution represents a consistent state in a UV-complete theory of gravity, it should correspond to a homogeneous state of some CFT on a sphere. For this state, the relative entropy to the vacuum state for any ball-shaped region must be positive. Taking the region to be a hemisphere, and using the expression (\ref{modHsphere}) for the modular Hamiltonian of a ball in the vacuum state on a sphere, the positivity of relative entropy (\ref{RE1}) yields \cite{Lashkari:2014kda}
\be
\label{REexample}
\Delta A \le 2 \pi G_N  M \ell \; .
\ee
where the left side represents the area of the surface bisecting the spherically-symmetric space relative to this area in pure AdS,\footnote{This surface will be extremal by symmetry. We are assuming it is the minimal-area extremal surface; if not, the results we derive apply to the minimal area extremal surface instead.} and $M$ is the mass of the spacetime. The right side is obtained from the modular Hamiltonian expectation value using the fact that for a translation-invariant stress-tensor on the sphere (which we assume to be of unit radius), we have
\be
\label{spherestress}
\Delta \langle T_{00} \rangle = {E_{CFT} \over 4 \pi} = {M \ell \over 4 \pi} \; .
\ee

The result (\ref{REexample}) tells us that there is a limit to how much a certain mass can deform the spacetime from pure $AdS$. This should restrict the equation of state that matter in a consistent theory of gravity can have. It would be interesting to understand whether this constraint also follows from one of the more standard energy conditions typically assumed in general relativity.

\subsubsection{Strong Subadditivity}

We can also investigate the necessary and sufficient conditions on a dual spacetime $M$ for the strong subadditivity constraints to be satisfied. According to Wall \cite{Wall:2012uf}, the null energy condition is a sufficient condition, but the necessary conditions for strong subadditivity may be weaker, and it is interesting to understand precisely what strong subadditivity tells us about the dual spacetime.

So far, there are some limited results \cite{Lashkari:2014kda,Bhattacharya:2014vja} for the gravitational interpretation of the constraints (\ref{EDlightlike}) in the case of two dimensional CFTs. For a geometry $M_\Psi$ dual to some holographic CFT state, the constraints are that (\ref{EDlightlike}) hold, where $S(x_1,x_2)$ is replaced by the area of the minimal-length geodesic with endpoints $x_1$ and $x_2$ on the boundary of $AdS$. In \cite{Lashkari:2014kda,Bhattacharya:2014vja}, it was found that for geometries dual to the vacuum states of Lorentz invariant field theories corresponding to RG-flows between UV and IR CFTs, and also for non-vacuum states of a CFT corresponding to small perturbations to AdS, the strong-subadditivity constraints are exactly equivalent to the conditions
\be
\label{SSAint}
\int_{\tilde{B}} G_{\mu \nu} k^\mu k^\nu \ge 0
\ee
where $k$ is a null vector field defined along $\tilde{B}$ and normalized so that the spatial scale factor in the geometry changes by the same rate at all points on the curve when we translate by $k$. This represents a version of the null energy condition averaged over a spatial geodesic.
Extending these results to more general geometries and to higher dimensions is an interesting open question.

\section{Which theories and which states are holographic?}

In the previous section, we considered constraints on spacetimes dual to states of holographic conformal field theories. We assumed that for these states (which we will call holographic), the entanglement entropies of spatial regions are all encoded in a single geometry $M$ via the holographic entanglement entropy formula. This represents a very special structure of entanglement:
the space of possible entanglement entropies (functions on a space of subsets of the AdS boundary) is much larger than the space of possible asymptotically AdS metrics (functions of a few spacetime coordinates), so only a tiny fraction of all quantum field theory states should be geometric in this way.

For a general conformal field theory, there is no reason to expect any of the states to have their entanglement represented geometrically. Even for holographic conformal field theories, not all states will have this property: taking a quantum superposition of a small number of states corresponding to different classical geometries, the resulting state will correspond to a quantum superposition of geometries rather than a single classical geometry. Thus, it is an interesting question to understand better which theories have these holographic states and which states of these theories are holographic.

In this section, we will mention a few of the special properties characterizing holographic states and holographic theories, specifically those related to entanglement structure.

\subsubsection*{A necessary condition for CFTs with Einstein gravity duals from entanglement}

For the vacuum state of any conformal field theory, the vacuum entanglement entropies for ball-shaped regions are given by a universal formula (e.g. equation (\ref{S2D}) for the two-dimensional case) that matches with the areas of extremal surfaces in pure Anti de Sitter spacetime. However, taking any more general region $A$, the CFT vacuum entanglement entropy for this region would not be expected to match with the gravity result for the area of the extremal surface $\tilde{A}$ in AdS. Requiring this agreement gives a necessary condition for a CFT to have a holographic dual whose classical limit is Einstein gravity coupled to matter (i.e. for which the entanglement functional is simply area). Since this condition must hold for all possible regions $\tilde{A}$, it should be quite constraining; perhaps it is even a sufficient condition for a CFT  to have an Einstein gravity dual.

One way to investigate this constraint in more detail is to start with the case of a ball-shaped region and work perturbatively in deformations to the shape of the region. For recent calculations along these lines, see \cite{Faulkner:2015csl}.

\subsubsection*{Entanglement entropy from one-point functions in holographic states}

Now consider a theory which has a holographic dual theory with some particular bulk equations of motion in the classical limit. Any classical spacetime $M_\Psi$ dual to a holographic state $|\Psi \rangle$ must be a solution to the bulk equations of motion. We can think of the bulk equations as determining a solution everywhere starting from some initial data on a Cauchy surface. In AdS, it is convenient to think instead of the bulk solution as being determined by evolution in the holographic radial direction, with ``initial data'' specified at the timelike boundary of AdS.\footnote{Here, the existence and uniqueness of solutions is less clear, but we can at least construct solutions perturbatively away from the boundary.} Thus, for geometries dual to holographic states, we can say that the bulk spacetime (at least in a perturbative sense) is encoded in the boundary behavior of the various fields.

The boundary behavior of the light bulk fields is determined according to AdS/CFT by the one-point functions of some associated low-dimension local operators in the CFT. For example, we have seen that the boundary behavior of the metric is determined by the expectation value of the stress-energy tensor as in (\ref{holoT}). Once we have the bulk spacetime solution, we can use the AdS/CFT correspondence to calculate many other non-local quantities in the CFT such as correlation functions and entanglement entropies. Thus, the assumption that a state is holographic allows us (via gravity calculations) to determine the entanglement entropies and other non-local properties of the state from the local data provided by the one-point functions:
\be
\label{Gravrules}
|\Psi \rangle \rightarrow \langle {\cal O}_\alpha(x^\mu)\rangle \rightarrow \phi_\alpha {\rm \; asymptotics}  \rightarrow \phi_\alpha (x^\mu, z) \rightarrow {\rm entanglement \; entropies \;} S(A)
\ee
where $\phi$ here indicates all light fields including the metric.

The recipe (\ref{Gravrules}) could be applied to any state, but for states that are not holographic, the results will be inconsistent with the actual CFT answers. For example, using this recipe, the vacuum one-point functions for any CFT will give us pure AdS spacetime, and we have seen in the previous subsection that entanglement entropies for more general regions will generally not match with those calculated by the holographic formula from this spacetime.

Thus, we have a stringent (in-principle) test for whether a CFT state has a dual description well-described by a classical spacetime: carry out the procedure in (\ref{Gravrules}) and compare the results with a direct CFT calculation of the entanglement entropies; if there is a mismatch for any region, the state is not holographic.

\subsubsection*{Quantum error correction}

An interesting feature of holographic states is that they share some features with states referred to by quantum information theorists as ``quantum error-correcting codes'' \cite{Almheiri:2014lwa}.
The basic feature of these states is that they encode some specific information non-locally, in such a way that the information can be recovered if we have access to any sufficiently large subsystem. Thus, even if a (sufficiently small) part of the our system is corrupted somehow, we can still recover the encoded information by looking at a complementary subsystem.

\begin{figure}
\centering
\includegraphics[width=0.4\textwidth]{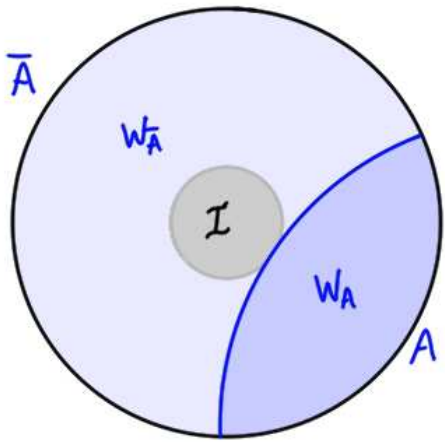}\\
\caption{The interior region ${\cal I}$ in a slice of AdS is contained in the entanglement wedge for any CFT subsystem larger than $\bar{A}$, but no CFT subsystem smaller than $A$.}
\label{center}
\end{figure}

To understand the connection with the AdS/CFT correspondence, recall that the density matrix $\rho_A$ for a spatial subsystem $A$ of the CFT encodes information about the entanglement wedge (call this $W_A$) in the bulk, as we discussed in section 4. Now, consider some ball ${\cal I}$ in the interior of AdS, as shown in figure \ref{center}. According to the duality between density matrices and bulk entanglement wedges, we can say that
\begin{itemize}
\item
The information about ${\cal I}$ is contained in the density matrix for any region larger than $\bar{A}$, since the region ${\cal I}$ will be contained in the corresponding entanglement wedge.
\item
The information about ${\cal I}$ is not contained in the density matrix for any region smaller than ${\cal A}$, since none of the corresponding entanglement wedges intersect ${\cal I}$
\end{itemize}
Thus, we have a behavior similar to a quantum error-correcting code; the information about the central region is ``protected'' from corruption of small subsystems. The latter property shows that this protection is not achieved by simply storing the information many times in different spatial subsystems (as in classical error-correction), but rather by distributing it nonlocally throughout the system.

An interesting consequence of this picture is that for a bulk observable localized in the region ${\cal I}$, there cannot just be a single CFT operator corresponding to this observable, since no nontrivial CFT operators are common to all subsystems larger than $\bar{A}$ (alternatively, there are no operators that commute with all operators in subsystems smaller than $A$). Instead, there must be a family of CFT operators corresponding to this observable, such that each CFT subsystem larger than $\bar{A}$ contains one of these operators. Since the operators in this family are different, they cannot have the same expectation values for all states of the CFT, but it is possible that they agree on some subset (or subspace) of the states for which we have nice classical dual geometry. In the language of quantum error correction, this is the ``code subspace.''

In summary, one of the special features of holographic states that distinguish them from general CFT states is that they have features reminiscent of quantum error-correcting codes. For a more detailed discussion, see \cite{Almheiri:2014lwa,Pastawski:2015qua}.

\subsubsection*{Entropy inequalities for holographic states}

In the previous section, we have discussed constraints on entanglement entropies that must hold in all multipartite quantum systems. We have seen that for holographic CFT states, some of these constraints can be related to properties of the dual geometries. For example, in static geometries, strong subadditivity inequalities can be shown to hold automatically from the Ryu-Takayanagi formula \cite{headrick2007holographic}. It turns out that by assuming the existence of a dual geometry for which the Ryu-Takayanagi formula holds, it is possible to prove additional constraints on entanglement entropies that are not generally true in all quantum systems. As an example, it can be shown \cite{Hayden:2011ag} that for any state dual to a static asymptotically AdS geometry, we have
\be
S(A \cup B \cup C) - S(A \cup B) - S(B \cup C) - S(A \cup C) + S(A) + S(B) + S(C) \le 0 \; ,
\ee
known as the negativity of tripartite information. Unlike subadditivity and strong subadditivity, this inequality is not generally true in quantum systems. Thus, it represents a necessary condition for a CFT state to be holographic (at least in the state case). A more general family of such constraints has been discussed in \cite{Bao:2015bfa}.

\section{Conclusion}

There are many interesting topics that I have not been able to mention here, due to limitations of time and my own expertise; this is certainly not meant to be a comprehensive review of the subject. The reader is encouraged to look at some of the references, and also to see which recent papers cite some of the key references in order to get a sense of what is going on currently. Also recommended are various online lectures, for example those from the ``It From Qubit'' summer school at Perimeter Institute, and from various recent conferences on gravity and entanglement at KITP and the Perimeter Institute. To conclude, I will briefly mention of a few more interesting topics that may be useful to be aware of.

\subsubsection*{Dynamics of entanglement}

It is interesting to understand how entanglement evolves in time-dependent states. For example, we can consider either a ``global quench'' in which the Hamiltonian of a system suddenly undergoes a translation-invariant perturbation, or a ``local quench'' for which the perturbation is localized to some finite region. In either case, the state of the system will evolve with time after the quench, and we can ask about the time-dependence of the entanglement entropy for regions of a certain size (and, in the case of the local quench, location). Many authors have considered this problem in quantum field theories and in holography; understanding the behavior in holographic theories implied by gravitational physics gives us further constraints on the properties of holographic CFTs and holographic states. For some recent work on this subject, including interesting connections to quantum chaos, see for example \cite{Hartman:2013qma,Liu:2013iza,Shenker:2013pqa,Mezei:2016wfz} and references therein.

\subsubsection*{Kinematic space}

In section 5.2, we saw that the linearized Einstein equations in the bulk could be understood in the CFT language as saying that the entanglement first law is satisfied for all ball-shaped regions in all Lorentz frames. This space of regions, or alternatively the space of domain of dependence regions $D_B$ for all balls $B$, has been called ``kinematic space''; in recent work \cite{Czech:2015qta,Czech:2016xec,deBoer:2016pqk}, there have been hints that various aspects of CFT physics and of AdS/CFT might be understood naturally using the language of kinematic space.

We can define a natural set of coordinates on kinematic space by noting that the domain of dependence $D_B$ of each ball-shaped region is bounded by the future light cone of some point $x_0$ and the past lightcone of some point $x_1$ in the future of $x_0$. Thus, we can associate points in kinematic space with pairs of points $X = (x_0,x_1)$ on the original spacetime with $x_1$ in the future of $x_0$; this is some $2d$-dimensional space. There is a conformally-invariant signature $(d,d)$ metric on this space \cite{Czech:2016xec,deBoer:2016pqk}, and also a natural partial ordering: we say $X_1 < X_2$ if $D_{X_1} \subset D_{X_2}$.

All regions $D_X$ are conformally equivalent. We can define a mapping from each $D_X$ to some standard region $D_0$ (e.g. the domain of dependence of a unit sphere in Minkowski space, or hyperbolic space times time). Then, given any state $|\Psi \rangle$ of the CFT, the density matrix $\rho_{D_X}$ maps to some density matrix $\rho_0$ for our standard region $D_0$. Thus, for every state of the CFT, we can associate a function $\rho_0(X)$ from kinematic space to the space of density matrices for the CFT on $D_0$. Starting from this construction, we can we define other natural functions on kinematic space, for example, the vacuum subtracted entanglement entropy $\Delta S(X)$. For states close to the vacuum, this has been shown to satisfy some simple differential equation on kinematic space \cite{Czech:2016xec,deBoer:2016pqk}. We can also consider the relative entropy to some reference state, $R(X)_|\Psi_0 \rangle$. Monotonicity of relative entropy implies that any such function obeys $R(X)_|\Psi_0 \rangle > R(Y)_|\Psi_0 \rangle$ for $X > Y$; i.e. these relative entropy functions are all monotonic on kinematic space with respect to the natural partial ordering.

It will be interesting to see whether additional new insights about CFTs and holography will be gained by thinking about these kinematic space structures.

\subsubsection*{Tensor Networks}

Holographic CFTs are strongly coupled theories with a large number of degrees of freedom. Thus, it is difficult to understand their quantum states directly; most of what we know about these states is based on calculations in the dual gravitational theory. It is interesting to ask whether we can find examples of states in simpler quantum systems which share qualitative features with the holographic CFT states. At the very least, constructing such states is a useful way of demonstrating that certain features deduced from the gravity side (e.g. a particular structure of entanglement) are even possible in a consistent quantum system.

Recently, starting with the work of Swingle \cite{Swingle:2009bg}, it has been understood that certain states associated with ``tensor networks'' share many features with states of holographic CFTs; thus studying the physics in the tensor network states may give us new insights into holography.

To define tensor network states, consider a quantum system with Hilbert space
\be
{\cal H} = {\cal H}_d \otimes {\cal H}_d \otimes {\cal H}_d \cdots \otimes {\cal H}_d
\ee
where we typically imagine that the dimension-$d$ tensor product factors correspond to subsystems arranged spatially, as in a spin chain. Given a set of basis elements $|i_1 \rangle \otimes |i_2 \rangle \otimes \cdots$, a general state of the system can be written as
\be
|\Psi \rangle = \sum_i \psi_{i_1 i_2 \dots i_N} |i_1 \rangle \otimes |i_2 \rangle \otimes \cdots |i_N \rangle\; .
\ee
Thus, any state of the system corresponds to some tensor $\psi_{i_1 i_2 \dots i_N}$.

\begin{figure}
\centering
\includegraphics[width=0.7\textwidth]{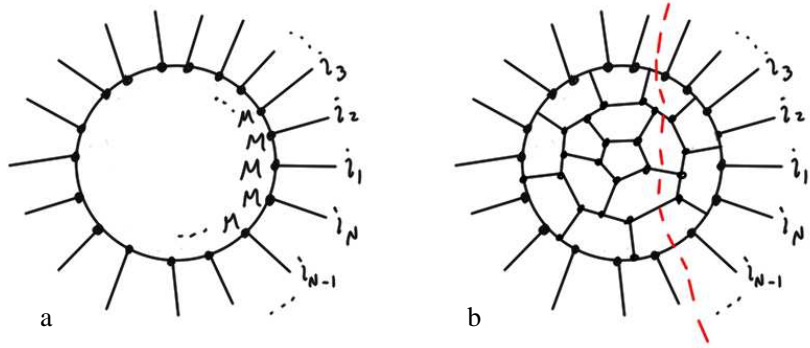}\\
\caption{Tensor network states for an $N$ part system. a) Tensor network defining matrix product states. b) More general tensor network, which provides a better representation of states approximating CFT states. Some network states (e.g. with random tensors with indices corresponding to large-dimension spaces) obey a Ryu-Takayanagi formula: the entanglement entropy of a subsystem (e.g. the legs to the right of the red dashed line) is proportional to to the minimal number of cuts required to separate this subsystem from the rest of the network. We can think of this as the ``area'' of the dashed red surface. }
\label{figTN}
\end{figure}

One is often interested in finding the ground state of a system for some particular choice of Hamiltonian, e.g. a local Hamilton that includes interactions between tensor product factors corresponding to nearby subsystems. It turns out that in many such cases, we can come up with a very good approximation to the ground state by building the tensor $\psi_{i_1 i_2 \dots i_N}$ out of simpler tensors. For example, consider a general three-index tensor $M_{i A}{}^B$, which could be used to describe a map
\be
{\cal M} : {\cal H}_D \to {\cal H}_D \otimes {\cal H}_d
\ee
as
\be
{\cal M} = \sum_{i,A,B} M_{i A}{}^B |i \rangle \otimes |A \rangle \langle B| \; .
\ee
From this tensor $M$, we can define a state of our original system (known as a matrix product state) as
\be
\label{MPS}
\psi_{i_1 i_2 \dots i_N} = \sum_{A_i} M_{i_1 A_1}{}^{A_2} M_{i_2 A_2}{}^{A_3} \cdots M_{i_N A_N}{}^{A_1} \; .
\ee
This construction can be represented diagrammatically as in figure \ref{figTN}a.\footnote{To be more precise, we could draw arrows on the edges of the graph such that an outgoing arrow represents a lower index and an incoming arrow represents an upper index. Alternatively, we can take all the vertices to represent tensors with lower indices (i.e. states) and edges to represent index contractions with some choice of tensor $M^{ij}$, giving $T^1_{\cdots i \cdots} M^{ij} T^2_{\cdots j \cdots}$. There is no particularly natural choice for $M^{ij}$; we can take $M^{ij} = \delta^{ij}$, but this is basis-dependent.}. The expression (\ref{MPS}) can be thought of as a variational ansatz used to approximate the true ground state. If the dimension $D$ of the auxiliary space is not too large, the number of parameters defining the state (\ref{MPS}) is far smaller than the number of parameters defining a general state of the system, so solving the variational problem is computationally much less demanding that finding the exact ground state. For systems with long-range correlations (e.g. those which approximate continuum conformal field theories), simple tensor network states (\ref{MPS}) generally do not provide a good approximation to the true ground state but we can consider states associated with more complicated networks such as the one shown in figure \ref{figTN}b. One might hope that certain tensor network states of this type, which provide good approximations to CFT ground states, could provide a useful toy model for holographic states. Specifically, we might hope that such states have a structure of entanglement similar to that in holographic theories, as suggested originally by \cite{Swingle:2009bg}.

In recent work \cite{Pastawski:2015qua,Hayden:2016cfa}, tensor network states have been constructed for which the entanglement structure is ``geometric'' just as in holographic theories. These states obey a Ryu-Takayanagi formula where the entropy of a subsystem $A$ is well-approximated by the ``area'' of a minimal surface in the network, defined to be the minimal number of network links that need to be cut in order to isolate $A$ from the rest of the system. This is illustrated in figure \ref{figTN}b. In particular, \cite{Hayden:2016cfa} has shown that this feature is generically true if the dimensions associated to legs in the network are large and the tensors are chosen randomly. Thus, if we choose a network for which the graph minimal ``areas'' behave in the same way as the minimal surface areas in $AdS$ or some other geometry (e.g. by choosing the graph to be a discretization of the given geometry), we obtain states whose entanglement structure is very similar to holographic CFT states. In the models of (\cite{Pastawski:2015qua}), the network states have also been shown to have features of quantum error-correcting codes, as expected for holographic states. Thus, certain tensor network states to seem to provide a nice explicit construction of states in simple quantum systems with qualitative features of holographic states.

\section*{Acknowledgements}

I would like to thank the organizers of the Jerusalem, CERN, TASI, and Trieste schools for inviting me to give these lectures, the students of the schools for many useful comments and questions that have helped me improve the presentation, and to Joseph Polchinski and Oliver De Wolfe for patiently encouraging me to write up these lecture notes. I would like to thank Stephen Pietromonaco and Jared Stang for pointing out typos in earlier versions of the notes.

\appendix

\section{Metrics}

In this section, we collect various representations of vacuum Anti de Sitter space and asymptotically Anti de Sitter black hole geometries that will be useful in our discussions.

To describe Anti-de-Sitter space $AdS_{d+1}$, we can start by considering the $(d+1)$-dimensional hyperboloid
\be
\label{AdSconstr}
-X_{-1}^2  - X_{0}^2 + X_{1}^2+ \dots + X_{d}^2 = -\ell^2
\ee
of $(d+2)$-dimensional space with metric
\be
ds^2 = -dX_{-1}^2  - dX_{0}^2 + dX_{1}^2+ \dots + dX_{d}^2 \; .
\ee
The resulting spacetime has signature $(d,1)$ but has closed timelike curves corresponding to orbits of the rotation symmetry in the $(X_{-1},X_0)$ plane. We can consider instead the universal cover of this space, obtained by letting the angular coordinate associated with this symmetry have a range $(-\infty,\infty)$ instead of $[0,2 \pi)$. The resulting spacetime is pure global Anti-de-Sitter spacetime, or ``global AdS'', a homogeneous and isotropic Lorentzian spacetime. For certain calculations, it is useful to work directly with these embedding space coordinates, taking into account the constraint (\ref{AdSconstr}).

We can alternatively represent global AdS via the metric
\be
\label{globalAdS}
ds^2 = \ell^2 (-\sinh^2 \rho dt^2 + \cosh^2 \rho d \rho^2 + \rho^2 d \Omega_{d-1}^2)
\ee
which arises by parameterizing the surface by
\bea
X_0 &=& \ell \cosh \rho \cos \tau \cr
X_{-1} &=& \ell \cosh \rho \sin \tau \cr
X_i &=& R \sinh \rho \Omega_i \; .
\eea
Here, $\Omega_i$ are combinations of the angular coordinates on $S^{d-1}$ whose squares sum to one (e.g. $(\cos\theta,\sin\theta \cos\phi,\sin\theta \sin \phi)$ for $S^2$.

Here, the spatial metric in (\ref{globalAdS}) describes $d$-dimensional hyperbolic space, the homogeneous and isotropic space with negative curvature. Note that the metric of AdS is not simply hyperbolic space times time, but rather a ``warped product,'' in which the time dilation factor relating coordinate time to proper time varies with space.

\begin{figure}
\centering
\includegraphics[width=0.4\textwidth]{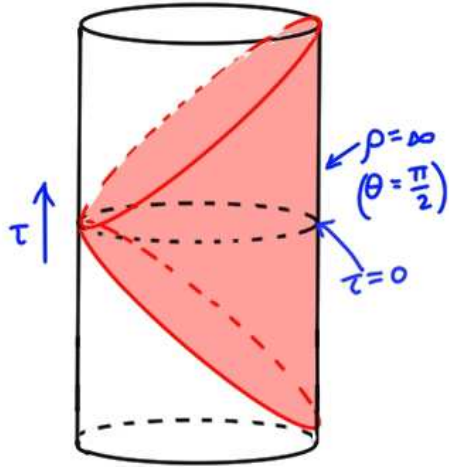}
\caption{Cylinder representation of Anti-de-Sitter space. Shaded wedge is the Poincar\'e patch of AdS.}
\label{figAds}
\end{figure}

Starting from any point in AdS, there is an infinite proper distance to $\rho=\infty$. However, a light signal can reach $\rho = \infty$ and return in finite proper time. Thus, there is a sense in which Anti de Sitter space has a boundary at $\rho = \infty$ that can be probed by light signals. To highlight this causal structure, it is useful to define yet another set of coordinates,via
\be
\tan \theta = \sinh \rho \; .
\ee
With this definition, the metric becomes
\be
\label{can}
ds^2 = {\ell^2 \over \cos^2 \theta}(-d \tau^2 + d \theta^2 + \sin^2 \theta d \Omega^2) \; ,
\ee
so that the boundary is at $\theta = \pi/2$.  Apart from the overall factor which has no effect on light-ray trajectories, the geometry is that of a Euclidean ball times time. Trajectories of light rays are the same in this space as in AdS, so this description is very useful in understanding the causal structure.  For this reason, AdS is often depicted as a (filled) cylinder with the vertical direction representing time, as shown in figure \ref{figAds}. It is important to keep in mind that while the cylinder appears to have points near the middle and points closer to the boundary, it is an infinite volume spacetime in which all points are equivalent.

The overall factor in front of the metric (\ref{can}) diverges at the boundary, so while the boundary has a well-defined causal structure, it does not a priori have a particular geometry associated with it. However, sometimes it is convenient to make a choice of boundary geometry consistent with this causal structure. In practice, a simple way to do this is to define a coordinate $z$ which vanishes at the boundary, such that the metric near the boundary of AdS takes the form
\be
\label{FGapp}
ds^2 = {\ell^2 \over z^2} \left( dz^2 + \Gamma_{\mu \nu}(z,x) dx^\mu dx^\nu \right) \; .
\ee
where $\Gamma_{\mu \nu}(z,x)$ has a finite limit $\Gamma^0_{\mu \nu}(x)$ as $z \to 0$. This choice is known as a Fefferman-Graham system of coordinates. In this representation, surfaces of constant $z$ then have a particular geometry induced by the metric in AdS. If we rescale the metric for these surfaces by $z^2$ in the limit $z \to 0$, we obtain a well-defined geometry in the limit, described by the metric $\Gamma^0_{\mu \nu}(x)$. We will refer to this as the boundary geometry. We emphasize that this boundary geometry is artificial and represents a choice among the class of geometries with the same causal structure. Such geometries are related by a conformal transformation (i.e. an overall spacetime-dependent rescaling of the metric). No experiment in AdS can be performed to determine which geometry in this class is {\it the} boundary geometry.

For pure AdS spacetime, the Fefferman-Graham representation of the metric with boundary geometry chosen to be $S^d \times R$ is
\be
ds^2 = {\ell^2 \over z^2} \left(dz^2 - {1 \over 4} (1 + z^2)^2 d \tau^2 + {1 \over 4} (1 - z^2)^2 d \Omega^2 \right) \; ,
\ee
related to the coordinates above by
\be
z = {\cos \theta \over 1 + \sin \theta} \qquad \qquad \cos \theta = {2 z \over 1 + z^2} \; .
\ee
This is a convenient representation for the dual spacetime for the vacuum state of a conformal field theory defined on a sphere times time.

We can also choose a Fefferman-Graham description with Minkowksi space boundary geometry, in which the metric takes the particularly simple form
\be
\label{MinkFG}
ds^2 = {\ell^2 \over z^2} (dz^2 + dx_\mu dx^\nu) \; .
\ee
These coordinates cover only a portion of AdS, the shaded region in figure \ref{figAds} known as the ``Poincar\'e patch''. The coordinates in (\ref{MinkFG}) are related to the global AdS coordinates in (\ref{globalAdS}) by
\be
{z \over \ell} = {1 \over \cosh \rho \cos \tau - \sinh \rho \cos \theta_1} \qquad {t \over z} = \cosh \rho \sin \tau \qquad {x^i \over z} =\Omega^i \sinh \rho
\ee
where $\Omega^i = \sin \theta_1 \cos \theta_2, ... , \sin \theta_1 \sin \theta_2 \cdots \sin \theta_{d-1}$ such that $\cos^2 \theta_1 + \Omega_i^2 = 1$. This is a very useful representation for the dual spacetime for the vacuum state of a holographic conformal field theory defined on Minkowski spacetime.

\subsubsection*{Asymptotically locally AdS geometries}

In the AdS/CFT correspondence, general holographic states of a CFT on a boundary spacetime ${\cal B}$ correspond to spacetimes which are asymptotically locally AdS, with boundary geometry ${\cal B}$. These may generally be represented in Fefferman-Graham coordinates (\ref{FGapp}), where $\Gamma(z=0)$ is taken to be the metric of ${\cal B}$. Spacetimes associated with different states are associated with different behavior for $\Gamma$ away from $z=0$.

Some important examples are the black hole geometries in $AdS_{d+1}$, corresponding to the CFT in a thermal state. For global AdS, the corresponding black hole is spherically symmetric and can be described by
\be
ds^2 = -f_S(r) dt^2 + {dr^2 \over f_S(r)} + r^2 d x_i^2
\ee
with
\be
f_S(r) = {r^2 \over \ell^2} + 1 - {\mu \over r^{d-2}}
\ee
where $\mu$ determines the mass and temperature of the black hole.

The thermal state of a CFT on Minkowski space describes a planar AdS black hole, described by metric
\be
ds^2 = -f_M(r) dt^2 + {dr^2 \over f_M(r)} + {r^2 \over \ell^2} d x_i^2
\ee
where
\be
f_M(r) = {r^2 \over \ell^2}  - {\mu \over r^{d-2}}
\ee
and the temperature is related to $\mu$ by
\be
\mu = {1 \over \ell^2} \left( {4 \pi \ell^2 T \over d} \right)^d
\ee
These can also be described in Fefferman-Graham coordinates; for example the metric of a 2+1 dimensional planar-AdS black hole (known as a BTZ black hole) can be rewritten as
\be
\label{BTZ}
ds^2 = {\ell^2 \over z^2} (dz^2 -(1 - \mu z^2 /2)^2 dt^2 + (1 + \mu z^2 /2)^2 dx^2) \; .
\ee
This solution (for the special case of 2+1 dimensions) is locally the same as pure AdS (as are all solutions to Einstein's equations in three dimensions without matter). By a coordinate transformation, this geometry can be mapped onto the region in the geometry (\ref{MinkFG}) which is the intersection of the causal past and the causal future of a diamond-shaped region which is the domain of dependence of the interval $\{x \in [-1,1], t=0 \}$ on the boundary.

\section{Conformal transformations on density matrices}\label{ConfTrans}

In this appendix, we briefly review a few relevant facts about conformal symmetries in field theory. We say that two spacetimes $M$ and $\tilde{M}$ with metrics $g(x),\tilde{g}(\tilde{x})$ are related by a conformal transformation if there is a map
\be
\label{ctrans}
x^\mu \to \tilde{x}^\mu = f^\mu(x)
\ee
between points of the two spacetimes such that the metrics are related as
\be
g_{\mu \nu} (\tilde{x}) d \tilde{x}^\mu d \tilde{x}^\nu = \Omega^2(x) g_{\mu \nu} (x) dx^\mu dx^\nu
\ee
where the overall spacetime-dependent scaling $\Omega(x)$ is known as the conformal factor. Conformal field theories on two spacetimes related by such a conformal transformation are equivalent, and there is a one-to-one mapping between the states of the two theories that we can write as
\be
\label{confmap}
\rho_{M} = U_\Omega \rho_{\tilde{M}} U_\Omega^\dagger \; .
\ee

In the special case where the two spacetimes are equivalent to one another (e.g. where they are both $R^{d,1}$ and the transformation represents a translation, boost, scaling, inversion or some combination of these), this transformation defines a map on the space of states of a single theory. The vacuum state of a CFT on $R^{d,1}$ or $S^d \times R$ is invariant under these transformations, which represent the spacetime symmetries of the CFT.

\subsubsection*{Examples}

There are several specific examples of conformal transformation that will be useful to us in the context of AdS/CFT. First, we can show that Minkowski space is conformally equivalent to a part of $S^d \times R$, namely the domain of dependence of a sphere minus a single point $p$, as shown in figure \ref{DptoRind}.\footnote{Alternatively, this is the set of all points on $S^d \times R$ spacelike separated from the point $p$.} Through this transformation, any state of the CFT on the sphere can be associated with a Minkowski space CFT state for each such domain of dependence region. The vacuum state of the sphere  CFT maps to the vacuum state of the CFT on Minkowski space.

Through this transformation, the domain of dependence of a ball-shaped region on the sphere maps to the domain of dependence of a ball-shaped region of Minkowski space. By a further conformal transformation mapping Minkowski space to itself, this domain of dependence region can be mapped to the domain of dependence of a half space (for example $x^1 > 0$). This region is also conformally equivalent to the static spacetime $H^d \times R$, where $H^d$ is hyperbolic space, the non-compact Euclidean manifold of constant negative curvature.

\section{Path integral representation of states and density matrices in quantum field theory}\label{AppPI}

Path integrals are a familiar tool in quantum field theory to represent various useful quantities such as the partition function or the generating functional for correlation functions. They can also be used to represent various states and density matrices. We begin with the basic result that transition amplitudes can be expressed as a path integral as
\be
\langle \phi_1(t_1) | e^{-i H t} | \phi_0(t_0) \rangle = {\cal N} \int_{\phi(t_0) = \phi_0}^{\phi(t_1) = \phi_1} [d \phi] e^{i S(\phi(t))} \; ,
\ee
where ${\cal N}$ is a normalization factor. The derivation, found in most field theory textbooks, involves splitting the time evolution operator into a product of infinitesimal pieces and inserting between each factor the identity operator, expressed as a sum over the projection operators associated with a complete basis of states.

\begin{figure}
\centering
\includegraphics[width=\textwidth]{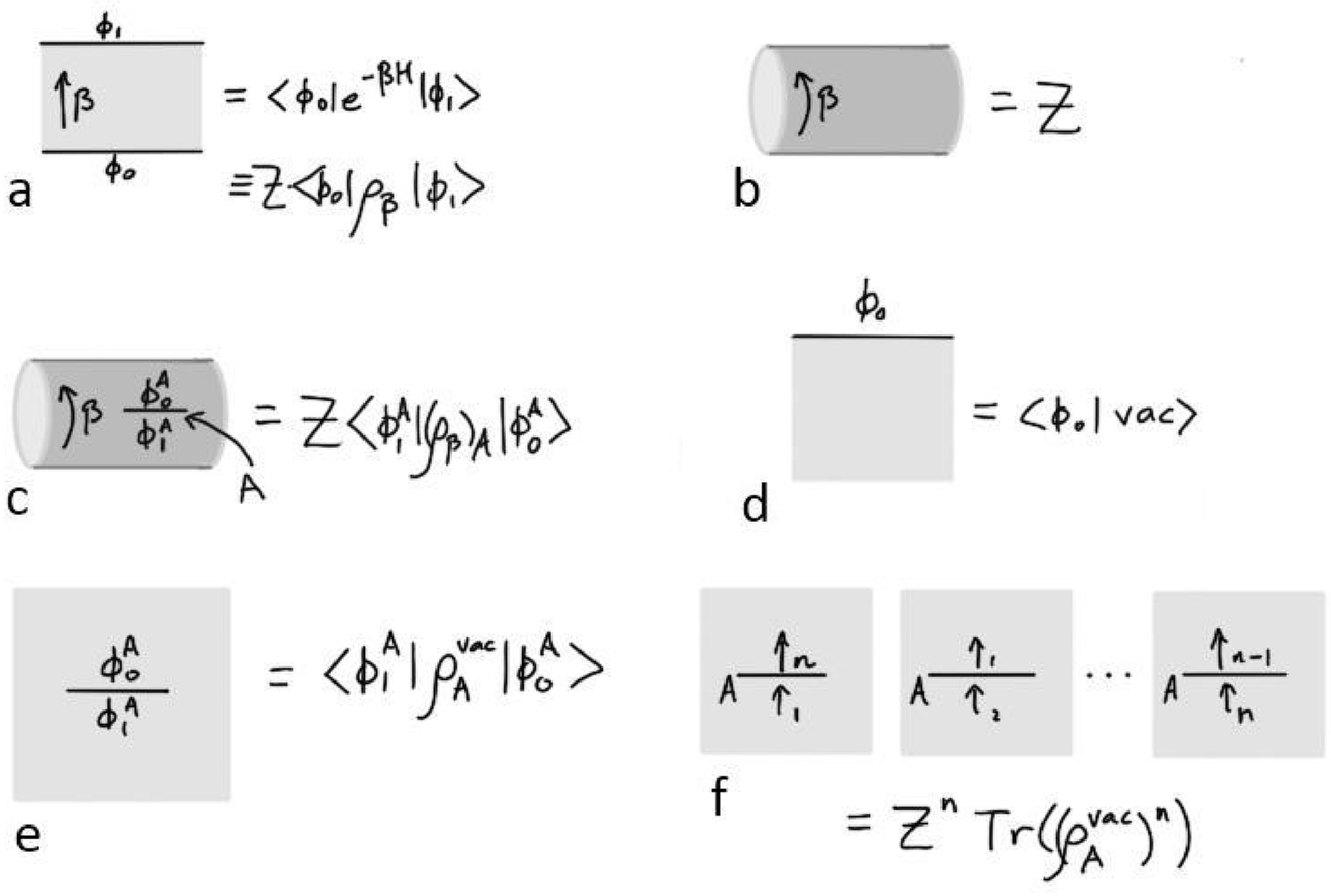}
\caption{Euclidean path integrals defining a) the thermal density matrix b) the partition function c) the reduced density matrix for region A in the thermal state d) the vacuum state e) the reduced density matrix for region A in the vacuum state f) $\tr(\rho_A^n)$ for the vacuum state.}
\label{pathint}
\end{figure}

Taking $t = -i \beta$, we obtain the corresponding Euclidean statement
\be
\label{euctime}
\langle \phi_1 | e^{- \beta H} | \phi_0 \rangle = {\cal N} \int_{\phi(0) = \phi_0}^{\phi(\beta) = \phi_1} [d \phi] e^{- S_{Euc}(\phi(\tau))} \; ,
\ee
where $S_{Euc}$ is the action analytically continued to Euclidean space. Up to a normalization factor, these are the matrix elements of the density matrix for the thermal state $e^{-\beta H}/Z$, i.e. the canonical ensemble. This path integral is depicted in figure \ref{pathint}a.

The partition function $Z_\beta = \tr(e^{-\beta H})$ is obtained by setting $\phi_1 = \phi_0$ and integrating over $\phi_0$, which is equivalent to defining the path integral on a space with periodic Euclidean time and period $\beta$,
\be
Z_\beta = \tr(e^{-\beta H})= {\cal N} \int_{\phi(0) = \phi(\beta)} [d \phi] e^{- S_{Euc}(\phi(\tau))} \; ,
\ee
as shown in figure \ref{pathint}b.
We can similarly define the partial trace over a region $\bar{A}$ to define the reduced density matrix for the region $A$ in the thermal state. This is
\be
\label{ThDM}
\langle \phi^A_1 | \rho_A^\beta | \phi^A_0 \rangle = {1 \over Z} \int^{ \phi^A(\beta) = \phi^A_1}_{\phi^A(0) = \phi^A_0 , \;  \phi^{\bar{A}}(0) = \phi^{\bar{A}}(\beta)} [d \phi] e^{- S_{Euc}(\phi(\tau))} \; ,
\ee
as in figure \ref{pathint}c.
A path integral representation of the vacuum state can be obtained by noting that
\be
\lim_{\beta \to \infty} e^{- \beta H} | \Psi \rangle = {\cal N} | vac \rangle
\ee
since, writing $| \Psi \rangle$ as a linear combination of energy eigenstates, all states with energy higher than the vacuum state energy obtain a relative coefficient $e^{-\beta (E - E_{vac})}$ that vanishes in the limit. In terms of the path integral, this gives
\be
\label{Vac}
 \langle \phi_0 | vac \rangle = {\cal N} \int_{\phi(0) = \phi_0} [d \phi(\tau < 0)] e^{- S_{Euc}(\phi(\tau))} \; ,
\ee
as depicted in figure \ref{pathint}d.
Starting from (\ref{ThDM}) and taking the limit $\beta \to \infty$ or working directly from (\ref{Vac}), we obtain a path integral expression for the density matrix associated with a region $A$ in the vacuum state (figure \ref{pathint}e),
\be
\label{vacDM}
\langle \phi^A_1 | \rho_A^{vac} | \phi^A_0 \rangle = {1 \over Z} \int_{ \phi^A(0^+) = \phi^A_0}^{\phi^A(0^-) = \phi^A_1} [d \phi] e^{- S_{Euc}(\phi(\tau))} \; ,
\ee
where the field is taken to be continuous across $\tau=0$ in the region $\bar{A}$. This is the path integral over Euclidean space with boundary conditions defined on either side of a cut at the spatial region $A$ at $\tau=0$. We will denote this spacetime with the cut by $M^A$.

It is now straightforward to describe a path integral calculation of the Renyi entropies (\ref{Renyi}) and entanglement entropy (\ref{entropy}). To calculated $\tr(\rho_A^n)$, we take
\bea
\label{rhoN}
\tr((Z \rho^{vac}_A)^n) &=& \int [d \phi^A_i] \langle \phi^A_0 | Z \rho_A^{vac} | \phi^A_n \rangle \cdots \langle \phi^A_2 | Z \rho_A^{vac} | \phi^A_1 \rangle \langle \phi^A_1 |Z \rho_A^{vac} | \phi^A_0 \rangle \cr
&=& \int_{M^A_n} [d \phi] e^{- S_{Euc}(\phi(\tau))}
\eea
where $M^A_n$ is a manifold obtained by taking $n$ copies of Euclidian space $M^A$ and cyclically gluing them together with the fields below the cut one each copy taken to be continuous with the fields above the cut in the next copy. This is depicted in figure \ref{pathint}f.

In order to calculate the entanglement entropy, the standard method (known as the ``replica trick'') is to find an analytic expression for the Renyi entropies $S_\alpha$ in (\ref{Renyi}) calculated using  (\ref{rhoN}) and then take the limit $\alpha \to 1$. Alternatively, we can define the entanglement entropy in terms of the integer Renyi entropies by various integral identities, for example
\be
S_A = \int_0^\infty da \sum_{n=0}^\infty {(-a)^n \over (n+1)!} (\tr(\rho_A^n) - 1)
\ee
which makes use of the identity
\be
\int_0^\infty {da \over a} (x e^{-a x} - x e^{-a}) = - x \ln x .
\ee
The calculation of entanglement entropy (\ref{S2D}) for an interval in 2D CFTs can be carried out directly using these path integral methods, as explained in \cite{Calabrese:2004eu}.

\subsubsection*{Vacuum density matrix for a half space}

Using the path integral technology above, we now derive a universal result for the density matrix of a half-space in the vacuum state of a Lorentz-invariant quantum field theory. Starting from the expression (\ref{vacDM}), and denoting the region $x^1 > 0$ by $R$, we have
\be
\label{HalfDM}
\langle \phi^R_1 | \rho_R^{vac} | \phi^R_0 \rangle = {1 \over Z} \int_{ \phi^R(0^+) = \phi^R_0}^{\phi^R(0^-) = \phi^R_1} [d \phi] e^{- S_{Euc}(\phi(\tau))} \; ,
\ee
Next, consider the change of variables to polar coordinates in the $x^1,\tau$ plane,$x^1 = r \cos(\theta), \tau = r \sin(\theta)$. This gives
\be
\label{HalfDM2}
\langle \phi^R_1 | \rho_R^{vac} | \phi^R_0 \rangle = {1 \over Z} \int_{ \phi^R(\theta=0) = \phi^R_0}^{\phi^R(\theta = 2 \pi) = \phi^R_1} [d \phi] e^{- S_{Euc}(\phi(\tau))} \; ,
\ee
Now, we notice that the expression on the right takes the same form as the expression (\ref{euctime}) with $\tau$ replaced by $\theta$ and $\beta$ replaced by $2 \pi$. The expression in (\ref{euctime}) defined the density matrix $e^{-\beta H}$, where $H$ was the Hamiltonian generating evolution in the variable $t = i \tau$. We conclude that the expression (\ref{HalfDM}) defines the density matrix $e^{-2 \pi H_\eta}$, where $H_\eta$ is the Hamiltonian generating evolution in the variable $\eta = i \theta$. For this analytically continued angle coordinate, the metric back in Lorentzian space becomes
\be
dr^2 + r^2 d \theta^2 \to dr^2 - r^2 d \eta^2 \; ,
\ee
and we can recognize the metric on the right as describing a Rindler wedge of the original Minkowski space, defined as the domain of dependence of the half-space $x^1 > 0$. Explicitly, we have $x^1 = r \cosh(\eta)$ and $t = r \sinh(\eta)$. Here $\eta$ defines "Rindler time," and the associated generator $H_\eta$ is the ordinary boost generator
\be
\label{Heta}
H_\eta = \int_{x^1 > 0} d^{d-1} x \left\{x^1 T_{00}\right\} \; .
\ee
The final result is that the density matrix for a half space is the thermal density matrix defined in terms of the boost generator,
\be
\label{rhoRind}
\rho_R = {1 \over Z} e^{- 2 \pi H_\eta} \; .
\ee
As $\eta$ can be viewed as the time coordinate for a family of observers on accelerated trajectories, this result is closely related to the result of Unruh that accelerated observers (who only have access to the physics in the Rindler wedge) experience thermal physics.

\subsubsection*{Modular Hamiltonians}

Starting from the result (\ref{Heta},\ref{rhoRind}) for the half-space vacuum density matrix, we can determine various explicit results for modular Hamiltonians in conformal field theory. For reference, we record the results here.

For a ball-shaped region $B$ of radius $R$ and center $x_0$ in Minkowski space, the modular Hamiltonian $H = -\log \rho_B$ is given by
\be
H = 2 \pi \int_B d^d x {R^2 - (\vec{x} - \vec{x}_0)^2 \over 2 R} T_{00} \; .
\ee
For a ball-shaped region $\theta < \theta_0$ on a sphere, we have
\be
\label{modHsphere}
H = 2 \pi \int_B d^d x {\cos(\theta) - \cos(\theta_0) \over \sin(\theta_0)} T_{00} \; .
\ee
In the special case of 1+1 dimensional CFTs the modular Hamiltonian can also be calculated for thermal state can be obtained from the vacuum state by a conformal transformation. For a spatial interval $[-R,R]$ in an unboosted thermal state with temperature $T = \beta^{-1}$, the modular Hamiltonian is
\be
\label{HmodTherm}
H = {2 \beta \over \sinh \left({2 \pi R  \over \beta}\right)} \int_{-R}^R d x \sinh\left({\pi (R-x) \over \beta} \right) \sinh\left({\pi (R+x) \over \beta} \right) T_{00}(x) \; ,
\ee
An explicit expression for the modular Hamiltonian of a boosted thermal state can be found in \cite{Lashkari:2014kda}.

\bibliographystyle{JHEP}

\bibliography{LECBIB}

\providecommand{\href}[2]{#2}\begingroup\raggedright\begin{thebibliography}{10}

\bibitem{'tHooft:1993gx}
G.~'t~Hooft, {\it {Dimensional reduction in quantum gravity}},  in {\em
  {Salamfest 1993:0284-296}}, pp.~0284--296, 1993.
\newblock \href{http://arxiv.org/abs/gr-qc/9310026}{{\tt gr-qc/9310026}}.

\bibitem{Susskind:1994vu}
L.~Susskind, {\it {The World as a hologram}},  {\em J. Math. Phys.} {\bf 36}
  (1995) 6377--6396, [\href{http://arxiv.org/abs/hep-th/9409089}{{\tt
  hep-th/9409089}}].

\bibitem{maldacena1997large}
J.~M. Maldacena, {\it The large n limit of superconformal field theories and
  supergravity},  {\em arXiv preprint hep-th/9711200} (1997).

\bibitem{aharony2000large}
O.~Aharony, S.~S. Gubser, J.~Maldacena, H.~Ooguri, and Y.~Oz, {\it Large n
  field theories, string theory and gravity},  {\em Physics Reports} {\bf 323}
  (2000), no.~3 183--386.

\bibitem{McGreevy:2009xe}
J.~McGreevy, {\it {Holographic duality with a view toward many-body physics}},
  {\em Adv. High Energy Phys.} {\bf 2010} (2010) 723105,
  [\href{http://arxiv.org/abs/0909.0518}{{\tt arXiv:0909.0518}}].

\bibitem{Witten:1998zw}
E.~Witten, {\it {Anti-de Sitter space, thermal phase transition, and
  confinement in gauge theories}},  {\em Adv. Theor. Math. Phys.} {\bf 2}
  (1998) 505--532, [\href{http://arxiv.org/abs/hep-th/9803131}{{\tt
  hep-th/9803131}}].

\bibitem{bekenstein1973black}
J.~D. Bekenstein, {\it Black holes and entropy},  {\em Physical Review D} {\bf
  7} (1973), no.~8 2333.

\bibitem{bardeen1973four}
J.~M. Bardeen, B.~Carter, and S.~W. Hawking, {\it The four laws of black hole
  mechanics},  {\em Communications in Mathematical Physics} {\bf 31} (1973),
  no.~2 161--170.

\bibitem{hawking1975particle}
S.~W. Hawking, {\it Particle creation by black holes},  {\em Communications in
  mathematical physics} {\bf 43} (1975), no.~3 199--220.

\bibitem{nielsen2010quantum}
M.~A. Nielsen and I.~L. Chuang, {\em Quantum computation and quantum
  information}.
\newblock Cambridge university press, 2010.

\bibitem{Maldacena:2001kr}
J.~M. Maldacena, {\it {Eternal black holes in anti-de Sitter}},  {\em JHEP}
  {\bf 04} (2003) 021, [\href{http://arxiv.org/abs/hep-th/0106112}{{\tt
  hep-th/0106112}}].

\bibitem{Balasubramanian:2014hda}
V.~Balasubramanian, P.~Hayden, A.~Maloney, D.~Marolf, and S.~F. Ross, {\it
  {Multiboundary Wormholes and Holographic Entanglement}},  {\em Class. Quant.
  Grav.} {\bf 31} (2014) 185015, [\href{http://arxiv.org/abs/1406.2663}{{\tt
  arXiv:1406.2663}}].

\bibitem{VanRaamsdonk:2009ar}
M.~Van~Raamsdonk, {\it {Comments on quantum gravity and entanglement}},
  \href{http://arxiv.org/abs/0907.2939}{{\tt arXiv:0907.2939}}.

\bibitem{VanRaamsdonk:2010pw}
M.~Van~Raamsdonk, {\it {Building up spacetime with quantum entanglement}},
  {\em Gen. Rel. Grav.} {\bf 42} (2010) 2323--2329,
  [\href{http://arxiv.org/abs/1005.3035}{{\tt arXiv:1005.3035}}]. [Int. J. Mod.
  Phys.D19,2429(2010)].

\bibitem{ryu2006holographic}
S.~Ryu and T.~Takayanagi, {\it Holographic derivation of entanglement entropy
  from the anti--de sitter space/conformal field theory correspondence},  {\em
  Physical review letters} {\bf 96} (2006), no.~18 181602.

\bibitem{hubeny2007covariant}
V.~E. Hubeny, M.~Rangamani, and T.~Takayanagi, {\it A covariant holographic
  entanglement entropy proposal},  {\em Journal of High Energy Physics} {\bf
  2007} (2007), no.~07 062.

\bibitem{Wall:2012uf}
A.~C. Wall, {\it {Maximin Surfaces, and the Strong Subadditivity of the
  Covariant Holographic Entanglement Entropy}},  {\em Class.Quant.Grav.} {\bf
  31} (2014), no.~22 225007, [\href{http://arxiv.org/abs/1211.3494}{{\tt
  arXiv:1211.3494}}].

\bibitem{Calabrese:2004eu}
P.~Calabrese and J.~L. Cardy, {\it {Entanglement entropy and quantum field
  theory}},  {\em J. Stat. Mech.} {\bf 0406} (2004) P06002,
  [\href{http://arxiv.org/abs/hep-th/0405152}{{\tt hep-th/0405152}}].

\bibitem{Hartman:2013mia}
T.~Hartman, {\it {Entanglement Entropy at Large Central Charge}},
  \href{http://arxiv.org/abs/1303.6955}{{\tt arXiv:1303.6955}}.

\bibitem{Faulkner:2013yia}
T.~Faulkner, {\it {The Entanglement Renyi Entropies of Disjoint Intervals in
  AdS/CFT}},  \href{http://arxiv.org/abs/1303.7221}{{\tt arXiv:1303.7221}}.

\bibitem{Liu:2012eea}
H.~Liu and M.~Mezei, {\it {A Refinement of entanglement entropy and the number
  of degrees of freedom}},  {\em JHEP} {\bf 04} (2013) 162,
  [\href{http://arxiv.org/abs/1202.2070}{{\tt arXiv:1202.2070}}].

\bibitem{Bhattacharya:2014vja}
J.~Bhattacharya, V.~E. Hubeny, M.~Rangamani, and T.~Takayanagi, {\it
  {Entanglement density and gravitational thermodynamics}},  {\em Phys. Rev.}
  {\bf D91} (2015), no.~10 106009, [\href{http://arxiv.org/abs/1412.5472}{{\tt
  arXiv:1412.5472}}].

\bibitem{casini2011towards}
H.~Casini, M.~Huerta, and R.~C. Myers, {\it Towards a derivation of holographic
  entanglement entropy},  {\em Journal of High Energy Physics} {\bf 2011}
  (2011), no.~5 1--41.

\bibitem{Faulkner:2014jva}
T.~Faulkner, {\it {Bulk Emergence and the RG Flow of Entanglement Entropy}},
  {\em JHEP} {\bf 05} (2015) 033, [\href{http://arxiv.org/abs/1412.5648}{{\tt
  arXiv:1412.5648}}].

\bibitem{Faulkner:2015csl}
T.~Faulkner, R.~G. Leigh, and O.~Parrikar, {\it {Shape Dependence of
  Entanglement Entropy in Conformal Field Theories}},  {\em JHEP} {\bf 04}
  (2016) 088, [\href{http://arxiv.org/abs/1511.0517}{{\tt arXiv:1511.0517}}].

\bibitem{Lewkowycz:2013nqa}
A.~Lewkowycz and J.~Maldacena, {\it {Generalized gravitational entropy}},  {\em
  JHEP} {\bf 1308} (2013) 090, [\href{http://arxiv.org/abs/1304.4926}{{\tt
  arXiv:1304.4926}}].

\bibitem{Dong:2016hjy}
X.~Dong, A.~Lewkowycz, and M.~Rangamani, {\it {Deriving covariant holographic
  entanglement}},  \href{http://arxiv.org/abs/1607.0750}{{\tt
  arXiv:1607.0750}}.

\bibitem{Wolfetal}
M.~M. Wolf, F.~Verstraete, M.~B. Hastings, and J.~I. Cirac, {\it Area laws in
  quantum systems: Mutual information and correlations},  {\em Phys. Rev.
  Lett.} {\bf 100} (Feb, 2008) 070502.

\bibitem{headrick2007holographic}
M.~Headrick and T.~Takayanagi, {\it Holographic proof of the strong
  subadditivity of entanglement entropy},  {\em Physical Review D} {\bf 76}
  (2007), no.~10 106013.

\bibitem{Wald:1993nt}
R.~M. Wald, {\it {Black hole entropy is the Noether charge}},  {\em Phys. Rev.}
  {\bf D48} (1993) 3427--3431, [\href{http://arxiv.org/abs/gr-qc/9307038}{{\tt
  gr-qc/9307038}}].

\bibitem{Camps:2013zua}
J.~Camps, {\it {Generalized entropy and higher derivative Gravity}},  {\em
  JHEP} {\bf 03} (2014) 070, [\href{http://arxiv.org/abs/1310.6659}{{\tt
  arXiv:1310.6659}}].

\bibitem{Dong:2013qoa}
X.~Dong, {\it {Holographic Entanglement Entropy for General Higher Derivative
  Gravity}},  {\em JHEP} {\bf 01} (2014) 044,
  [\href{http://arxiv.org/abs/1310.5713}{{\tt arXiv:1310.5713}}].

\bibitem{Faulkner:2013ana}
T.~Faulkner, A.~Lewkowycz, and J.~Maldacena, {\it {Quantum corrections to
  holographic entanglement entropy}},  {\em JHEP} {\bf 1311} (2013) 074,
  [\href{http://arxiv.org/abs/1307.2892}{{\tt arXiv:1307.2892}}].

\bibitem{Engelhardt:2013tra}
N.~Engelhardt and A.~C. Wall, {\it {Extremal Surface Barriers}},  {\em JHEP}
  {\bf 03} (2014) 068, [\href{http://arxiv.org/abs/1312.3699}{{\tt
  arXiv:1312.3699}}].

\bibitem{Balasubramanian:2014sra}
V.~Balasubramanian, B.~D. Chowdhury, B.~Czech, and J.~de~Boer, {\it
  {Entwinement and the emergence of spacetime}},  {\em JHEP} {\bf 01} (2015)
  048, [\href{http://arxiv.org/abs/1406.5859}{{\tt arXiv:1406.5859}}].

\bibitem{Lin:2016fqk}
J.~Lin, {\it {A Toy Model of Entwinement}},
  \href{http://arxiv.org/abs/1608.0204}{{\tt arXiv:1608.0204}}.

\bibitem{Radicevic:2016tlt}
D.~Radicevic, {\it {Entanglement Entropy and Duality}},
  \href{http://arxiv.org/abs/1605.0939}{{\tt arXiv:1605.0939}}.

\bibitem{Harlow:2016vwg}
D.~Harlow, {\it {The Ryu-Takayanagi Formula from Quantum Error Correction}},
  \href{http://arxiv.org/abs/1607.0390}{{\tt arXiv:1607.0390}}.

\bibitem{Czech:2012be}
B.~Czech, J.~L. Karczmarek, F.~Nogueira, and M.~Van~Raamsdonk, {\it {Rindler
  Quantum Gravity}},  {\em Class. Quant. Grav.} {\bf 29} (2012) 235025,
  [\href{http://arxiv.org/abs/1206.1323}{{\tt arXiv:1206.1323}}].

\bibitem{Swingle:2009bg}
B.~Swingle, {\it {Entanglement Renormalization and Holography}},  {\em Phys.
  Rev.} {\bf D86} (2012) 065007, [\href{http://arxiv.org/abs/0905.1317}{{\tt
  arXiv:0905.1317}}].

\bibitem{Donnelly:2011hn}
W.~Donnelly, {\it {Decomposition of entanglement entropy in lattice gauge
  theory}},  {\em Phys. Rev.} {\bf D85} (2012) 085004,
  [\href{http://arxiv.org/abs/1109.0036}{{\tt arXiv:1109.0036}}].

\bibitem{Casini:2013rba}
H.~Casini, M.~Huerta, and J.~A. Rosabal, {\it {Remarks on entanglement entropy
  for gauge fields}},  {\em Phys. Rev.} {\bf D89} (2014), no.~8 085012,
  [\href{http://arxiv.org/abs/1312.1183}{{\tt arXiv:1312.1183}}].

\bibitem{Bousso:2012sj}
R.~Bousso, S.~Leichenauer, and V.~Rosenhaus, {\it {Light-sheets and AdS/CFT}},
  {\em Phys. Rev.} {\bf D86} (2012) 046009,
  [\href{http://arxiv.org/abs/1203.6619}{{\tt arXiv:1203.6619}}].

\bibitem{Czech:2012bh}
B.~Czech, J.~L. Karczmarek, F.~Nogueira, and M.~Van~Raamsdonk, {\it {The
  Gravity Dual of a Density Matrix}},  {\em Class. Quant. Grav.} {\bf 29}
  (2012) 155009, [\href{http://arxiv.org/abs/1204.1330}{{\tt
  arXiv:1204.1330}}].

\bibitem{Hubeny:2012wa}
V.~E. Hubeny and M.~Rangamani, {\it {Causal Holographic Information}},  {\em
  JHEP} {\bf 06} (2012) 114, [\href{http://arxiv.org/abs/1204.1698}{{\tt
  arXiv:1204.1698}}].

\bibitem{Entwedge}
M.~Headrick, V.~E. Hubeny, A.~Lawrence, and M.~Rangamani, {\it {Causality \&
  holographic entanglement entropy}},  {\em JHEP} {\bf 12} (2014) 162,
  [\href{http://arxiv.org/abs/1408.6300}{{\tt arXiv:1408.6300}}].

\bibitem{Dong:2016eik}
X.~Dong, D.~Harlow, and A.~C. Wall, {\it {Bulk Reconstruction in the
  Entanglement Wedge in AdS/CFT}},  \href{http://arxiv.org/abs/1601.0541}{{\tt
  arXiv:1601.0541}}.

\bibitem{AMPS}
A.~Almheiri, D.~Marolf, J.~Polchinski, and J.~Sully, {\it {Black Holes:
  Complementarity or Firewalls?}},  {\em JHEP} {\bf 02} (2013) 062,
  [\href{http://arxiv.org/abs/1207.3123}{{\tt arXiv:1207.3123}}].

\bibitem{Maldacena:2013xja}
J.~Maldacena and L.~Susskind, {\it {Cool horizons for entangled black holes}},
  {\em Fortsch. Phys.} {\bf 61} (2013) 781--811,
  [\href{http://arxiv.org/abs/1306.0533}{{\tt arXiv:1306.0533}}].

\bibitem{VanRaamsdonk:2013sza}
M.~Van~Raamsdonk, {\it {Evaporating Firewalls}},  {\em JHEP} {\bf 11} (2014)
  038, [\href{http://arxiv.org/abs/1307.1796}{{\tt arXiv:1307.1796}}].

\bibitem{Marolf:2013dba}
D.~Marolf and J.~Polchinski, {\it {Gauge/Gravity Duality and the Black Hole
  Interior}},  {\em Phys. Rev. Lett.} {\bf 111} (2013) 171301,
  [\href{http://arxiv.org/abs/1307.4706}{{\tt arXiv:1307.4706}}].

\bibitem{Marolf:2012xe}
D.~Marolf and A.~C. Wall, {\it {Eternal Black Holes and Superselection in
  AdS/CFT}},  {\em Class. Quant. Grav.} {\bf 30} (2013) 025001,
  [\href{http://arxiv.org/abs/1210.3590}{{\tt arXiv:1210.3590}}].

\bibitem{Papadodimas:2015jra}
K.~Papadodimas and S.~Raju, {\it {Remarks on the necessity and implications of
  state-dependence in the black hole interior}},  {\em Phys. Rev.} {\bf D93}
  (2016), no.~8 084049, [\href{http://arxiv.org/abs/1503.0882}{{\tt
  arXiv:1503.0882}}].

\bibitem{blanco2013relative}
D.~D. Blanco, H.~Casini, L.-Y. Hung, and R.~C. Myers, {\it Relative entropy and
  holography},  {\em Journal of High Energy Physics} {\bf 2013} (2013), no.~8
  1--65.

\bibitem{Lashkari:2014kda}
N.~Lashkari, C.~Rabideau, P.~Sabella-Garnier, and M.~Van~Raamsdonk, {\it
  {Inviolable energy conditions from entanglement inequalities}},  {\em JHEP}
  {\bf 06} (2015) 067, [\href{http://arxiv.org/abs/1412.3514}{{\tt
  arXiv:1412.3514}}].

\bibitem{Casini:2004bw}
H.~Casini and M.~Huerta, {\it {A Finite entanglement entropy and the
  c-theorem}},  {\em Phys.Lett.} {\bf B600} (2004) 142--150,
  [\href{http://arxiv.org/abs/hep-th/0405111}{{\tt hep-th/0405111}}].

\bibitem{Myers:2010tj}
R.~C. Myers and A.~Sinha, {\it {Holographic c-theorems in arbitrary
  dimensions}},  {\em JHEP} {\bf 01} (2011) 125,
  [\href{http://arxiv.org/abs/1011.5819}{{\tt arXiv:1011.5819}}].

\bibitem{Casini:2016fgb}
H.~Casini, I.~S. Landea, and G.~Torroba, {\it {The g-theorem and quantum
  information theory}},  \href{http://arxiv.org/abs/1607.0039}{{\tt
  arXiv:1607.0039}}.

\bibitem{Casini:2015woa}
H.~Casini, M.~Huerta, R.~C. Myers, and A.~Yale, {\it {Mutual information and
  the F-theorem}},  {\em JHEP} {\bf 10} (2015) 003,
  [\href{http://arxiv.org/abs/1506.0619}{{\tt arXiv:1506.0619}}].

\bibitem{faulkner2014gravitation}
T.~Faulkner, M.~Guica, T.~Hartman, R.~C. Myers, and M.~Van~Raamsdonk, {\it
  {Gravitation from Entanglement in Holographic CFTs}},  {\em JHEP} {\bf 03}
  (2014) 051, [\href{http://arxiv.org/abs/1312.7856}{{\tt arXiv:1312.7856}}].

\bibitem{Lashkari:2013koa}
N.~Lashkari, M.~B. McDermott, and M.~Van~Raamsdonk, {\it {Gravitational
  dynamics from entanglement 'thermodynamics'}},  {\em JHEP} {\bf 04} (2014)
  195, [\href{http://arxiv.org/abs/1308.3716}{{\tt arXiv:1308.3716}}].

\bibitem{Iyer:1994ys}
V.~Iyer and R.~M. Wald, {\it {Some properties of Noether charge and a proposal
  for dynamical black hole entropy}},  {\em Phys. Rev.} {\bf D50} (1994)
  846--864, [\href{http://arxiv.org/abs/gr-qc/9403028}{{\tt gr-qc/9403028}}].

\bibitem{swingle2014universality}
B.~Swingle and M.~Van~Raamsdonk, {\it Universality of gravity from
  entanglement},  {\em arXiv preprint arXiv:1405.2933} (2014).

\bibitem{hollands2013stability}
S.~Hollands and R.~M. Wald, {\it Stability of black holes and black branes},
  {\em Communications in Mathematical Physics} {\bf 321} (2013), no.~3
  629--680.

\bibitem{Lashkari:2015hha}
N.~Lashkari and M.~Van~Raamsdonk, {\it {Canonical Energy is Quantum Fisher
  Information}},  \href{http://arxiv.org/abs/1508.0089}{{\tt arXiv:1508.0089}}.

\bibitem{Lashkari:2016idm}
N.~Lashkari, J.~Lin, H.~Ooguri, B.~Stoica, and M.~Van~Raamsdonk, {\it
  {Gravitational Positive Energy Theorems from Information Inequalities}},
  \href{http://arxiv.org/abs/1605.0107}{{\tt arXiv:1605.0107}}.

\bibitem{Almheiri:2014lwa}
A.~Almheiri, X.~Dong, and D.~Harlow, {\it {Bulk Locality and Quantum Error
  Correction in AdS/CFT}},  {\em JHEP} {\bf 04} (2015) 163,
  [\href{http://arxiv.org/abs/1411.7041}{{\tt arXiv:1411.7041}}].

\bibitem{Pastawski:2015qua}
F.~Pastawski, B.~Yoshida, D.~Harlow, and J.~Preskill, {\it {Holographic quantum
  error-correcting codes: Toy models for the bulk/boundary correspondence}},
  {\em JHEP} {\bf 06} (2015) 149, [\href{http://arxiv.org/abs/1503.0623}{{\tt
  arXiv:1503.0623}}].

\bibitem{Hayden:2011ag}
P.~Hayden, M.~Headrick, and A.~Maloney, {\it {Holographic Mutual Information is
  Monogamous}},  {\em Phys.Rev.} {\bf D87} (2013), no.~4 046003,
  [\href{http://arxiv.org/abs/1107.2940}{{\tt arXiv:1107.2940}}].

\bibitem{Bao:2015bfa}
N.~Bao, S.~Nezami, H.~Ooguri, B.~Stoica, J.~Sully, and M.~Walter, {\it {The
  Holographic Entropy Cone}},  {\em JHEP} {\bf 09} (2015) 130,
  [\href{http://arxiv.org/abs/1505.0783}{{\tt arXiv:1505.0783}}].

\bibitem{Hartman:2013qma}
T.~Hartman and J.~Maldacena, {\it {Time Evolution of Entanglement Entropy from
  Black Hole Interiors}},  {\em JHEP} {\bf 05} (2013) 014,
  [\href{http://arxiv.org/abs/1303.1080}{{\tt arXiv:1303.1080}}].

\bibitem{Liu:2013iza}
H.~Liu and S.~J. Suh, {\it {Entanglement Tsunami: Universal Scaling in
  Holographic Thermalization}},  {\em Phys. Rev. Lett.} {\bf 112} (2014)
  011601, [\href{http://arxiv.org/abs/1305.7244}{{\tt arXiv:1305.7244}}].

\bibitem{Shenker:2013pqa}
S.~H. Shenker and D.~Stanford, {\it {Black holes and the butterfly effect}},
  {\em JHEP} {\bf 03} (2014) 067, [\href{http://arxiv.org/abs/1306.0622}{{\tt
  arXiv:1306.0622}}].

\bibitem{Mezei:2016wfz}
M.~Mezei and D.~Stanford, {\it {On entanglement spreading in chaotic systems}},
   \href{http://arxiv.org/abs/1608.0510}{{\tt arXiv:1608.0510}}.

\bibitem{Czech:2015qta}
B.~Czech, L.~Lamprou, S.~McCandlish, and J.~Sully, {\it {Integral Geometry and
  Holography}},  {\em JHEP} {\bf 10} (2015) 175,
  [\href{http://arxiv.org/abs/1505.0551}{{\tt arXiv:1505.0551}}].

\bibitem{Czech:2016xec}
B.~Czech, L.~Lamprou, S.~McCandlish, B.~Mosk, and J.~Sully, {\it {A
  Stereoscopic Look into the Bulk}},  {\em JHEP} {\bf 07} (2016) 129,
  [\href{http://arxiv.org/abs/1604.0311}{{\tt arXiv:1604.0311}}].

\bibitem{deBoer:2016pqk}
J.~de~Boer, F.~M. Haehl, M.~P. Heller, and R.~C. Myers, {\it {Entanglement,
  Holography and Causal Diamonds}},  \href{http://arxiv.org/abs/1606.0330}{{\tt
  arXiv:1606.0330}}.

\bibitem{Hayden:2016cfa}
P.~Hayden, S.~Nezami, X.-L. Qi, N.~Thomas, M.~Walter, and Z.~Yang, {\it
  {Holographic duality from random tensor networks}},
  \href{http://arxiv.org/abs/1601.0169}{{\tt arXiv:1601.0169}}.

\end{thebibliography}\endgroup
\end{document}